\documentclass[structabstract]{aa}

\usepackage{afterpage}

\usepackage{graphicx}
\usepackage{epstopdf}

\usepackage{txfonts}
\usepackage{natbib}
\bibpunct{(}{)}{;}{a}{}{,}

\usepackage{mathbbol}

\begin{document}

\title{3D LTE spectral line formation with scattering in red giant stars}

\author{
W. Hayek\inst{1,2,3}
\and M. Asplund\inst{1}
\and R. Collet\inst{1}
\and {\AA}. Nordlund\inst{4}
}
\institute{
Max Planck Institut f\"ur Astrophysik, Karl-Schwarzschild-Str. 1, D-85741 Garching, Germany\\
\email{[hayek,asplund,remo]@mpa-garching.mpg.de}
\and
Research School of Astronomy \& Astrophysics, Cotter Road, Weston Creek 2611, Australia\\
\and
Astrophysics Group, School of Physics, University of Exeter, Stocker Road, Exeter, EX4 4QL\\
\and
Niels Bohr Institute, University of Copenhagen, Juliane Maries Vej 30, DK-2100 K{\o}benhavn {\O}, Denmark\\
\email{aake@nbi.dk}
}

\date{\today}

\abstract
{}
{We investigate the effects of coherent isotropic continuum scattering on the formation of spectral lines in local thermodynamic equilibrium (LTE) using 3D hydrodynamical and 1D hydrostatic model atmospheres of red giant stars.}
{Detailed radiative transfer with coherent and isotropic continuum scattering is computed for 3D hydrodynamical and 1D hydrostatic models of late-type stellar atmospheres using the \texttt{SCATE} code. Opacities are computed in LTE, while a coherent and isotropic scattering term is added to the continuum source function. We investigate the effects of scattering by comparing continuum flux levels, spectral line profiles and curves of growth for different species with calculations that treat scattering as absorption.}
{Rayleigh scattering is the dominant source of scattering opacity in the continuum of red giant stars. Photons may escape from deeper, hotter layers through scattering, resulting in significantly higher continuum flux levels beneath a wavelength of $\lambda\lesssim5000$\,{\AA}. The magnitude of the effect is determined by the importance of scattering opacity with respect to absorption opacity; we observe the largest changes in continuum flux at the shortest wavelengths and lowest metallicities; intergranular lanes of 3D models are more strongly affected than granules. Continuum scattering acts to increase the profile depth of LTE lines: continua gain more brightness than line cores due to their larger thermalization depth in hotter layers. We thus observe the strongest changes in line depth for high-excitation species and ionized species, which contribute significantly to photon thermalization through their absorption opacity near the continuum optical surface. Scattering desaturates the line profiles, leading to larger abundance corrections for stronger lines, which reach $-0.5$\,dex at 3000\,{\AA} for \ion{Fe}{II} lines in 3D with excitation potential $\chi=2$\,eV at $\mathrm{[Fe/H]}=-3.0$. The corrections are less severe for low-excitation lines, longer wavelengths, and higher metallicity. Velocity fields increase the effects of scattering by separating emission from granules and intergranular lanes in wavelength. 1D calculations exhibit similar scattering abundance corrections for weak lines, but those for strong lines are generally smaller compared to 3D models and depend on the choice of microturbulence.}
{Continuum scattering should be taken into account for computing realistic spectral line profiles at wavelengths $\lambda\lesssim4000$\,{\AA} in metal-poor giant stars. Profile shapes are strongly affected by velocity fields and horizontal inhomogeneities, requiring a treatment based on 3D hydrodynamical rather than classical 1D hydrostatic model atmospheres.}


\maketitle

\section{Introduction}

Spectral line formation is an important discipline in the field of stellar astronomy. It is used as a diagnostic tool for numerical models of stellar atmospheres by comparing their predictions with observations, as well as for measuring quantities of astrophysical interest, such as stellar parameters, chemical abundances, surface velocity fields and many more. Spectral lines sample the physical conditions in stellar atmospheres in a wide height range and can thereby reveal useful information about the atmospheric structure.

The physics of line formation involves many different processes that require a detailed treatment of the interaction between radiation and matter. Beside the transition itself, these processes include, e.g., Doppler-shifts through thermal and non-thermal particle motion, collisional interaction, photoionization etc. The physical state of matter depends in general not only on local conditions, such as the gas temperature and pressure, but also on conditions in other parts of the atmosphere that are connected through the radiation field. Ignoring such non-local coupling enables the approximation of local thermodynamic equilibrium (LTE), which allows a tremendous simplification of line formation computations and is therefore frequently used. However, LTE is in general a bad approximation, and it can be only applied to specific cases in which its validity has been proven or when only estimated quantities are needed. A more accurate non-LTE treatment involves detailed atomic models, which require a significant amount of additional physical data \citep[see, e.g., the discussion in][]{Asplund:2005}. Owing to its much lower complexity and computational demands, 3D LTE line formation is nevertheless a useful tool for cases that are known to show weak or vanishing departures from LTE, such as the often-used forbidden [\ion{O}{I}] line at 6300\,{\AA}.

Late-type stellar atmospheres are bounded by the underlying convection zone, which is responsible for horizontal inhomogeneities in gas temperature and pressure, and flow fields that affect the line profile shape through Doppler-shifts \citep[e.g.,][]{Nordlundetal:2009}. Accurate line formation computations need to take this 3D structure of the atmosphere into account to obtain correct line profiles. The complexity of 3D calculations is rewarded by the elimination of additional parameters such as microturbulence and macroturbulence \citep[e.g.][]{Asplundetal:2000b}.

Metal-poor halo stars are particularly interesting targets for stellar abundance analyses: they are among the oldest objects in the Galaxy and largely preserve the chemical composition of the ISM in their envelopes, allowing detailed studies of nucleosynthesis processes and Galactic chemical evolution \citep[e.g.,][]{Beersetal:2005}. Most abundance studies of metal-poor stars rely either on 1D LTE or on 1D non-LTE syntheses. The effects of 3D line formation have not yet been fully explored, although, e.g., \citet{Asplundetal:1999}, \citet{Colletetal:2007} and \citet{Beharaetal:2010} have shown that the temperature stratification and inhomogeneous structure of 3D model atmospheres have an important effect on line formation in metal-poor dwarf stars and giant stars. Molecular species, low-excitation lines and atomic minority species are thus particularly affected by 3D effects through their high temperature-sensitivity; molecules are important instruments for determining CNO abundances of such stars. Full 3D non-LTE studies of lithium abundances in metal-poor stars have been presented by \citet{Asplundetal:2003}, \citet{Barklemetal:2003}, \citet{Cayreletal:2007} and \citet{Sbordoneetal:2010}, following earlier work of \citet{Kiselman:1997} and \citet{Uitenbroek:1998} for the Sun.

An important issue for the accuracy of abundance analyses of metal-poor giants is the role of Rayleigh scattering opacity for continuum formation in the blue and UV wavelength regions, where transitions of many important elements are found. Such scattering processes can increase the thermalization depth of photons, which may escape from deeper, hotter layers in the star, resulting in higher continuum flux. Treating continuum scattering as absorption instead of computing radiative transfer with scattering in 1D chemical abundance analyses leads to significantly larger abundances and trends with transition wavelength \citep{Cayreletal:2004,Bihainetal:2004,Laietal:2008}.

We investigate the effects of coherent scattering in the background continuum by computing 3D LTE line profiles for a selection of typical transitions, including neutral and singly ionized iron, as well as various molecular species. Section~\ref{sec:theory} discusses important aspects of the underlying radiative transfer model for spectral line formation, which builds on 3D time-dependent radiation-hydrodynamical model atmospheres of red giant stars discussed in Sect.~\ref{sec:3Datmo}. We analyze continuum formation with scattering in Sect.~\ref{sec:3Dcont}, investigate its effects on 3D LTE line formation in Sect.~\ref{sec:3Dline} and derive 3D$-$3D abundance corrections for typical spectral lines in Sect.~\ref{sec:scatcorr3D}. 3D$-$1D abundance corrections computed with continuum scattering and treating scattering as absorption are compared in Sect.~\ref{sec:scatcorr3D1D}; we conclude in Sect.~\ref{sec:conclusion}.

\section{Line formation with continuum scattering}\label{sec:theory}

We solve the 3D time-independent radiative transfer problem for stellar spectral lines and a background continuum with coherent isotropic scattering. Excitation level populations, ionization and molecule formation needed to derive gas opacities are computed assuming local thermodynamic equilibrium (LTE); the line profiles include collisional broadening by neutral hydrogen and Doppler-shifts through macroscopic velocity fields (see Sect.~\ref{sec:3Dlinforcode} and Appendix~\ref{sec:code} for a description of the numerical methods).

The time-independent radiative transfer equation evaluates the monochromatic specific intensity $I_{\nu}$ in direction $\vec{\hat{n}}$ at frequency $\nu$ in the observer's frame:
\begin{equation}
\frac{dI_{\nu}(\tau_{\nu})}{d\tau_{\nu}}=-I_{\nu}(\tau_{\nu})+S_{\nu}(\tau_{\nu}),
\label{eqn:RT}
\end{equation}
where $S_{\nu}$ is the monochromatic source function. The optical depth $d\tau_{\nu}$ of a photon path length $ds$ on a light ray is defined as
\begin{equation}
d\tau_{\nu}\equiv\left(\chi_{\nu}^{\mathrm{l}}+\chi^{\mathrm{c}}\right)ds,
\label{eqn:tau}
\end{equation}
with the line opacity $\chi_{\nu}^{\mathrm{l}}$ and the continuum opacity $\chi^{\mathrm{c}}$. The source function $S_{\nu}$ at optical depth $\tau_{\nu}$ includes contributions from spectral lines and continuum processes:
\begin{equation}
S_{\nu}=\frac{\chi_{\nu}^{\mathrm{l}}}{\chi_{\nu}^{\mathrm{l}}+\chi^{\mathrm{c}}}B+\frac{\chi^{\mathrm{c}}}{\chi_{\nu}^{\mathrm{l}}+\chi^{\mathrm{c}}}\left(\left[1-\epsilon^{\mathrm{c}}\right]J_{\nu}+\epsilon^{\mathrm{c}}B\right);
\label{eqn:sfct}
\end{equation}
$B$ is the Planck function, $\epsilon^{\mathrm{c}}$ is the continuum photon destruction probability and $J_{\nu}$ is the monochromatic mean intensity.
Note that frequency subscripts have been dropped in Eq.~(\ref{eqn:RT}) through Eq.~(\ref{eqn:sfct}) for all quantities that depend only weakly on frequency across a line profile and are therefore assumed constant in our computation.

The line opacity $\chi_{\nu}^{\mathrm{l}}$ may include contributions from several transitions of different species. At each point in the atmosphere, the individual line profiles are evaluated in the local rest frame of the gas around the center frequency $\nu'_{0}$ of the transition. In the observer's frame, the profiles appear Doppler-shifted through the non-relativistic velocity field $\vec{u}$ predicted by the 3D hydrodynamical model atmosphere. The line center frequency $\nu_{0}$ is then given by
\begin{equation}
\nu_{0}=\left(1+\frac{\vec{\hat{n}\cdot u}}{c}\right)\nu'_{0}
\label{eqn:Doppler}
\end{equation}
in the observer's frame for a light ray in direction $\vec{\hat{n}}$ \citep[see, e.g.,][]{Mihalasetal:1984}; $c$ is the speed of light. Relative to the frequency $\nu$ of photons that stream in the same direction as the flowing gas, the absorption profile appears blue-shifted; the same holds for line emission from the observer's point of view. In the solar atmosphere, upflowing gas in the bright granules dominates the emitted radiative flux \citep{Steinetal:1998}, and the resulting Doppler-shifts along the line of sight distort profile bisectors dominantly towards higher frequencies \citep[see the discussion in Sect.~\ref{sec:3Dline} and, e.g.,][]{Asplundetal:2000b}.

The opacity quotients $\chi_{\nu}^{\mathrm{l}}/\left(\chi_{\nu}^{\mathrm{l}}+\chi^{\mathrm{c}}\right)$ and $\chi^{\mathrm{c}}/\left(\chi_{\nu}^{\mathrm{l}}+\chi^{\mathrm{c}}\right)$ in Eq.~(\ref{eqn:sfct}) yield the probabilities that a photon was created by a line transition or by a continuous process. The direction-dependence of line opacity in the observer's frame through Doppler-shifts (Eq.~(\ref{eqn:Doppler})) induces a direction-dependence of these emission probabilities, and the combined source function $S_{\nu}$ becomes anisotropic.

The continuum photon destruction probability $\epsilon^{\mathrm{c}}$ is defined by the ratio
\begin{equation}
\epsilon^{\mathrm{c}}\equiv\frac{\kappa^{\mathrm{c}}}{\kappa^{\mathrm{c}}+\sigma^{\mathrm{c}}},
\end{equation}
with the continuous absorption coefficient $\kappa^{\mathrm{c}}$ and the continuous scattering coefficient $\sigma^{\mathrm{c}}$. In analogy to the above described opacity quotients, $\epsilon^{\mathrm{c}}$ can be interpreted as the probability that a photon was fed into the beam from the thermal pool rather than from a scattering event. In late-type stellar atmospheres, Rayleigh scattering on \ion{H}{I} atoms and electron scattering mainly contribute to continuous scattering opacity, while continuous absorption includes many bound-free and free-free transitions of hydrogen, helium and metals \citep[see, e.g., the discussion in][]{Hayeketal:2010}.

Continuous scattering may be treated as a coherent mechanism to very good approximation when the radiation field $J_{\nu}$ varies only slowly with frequency. In the presence of spectral lines, $J_{\nu}$ exhibits much stronger frequency-dependence, and scattering leads to a complex coupling of the radiation field in frequency and angle through Doppler-shifts \citep[see, e.g., discussions in][]{Mihalas:1978,Peraiah:2001}. \cite{Aueretal:1968a, Aueretal:1968b} analyzed the effects of non-coherent electron scattering on line formation in early-type stars. They assumed redistribution through thermal motion of the electrons, which occurs across a very wide frequency band compared to thermal redistribution in the line: the ratio between the thermal profile widths of electrons with mass $m_{\mathrm{e}}$ and line-forming atoms with mass $m_{\mathrm{atom}}$ is given by $\sqrt{m_{\mathrm{atom}}/m_{\mathrm{e}}}\approx42.7\sqrt{A}$, where $A$ is the atomic mass number. In late-type stars, Rayleigh scattering dominates scattering opacity at the continuum optical surface in the blue and UV wavelength regions, and the ratio of Doppler widths scales only with $\sqrt{A}\lesssim15$. Although electron scattering is an important continuous opacity source in early-type stars, \cite{Aueretal:1968b} find a relatively small effect of non-coherence on the wings of \ion{He}{II} profiles. We therefore expect the assumption of coherent scattering to be a reasonable approximation for the rather weakly scattering background continua of red giant stars.

Rayleigh scattering and electron scattering cross-sections depend on the scattering angle $\theta$ through the angular redistribution function $(1+\cos^{2}\theta)$, which may be neglected to good approximation \citep[see, e.g.,][]{Mihalas:1978}.

\section{3D radiation-hydrodynamical model atmospheres}\label{sec:3Datmo}

\begin{table*}[!htdp]
\caption{Stellar parameters of the 3D radiation-hydrodynamical model atmospheres.}
\centering
\begin{tabular}{cccc}
\hline\hline
$\left<T_{\mathrm{eff}}\right>$ [K] & $\log g$ [cgs] & $\mathrm{[Fe/H]}$ & $\mathrm{[}\alpha\mathrm{/Fe]}$\\
\hline
$5100$ & $2.2$ & $-3.0$ & $+0.4$\\
$5051$ & $2.2$ & $-2.0$ & $\phantom{+}0.0$\\
$4730$ & $2.2$ & $-1.0$ & $\phantom{+}0.0$\\
$5063$ & $2.2$ & $\phantom{-}0.0$ & $\phantom{+}0.0$\\
\hline
\end{tabular}
\label{tab:stellarparams}
\end{table*}

Various studies have shown that the formation of spectral lines in the atmospheres of late-type stars can be severely affected by the presence of inhomogeneities in the temperature structure and velocity fields \citep[e.g.,][]{Nordlund:1980,Dravinsetal:1990,Asplundetal:2000b,Steffenetal:2002,AllendePrietoetal:2002,Caffauetal:2008}. Synthetic line formation computations based on the current generation of 3D radiation-hydrodynamical models are capable of providing realistic predictions of the observations; see \citet{Pereiraetal:2009b,Pereiraetal:2009a} for the important case of the Sun. Classical 1D hydrostatic model atmospheres, which are still widely used, simply cannot achieve this degree of realism.

An important aspect in this context is the coupling of radiative transfer and hydrodynamics in 3D models, which leads to significantly cooler gas temperatures above the surface of metal-poor stars than predicted by 1D hydrostatic models \citep{Asplundetal:1999}. Low metallicity strongly reduces radiative heating in these objects, and the photospheric temperature stratification steepens towards an adiabatic gradient, while the assumption of radiative equilibrium that is inherent to 1D models keeps the gradient shallow. The lower temperatures of 3D models have a strong impact on the formation of molecules and thus on the predicted molecular line strengths \citep{Asplundetal:2001,Colletetal:2007,Beharaetal:2010}; molecules are frequently used for determining the abundances of carbon, nitrogen and oxygen \citep[e.g.,][]{Spiteetal:2006}.

We base our analysis on 3D radiation-hydrodynamical model atmospheres of red giant stars with similar effective temperatures, a surface gravity of $\log g=2.2$ (in cgs units) and metallicities\footnote{Metallicity is defined here in spectroscopic notation as the logarithmic iron abundance relative to the Sun, $\mathrm{[Fe/H]}\equiv\log_{10}(N_{\mathrm{Fe}}/N_{\mathrm{H}})_{\ast}-\log_{10}(N_{\mathrm{Fe}}/N_{\mathrm{H}})_{\sun}$.} between $\mathrm{[Fe/H]}=-3.0$ and $\mathrm{[Fe/H]}=0.0$ (see Table~\ref{tab:stellarparams}); the chemical abundance mixture adopts the solar composition of \citet{Asplundetal:2009}, scaled by metallicity and with a $+0.4$\,dex enhancement of $\alpha$-elements for the most metal-poor model.

The simulations were created with the \texttt{StaggerCode} \citep{Nordlundetal:1995}, which solves the coupled equations of compressible hydrodynamics and time-independent radiative transfer, computing radiative heating rates in LTE. Line-blanketing is approximated using the opacity binning method \citep{Nordlund:1982,Skartlien:2000} with 4 opacity bins. \citet{Colletetal:2011} demonstrate that the effects of coherent continuum scattering on the temperature structure of metal-poor giant stars with $\mathrm{[Fe/H]}=-3.0$ and $\mathrm{[Fe/H]}=-2.0$ may be reproduced to good approximation by removing the scattering contribution from the opacities in the optically thin parts of the atmosphere and using a Planck source function, which we applied to the models used for the analysis to reduce computational load. The deviations in average temperature from simulations with coherent scattering radiative transfer remain below $\lesssim50$\,K in the photospheres of the two most metal-poor giants; see \citet{Colletetal:2011} for further details.

The effective temperature of 3D radiation-hydrodynamical model atmospheres is an observable rather than a parameter: it is determined by the entropy of the inflowing gas at the bottom of the simulation, but exhibits some temporal variation due to the convective motions. The time-averaged $\left<T_{\mathrm{eff}}\right>$ therefore varies between the different model atmospheres, which is not critical for our differential investigation of continuum scattering effects. After the initial scaling of the model atmosphere to reach the desired stellar parameters, each simulation was run until the atmospheric stratification reached a quasi-steady state with little temporal variation.

\begin{figure}[htbp]
\centering
\includegraphics[width=\linewidth]{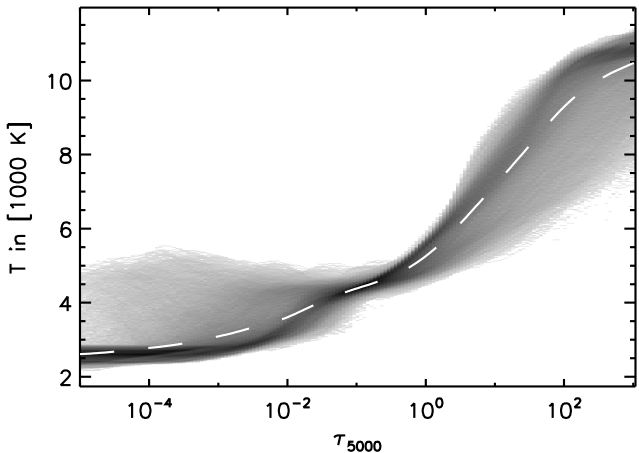}
\caption{Temperature distribution as a function of vertical continuum optical depth $\tau$ at 5000\,{\AA}, computed for an arbitrary snapshot of a 3D radiation-hydrodynamical model atmosphere with $\left<T_{\mathrm{eff}}\right>=5100$\,K, $\log g=2.2$ (cgs) and $\mathrm{[Fe/H]}=-3.0$. The dashed line indicates the average temperature at each optical depth.}
\label{fig:3Dmodel}
\end{figure}

Figure~\ref{fig:3Dmodel} shows the temperature distribution of an arbitrary snapshot of the $\mathrm{[Fe/H]}=-3.0$ model as a function of optical depth at 5000\,{\AA}. The atmospheres cover several pressure scale heights above and below the continuum optical surface of the stars to include the relevant line-forming regions and to avoid boundary effects on the granulation flow. The simulation domains have a resolution of $240\times240\times230$ grid points and assume periodic horizontal boundaries. Synthetic curves of growth were computed using time-series of simulation snapshots that span several periods of the fundamental pressure oscillation mode; a representative sample of line profiles is thereby obtained that includes statistical variation in the predicted line strengths through convective motions.

Recent calculations of OH line profiles by \citet{GonzalezHernandezetal:2010} based on 3D model atmospheres and radiative transfer with 12 opacity bins yield smaller 3D$-$1D LTE abundance corrections compared to their 4 bin models. The refined opacity treatment leads to a shallower temperature gradient, which reduces molecular equilibrium populations in the atmosphere. The OH abundances produced by the 12 bin 3D models are thus closer to those derived from 1D flux-equilibrium models. We test the importance of the opacity treatment for the most metal-poor giant star where 3D abundance effects are largest \citep[see][]{Colletetal:2007} using a 3D model constructed with a 12 bin opacity table. The opacities are sorted by Rosseland optical depth of the monochromatic optical surfaces and by wavelength, similar to \citet{GonzalezHernandezetal:2010}. The time-averaged effective temperature of the model is $\left<T_{\mathrm{eff}}\right>=5058$\,K and thus very close to the 4 bin model (see Table~\ref{tab:stellarparams}).

We also include \texttt{MARCS} model atmospheres with the same stellar parameters as the time-averaged 3D models (see Table \ref{tab:stellarparams}) in our analysis to compare the effects of scattering between 3D hydrodynamical and 1D hydrostatic atmospheres. Each 1D model is converted into a 3D box with zero gas velocity, horizontally homogeneous layers and with the same $T$-$\tau_{5000}$ relation; line formation is computed using the exact same method as in the 3D case.

\section{3D line formation computations}\label{sec:3Dlinforcode}

Synthetic continuum flux and line flux profiles are calculated using the \texttt{SCATE} code. The program first computes the monochromatic continuum source function with coherent isotropic scattering using a short characteristics-based radiative transfer solver with Gauss-Seidel-type approximate $\Lambda$-iteration \citep{TrujilloBuenoetal:1995}. A detailed description of the implementation can be found in Appendix~\ref{sec:code} and in \citet{Hayeketal:2010}. Continuum opacities and photon destruction probabilities are looked up in precomputed tables, line opacities are calculated during runtime. All quantities assume the LTE approximation.

The numerical method takes Doppler-shifts of line profiles into account, which influence absorption and line-to-continuum photon emission probabilities as a function of ray direction, gas velocity and frequency according to Eq.~(\ref{eqn:Doppler}). Including this inherent coupling is essential for correctly predicting the impact of scattering on the local radiation field and consequently on the profile shapes (see Sect.~\ref{sec:3Dline}). A minimum resolution in solid angle and frequency is furthermore needed to reproduce the effects of Doppler-shifts. For the case of continuum scattering with LTE lines in late-type stellar atmospheres, where gas flow reaches only moderate velocities, \citet{Carlson:1963} quadrature with 24 ray directions provides sufficient accuracy; computing radiative transfer with 48 angles changes the spatially averaged flux level by $\lesssim0.3$\,\%. Carlson quadrature also has the advantage of rotational invariance, which avoids directional bias for determining the local mean radiation field $J_{\nu}$. Line profiles for synthetic curves of growth are computed with typically 40 frequency points. We use linear (non-logarithmic) interpolation of quantities between the grid of the model atmosphere and the characteristics grid, as well as for the source function integral in the optically thin regime. Second-order interpolation is applied to the source function integral in optically thick regions to correctly recover the diffusion approximation.

Once the continuum source function has converged, it is passed to a second radiative transfer solver, which computes outgoing specific intensities on characteristics that span across the entire 3D atmosphere cube. This method has the advantage of reduced numerical diffusion with respect to the short characteristics method when angle-resolved surface intensities are needed. Local cubic logarithmic interpolation translates the relevant quantities onto the tilted characteristics grid. The radiative transfer equation is then solved along vertical columns using the \citet{Feautrier:1964} method to obtain surface intensities.

The grid resolution of the model atmosphere is an important issue for 3D line formation calculations \citep[see the discussion in][]{Asplundetal:2000a}. The axis spacing in the vertical dominates the accuracy of the solution, due to the strong temperature gradients near the stellar surface. Our calculations are based on a mesh with $120\times120\times230$ grid points, and we refine the vertical grid by automatic insertion of additional layers where needed to obtain robust intensity and flux profiles. Computation of the scattering source function excludes the deepest layers of the 3D model atmosphere, which are optically thick and dominated by local thermal radiation. The Feautrier-type solver starts at a continuum optical depth $\log_{10}\tau^{\mathrm{c}}\approx2$ on each individual ray to integrate the entire contribution function of continuum and line emission.

Stellar spectroscopy measures the radiative flux integrated over the stellar disk in most cases of interest. The monochromatic flux $F_{\nu,z}$ that leaves the surface on the visible hemisphere into the observer's direction is given by
\begin{equation}
F_{\nu,z}=\int_{\mu=0}^{1}\int_{\phi=0}^{2\pi}\left<I_{\nu}\right>(\mu,\phi)\mu d\mu d\phi,
\label{eqn:flux}
\end{equation}
where $\mu\equiv\cos\theta$ is the projection factor, $\theta$ is the polar angle, $\phi$ is the azimuth angle and $\left<I_{\nu}\right>$ is the horizontal average surface intensity of the simulation cube. The 3D hydrodynamical model atmospheres are interpreted as local statistical representations of stellar surface convection. Each angle pair $(\mu,\phi)$ for which radiative transfer is computed in the 3D cube corresponds to a different position on the disk. The flux integral (Eq.~(\ref{eqn:flux})) is thus equivalent to an integral over the stellar surface as seen by the observer; $F_{\nu,z}$ thereby automatically includes the limb darkening effect. Including stellar rotation in 3D spectral line formation computations requires additional consideration \citep{Ludwig:2007}, we therefore assume zero rotation ($v\sin i=0$) for simplicity.

We approximate the flux integral (Eq.~(\ref{eqn:flux})) using Gauss-Legendre quadrature for the polar angle and the trapezoid rule for the azimuth angle; $4\times4$ ray angles reproduce the surface flux with good accuracy. Doubling the number of polar angles or azimuth angles changes the spatial average flux level by~$\lesssim0.4$\,\%.

\section{The effects of scattering on the continuum flux}\label{sec:3Dcont}

\begin{figure*}[htbp]
\centering
\includegraphics[width=7cm]{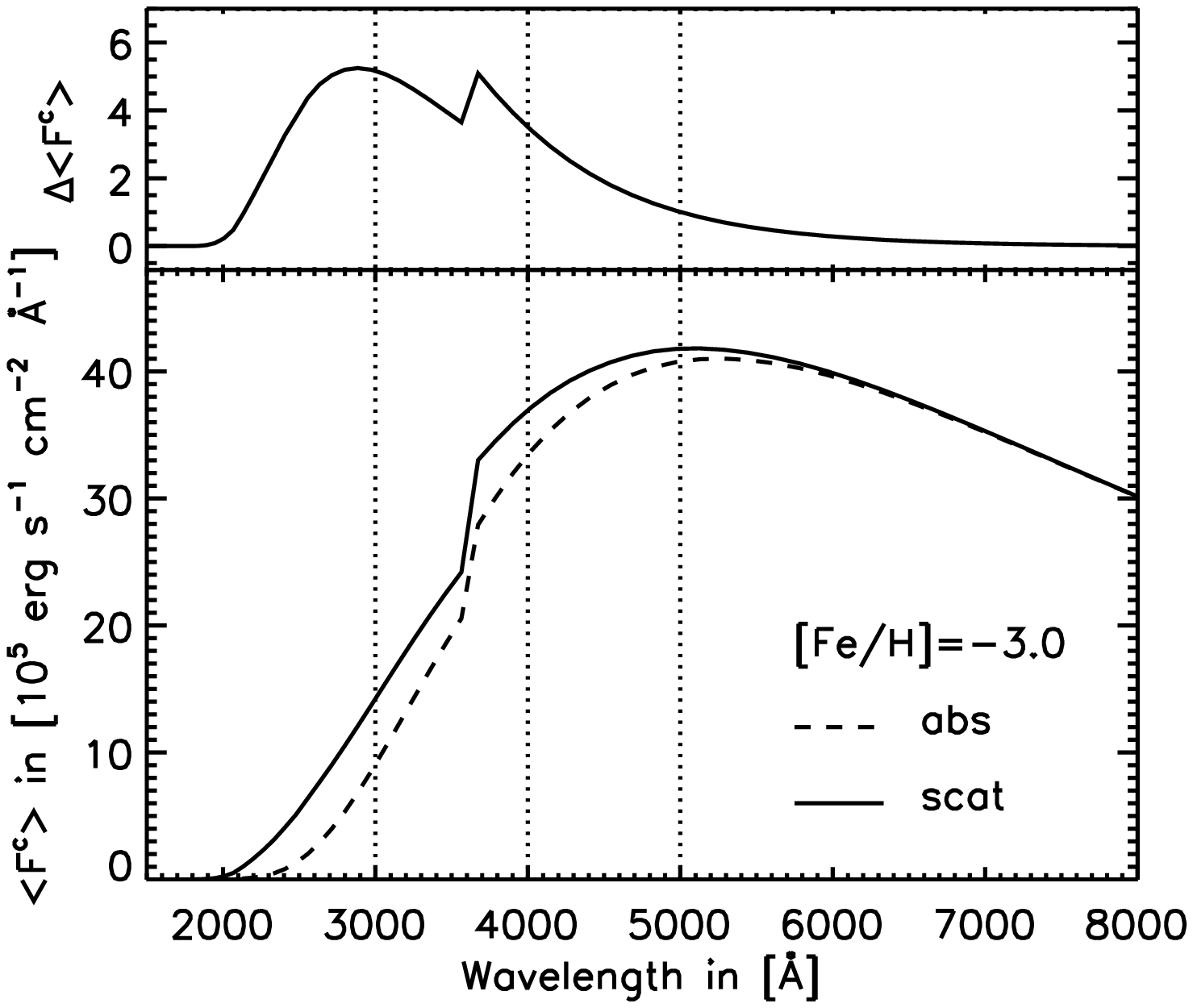}
\includegraphics[width=7cm]{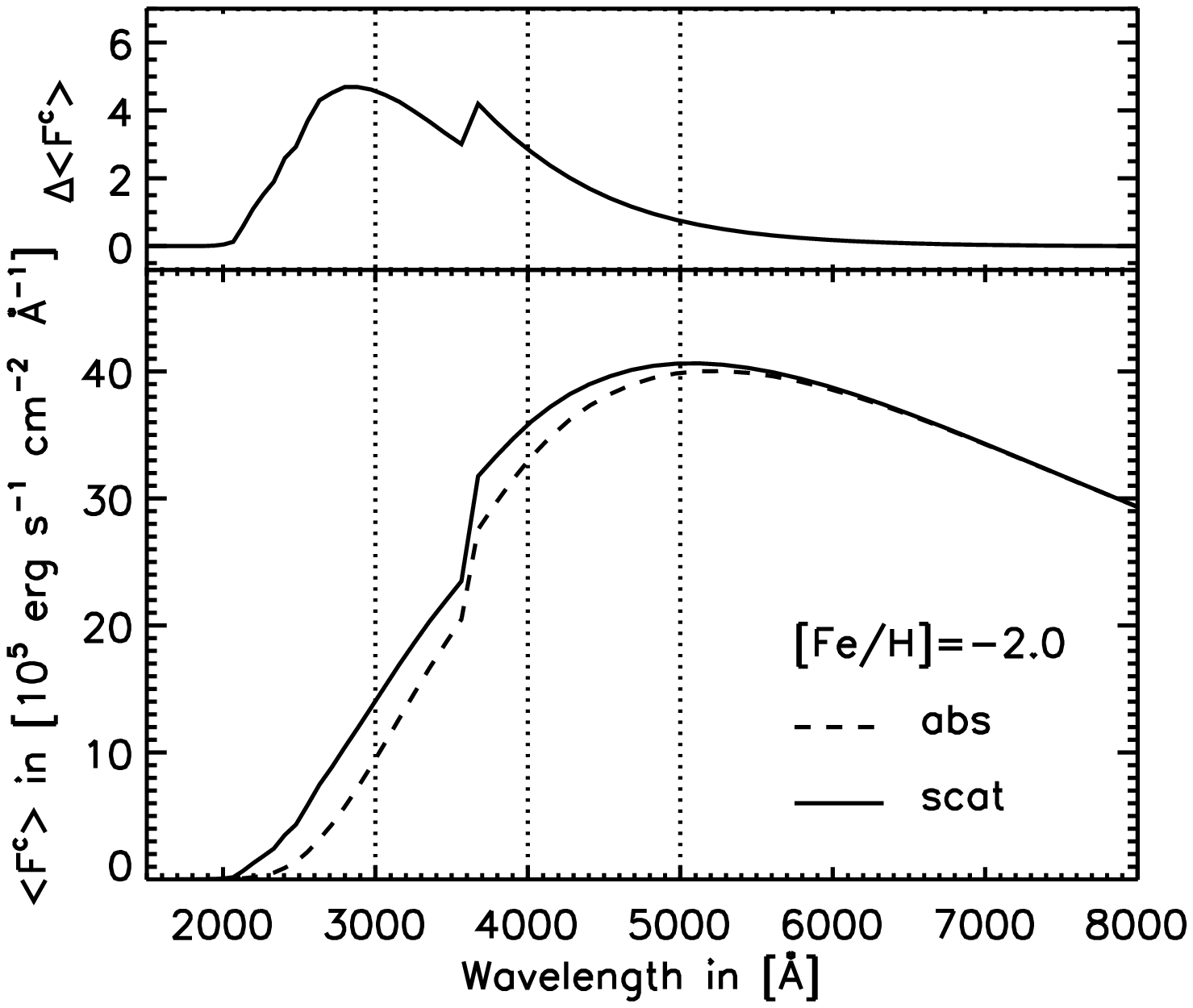}
\includegraphics[width=7cm]{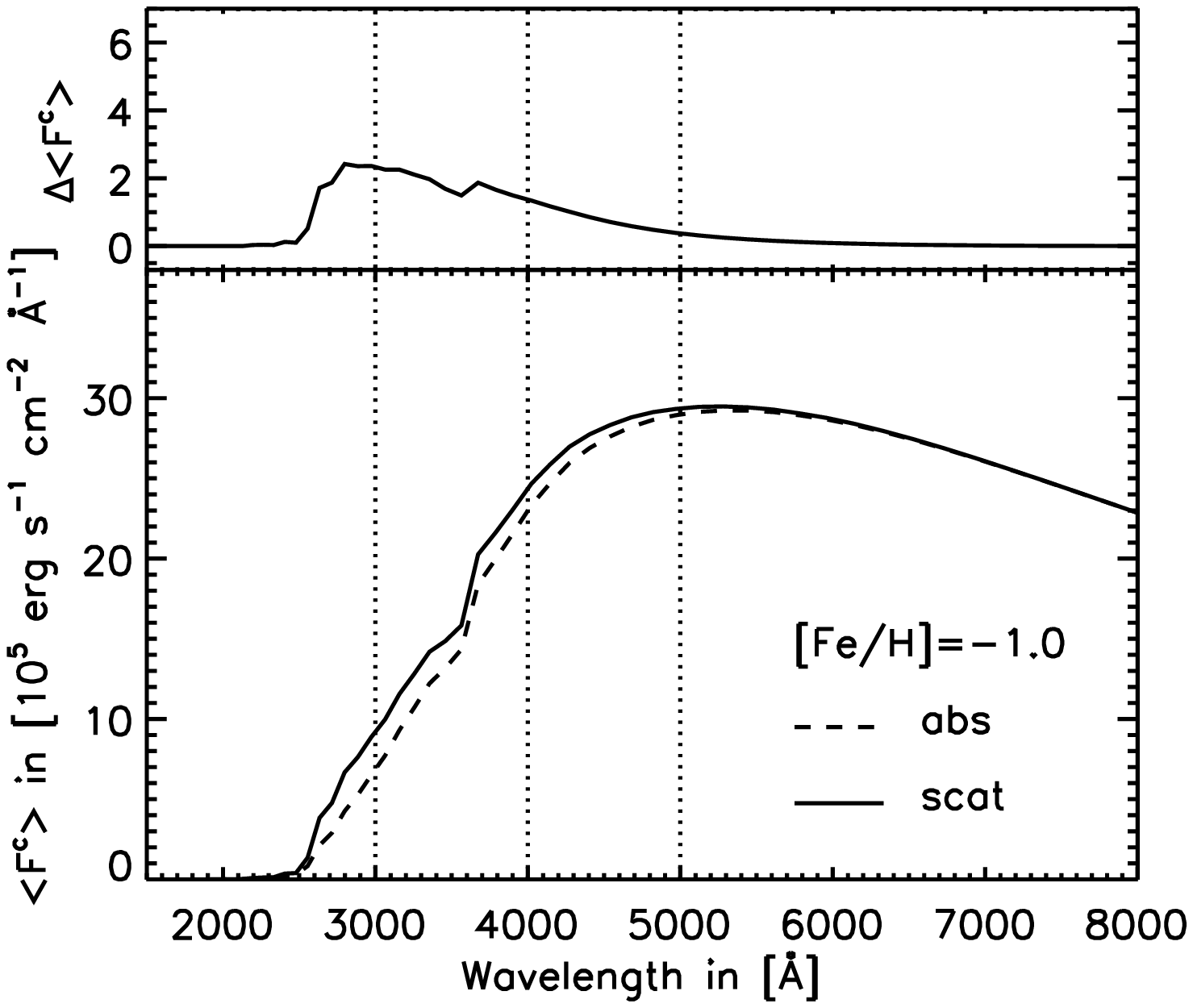}
\includegraphics[width=7cm]{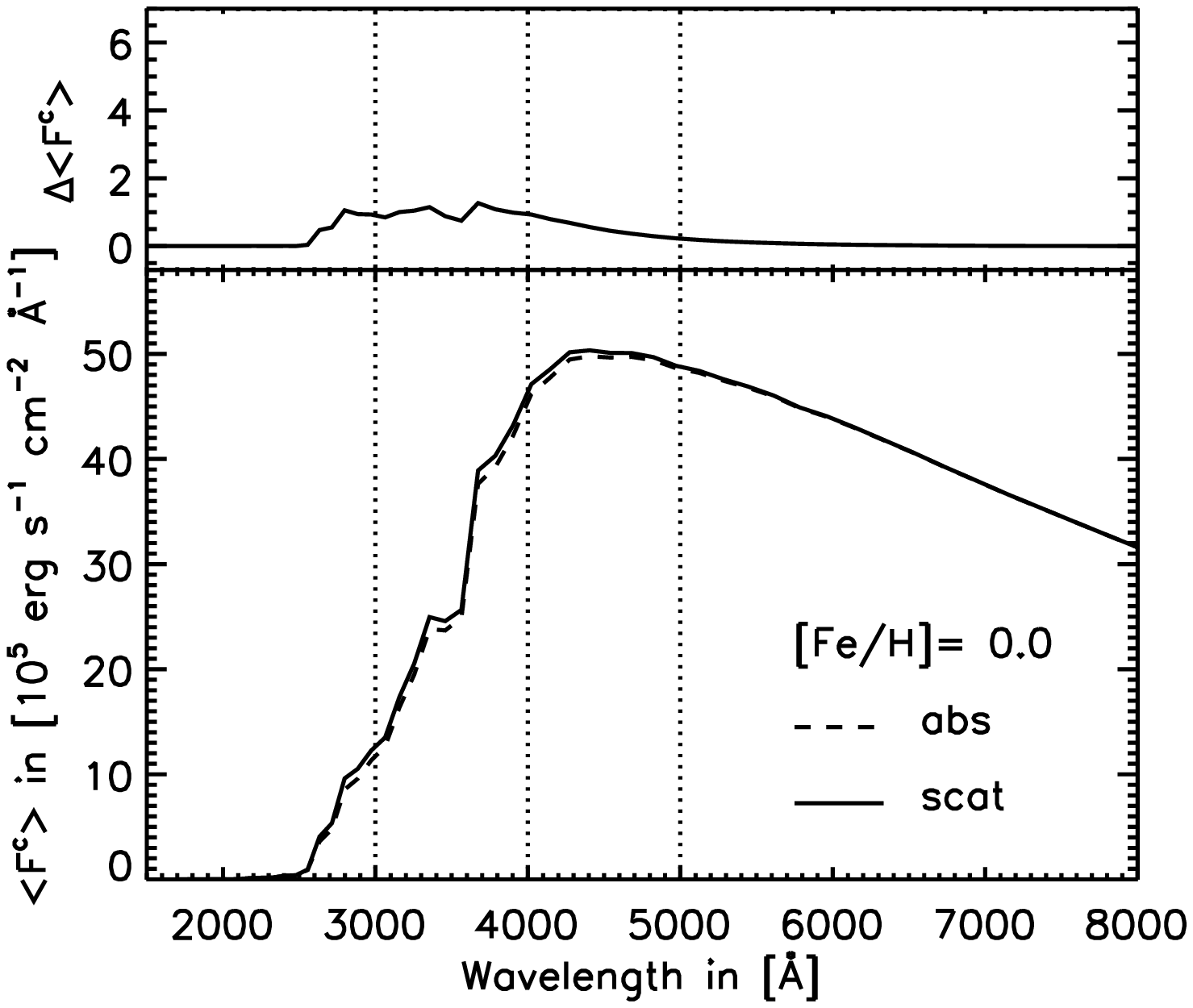}
\caption{Spatial and temporal averages of the continuum flux distribution $\left<F^{\mathrm{c}}\right>$ as a function of wavelength, computed for time sequences of the 3D model atmospheres with metallicity $-3.0\le\mathrm{[Fe/H]}\le0.0$ (upper left to lower right), treating scattering as absorption (dashed lines) and as coherent scattering (solid lines). The upper panel of each plot shows the deviation $\Delta\left<F^{\mathrm{c}}\right>$ of the coherent scattering cases from the continuum flux distribution where scattering is treated as absorption; vertical dotted lines indicate $\lambda=3000$\,{\AA}, 4000\,{\AA} and 5000\,{\AA} where Fe line profiles were computed. Note the Balmer jump at 3647\,{\AA}.}
\label{fig:contflx}
\end{figure*}

\begin{figure*}[htbp]
\centering
\includegraphics[width=7cm]{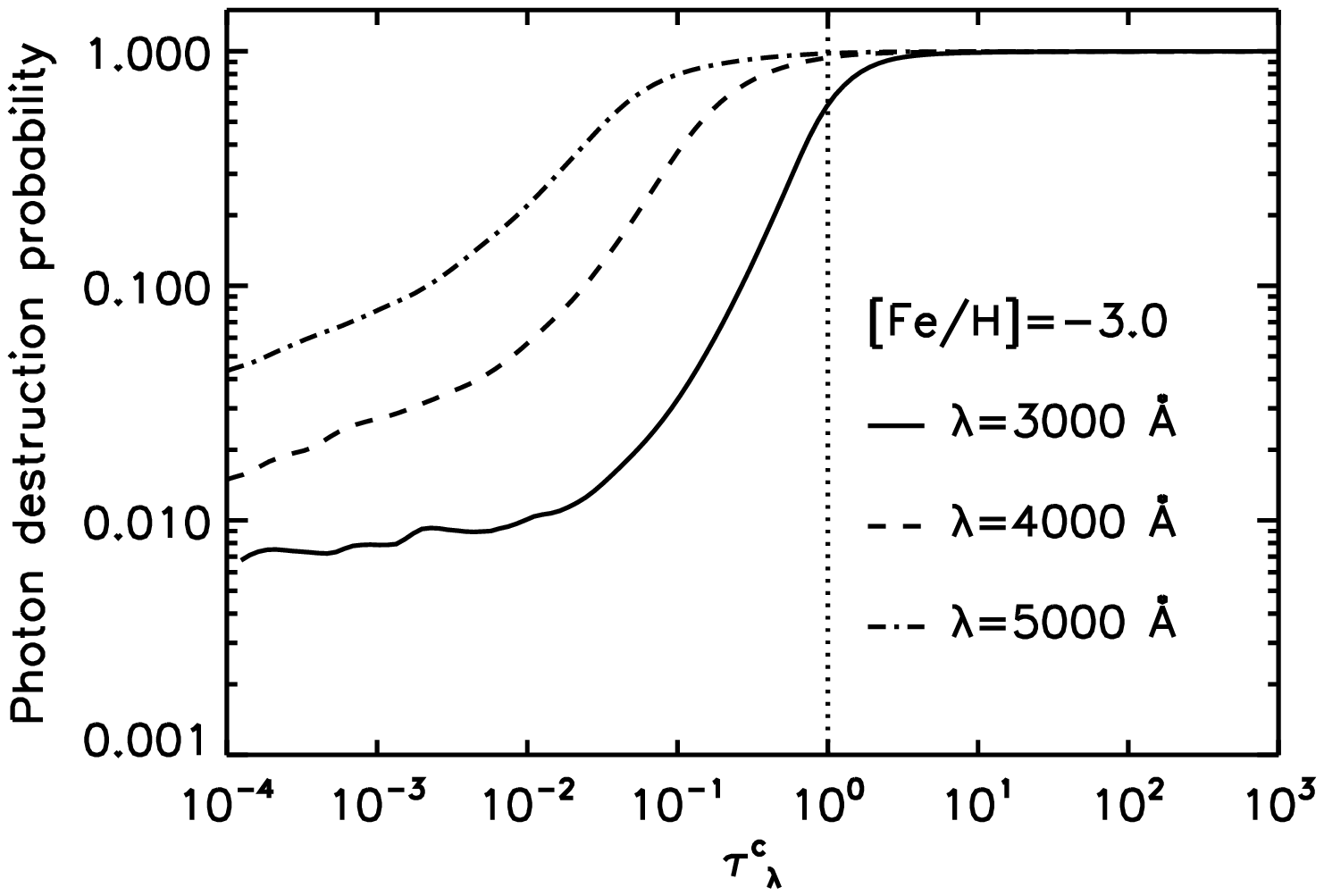}
\includegraphics[width=7cm]{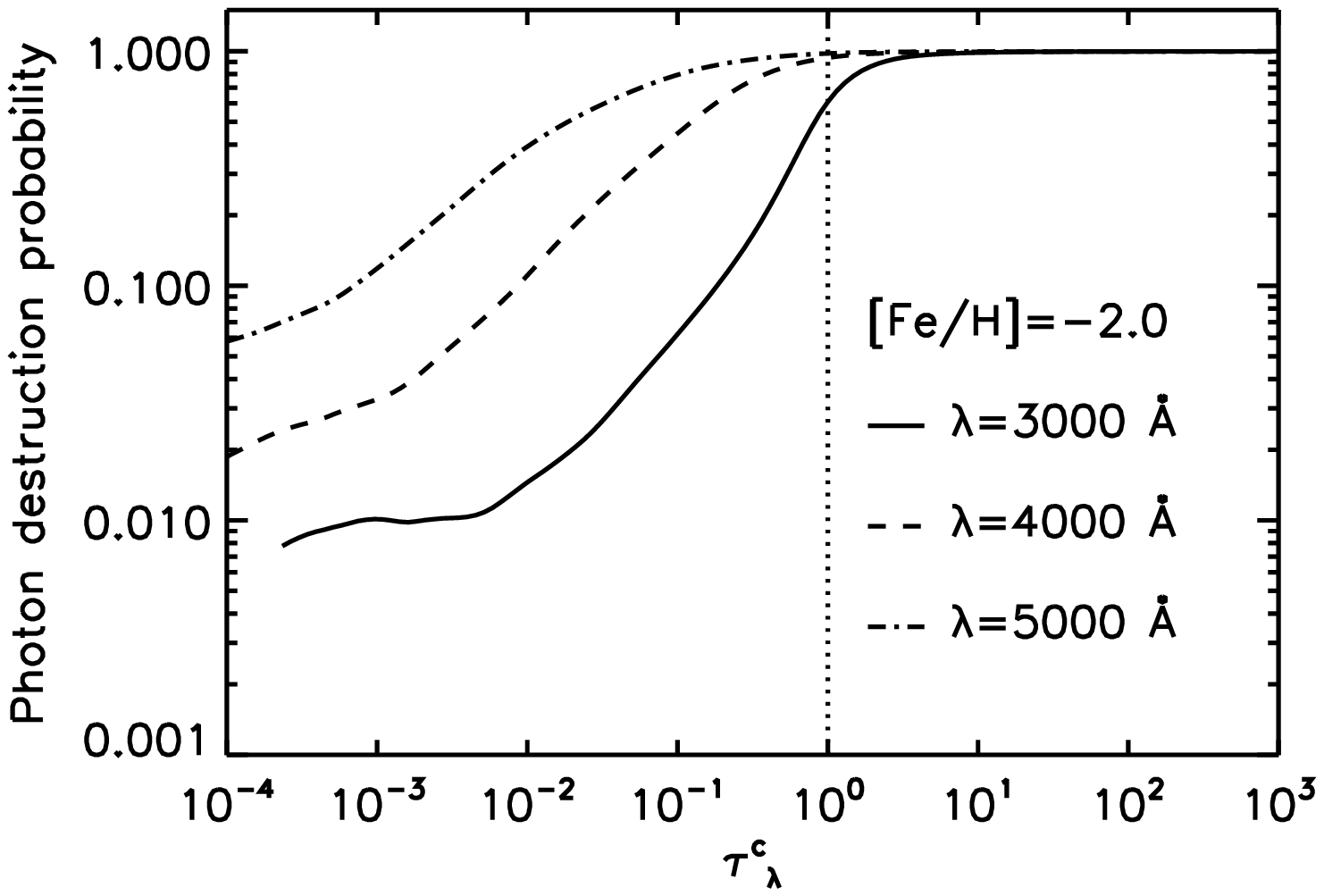}
\includegraphics[width=7cm]{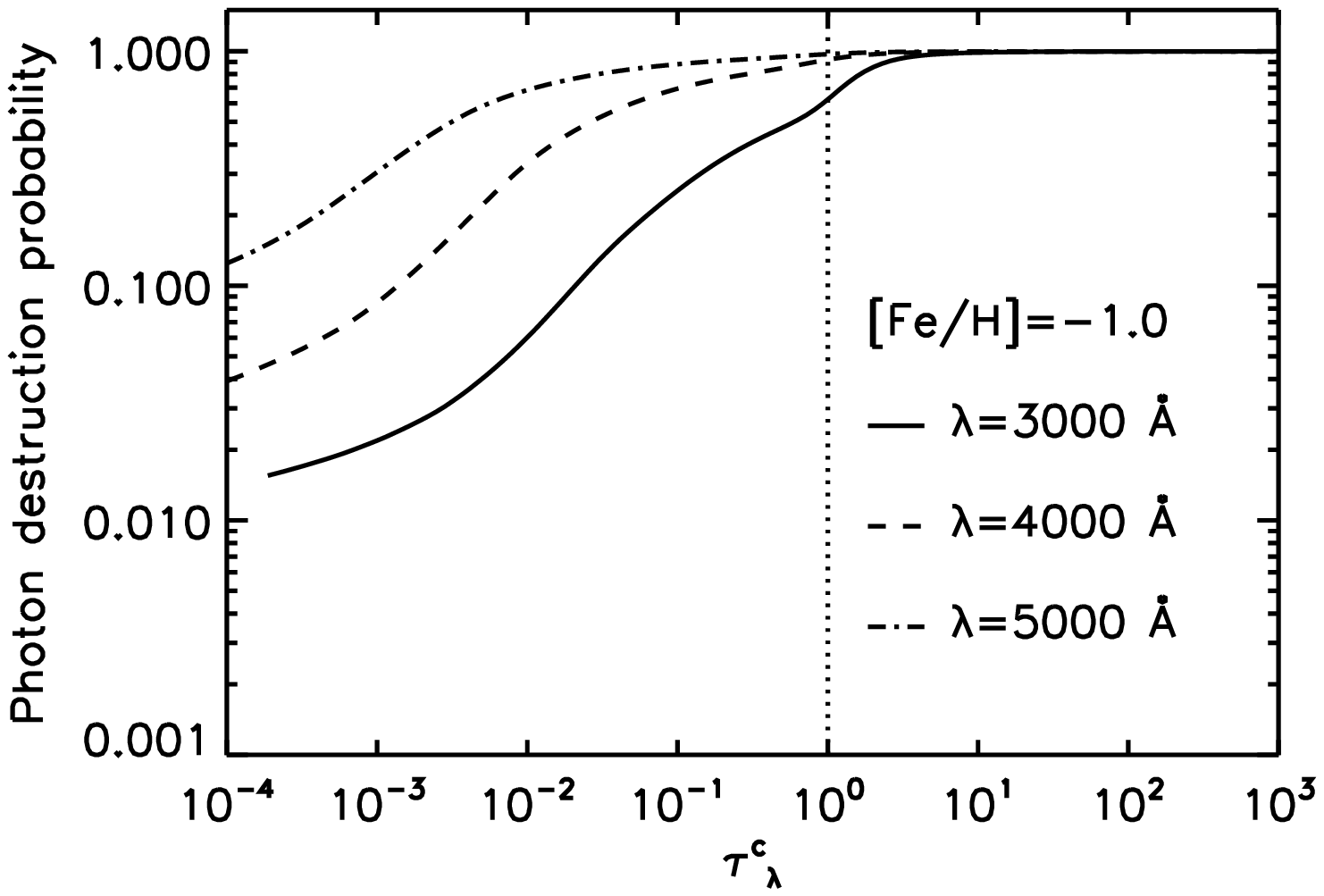}
\includegraphics[width=7cm]{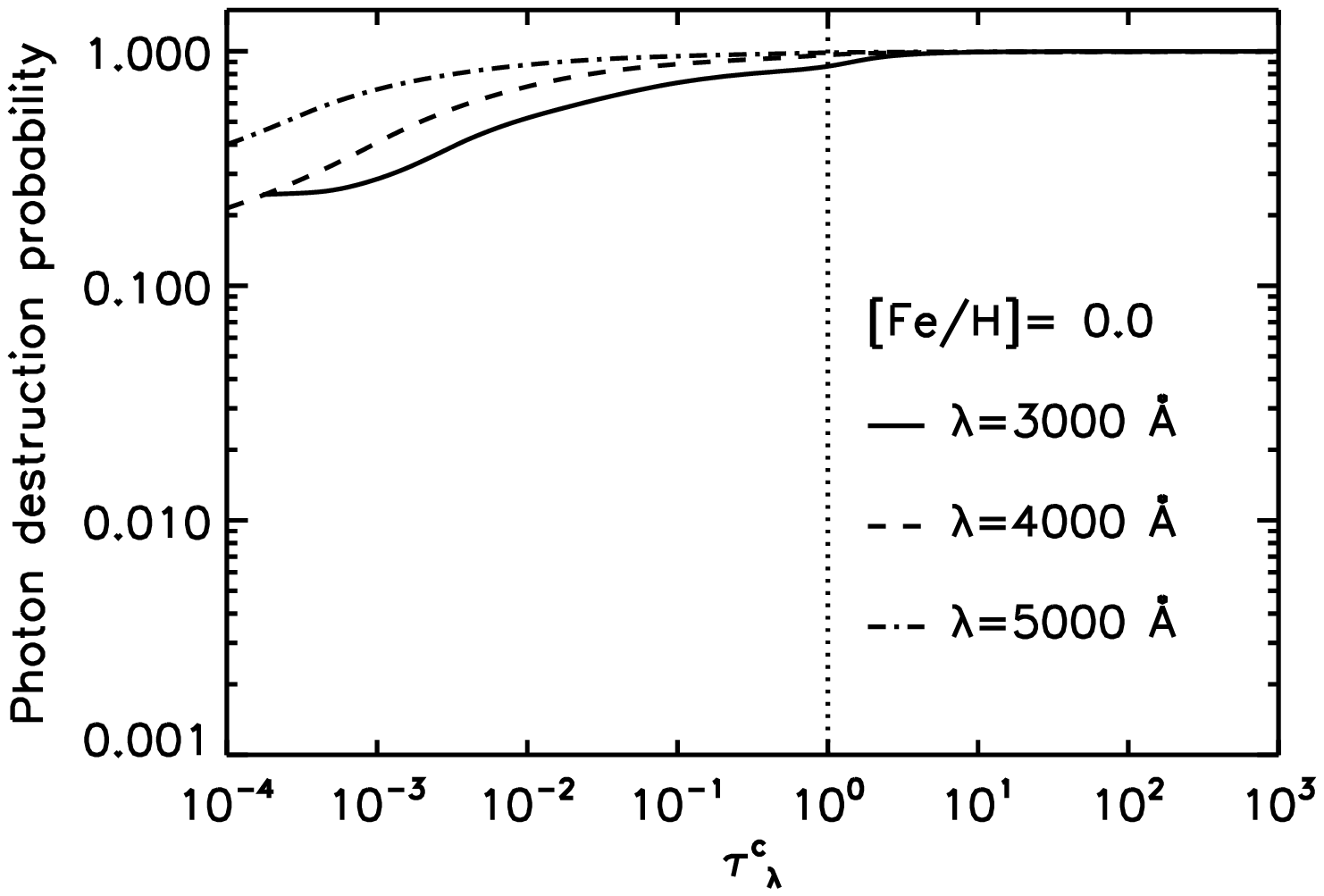}
\caption{Continuum photon destruction probabilities for arbitrary snapshots of the 3D model atmospheres with $\mathrm{[Fe/H]}=-3.0$ (upper left panel), $\mathrm{[Fe/H]}=-2.0$ (upper right panel), $\mathrm{[Fe/H]}=-1.0$ (lower left panel), and $\mathrm{[Fe/H]}=0.0$ (lower right panel), averaged over surfaces of constant monochromatic continuum optical depth $\tau^{\mathrm{c}}_{\lambda}$ for wavelengths $\lambda=3000$\,{\AA} (solid lines), $\lambda=4000$\,{\AA} (dashed lines) and $\lambda=5000$\,{\AA} (dot-dashed lines). The monochromatic optical surfaces at $\tau^{\mathrm{c}}_{\lambda}=1$ are marked by vertical dotted lines.}
\label{fig:epstau}
\end{figure*}

\begin{figure*}[htbp]
\centering
\includegraphics[width=12cm]{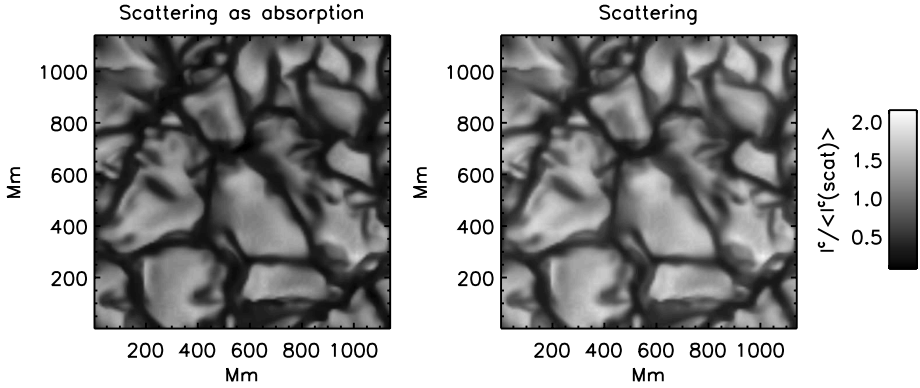}
\caption{Continuum surface intensities in the disk center at 3000\,{\AA} computed for an arbitrary snapshot of the 3D model with $\mathrm{[Fe/H]}=-3.0$, treating scattering as absorption (left panel) and as coherent scattering (right panel). Intensities are normalized to the average continuum intensity with coherent scattering.}
\label{fig:surfint}
\end{figure*}

\begin{figure*}[htbp]
\centering
\includegraphics[width=7cm]{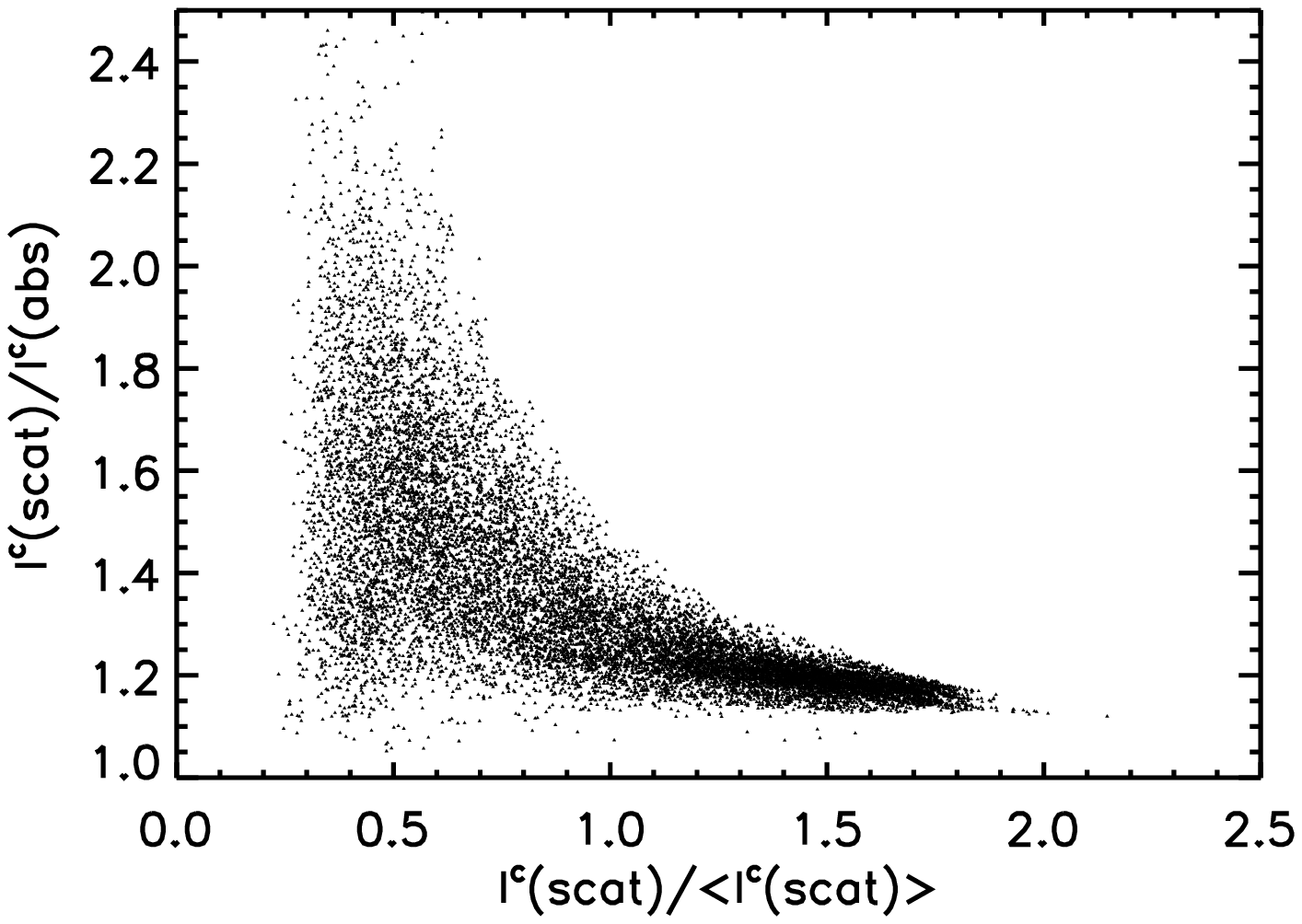}
\includegraphics[width=7cm]{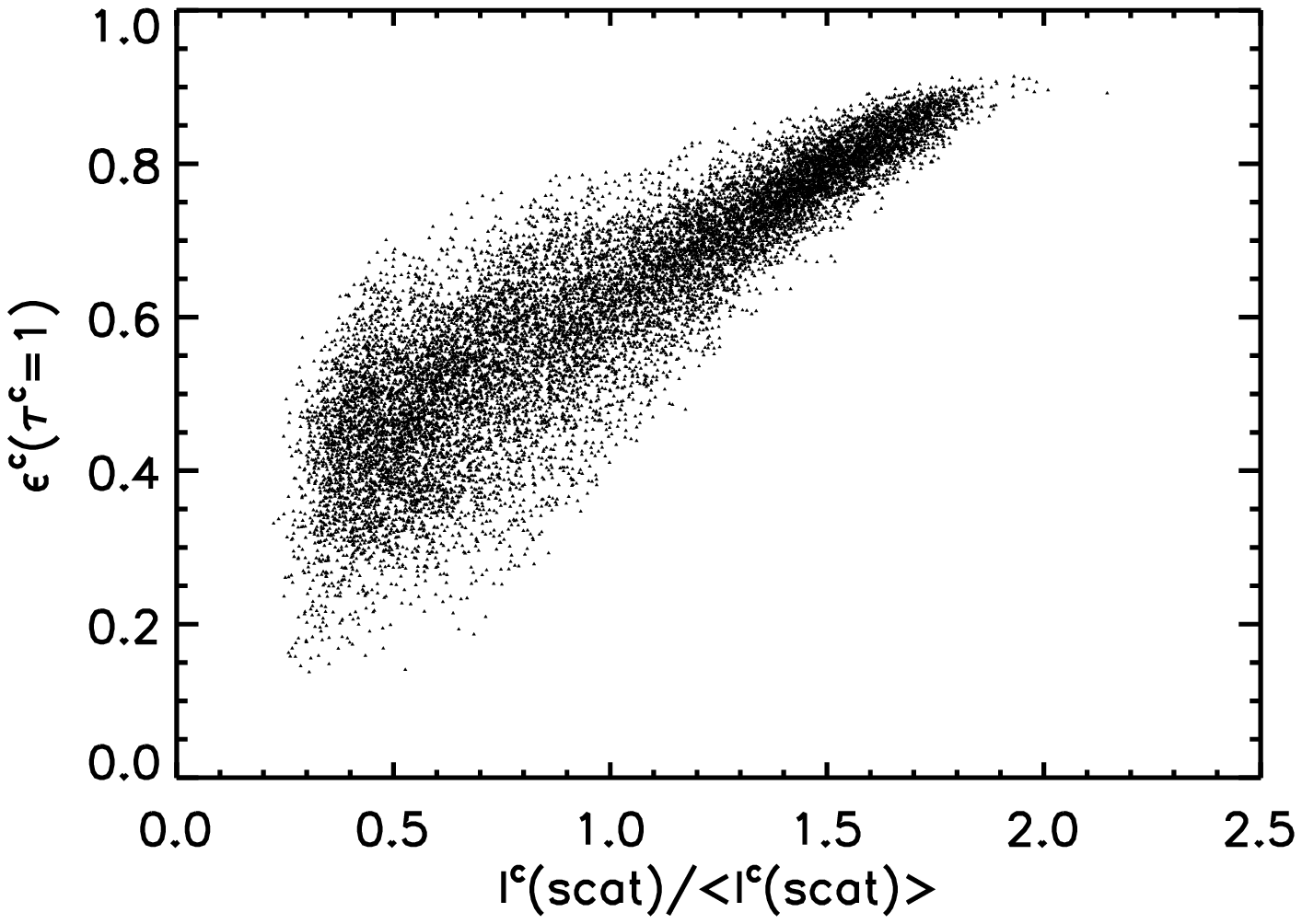}
\caption{\textit{Left:} Correlation between the continuum intensity ratio $I^{\mathbf{c}}(\mathrm{scat})/I^{\mathbf{c}}(\mathrm{abs})$ and the normalized continuum intensities with scattering $I^{\mathbf{c}}(\mathrm{scat})/\left<I^{\mathbf{c}}(\mathrm{scat})\right>$ at 3000\,{\AA}, computed for an arbitrary snapshot of the 3D model with $\mathrm{[Fe/H]}=-3.0$. \textit{Right:} Correlation between the continuum photon destruction probabilities $\epsilon^{\mathrm{c}}$ at the local monochromatic continuum optical surface $\tau^{\mathrm{c}}=1$ and the normalized continuum intensities with scattering $I^{\mathbf{c}}(\mathrm{scat})/\left<I^{\mathbf{c}}(\mathrm{scat})\right>$ at 3000\,{\AA} for the same model.}
\label{fig:contcorr}
\end{figure*}

We first seek to determine the wavelength range in which continuum scattering contributes sufficient opacity to influence continuum flux levels. Radiative transfer is computed in 3D with scattering opacity treated as absorption and as coherent scattering for a set of wavelength points with logarithmic distribution in the range $1500$\,{\AA}\,$\le\lambda\le8000$\,{\AA}, taking only continuum opacity into account; Fig.~\ref{fig:contflx} shows the resulting flux distributions for the different model atmospheres (see Table~\ref{tab:stellarparams} for stellar parameters). It is clear that continuum scattering is only significant in the UV and blue regions in all cases, where Rayleigh scattering on \ion{H}{I} atoms is mostly responsible for the increased flux levels compared with the case of treating scattering as absorption; the $\lambda^{-4}$-dependence of the Rayleigh cross-section and thermalizing bound-free transitions of primarily various metals at shorter wavelengths limit the spectral range. Scattering is therefore completely negligible in the infrared.

Continuum scattering evidently increases the thermalization depth in the UV, as a significant fraction $(1-\epsilon^{\mathrm{c}})$ of outward streaming photons from deeper, hotter layers is scattered instead of absorbed. Figure~\ref{fig:epstau} shows the vertical variation of photon destruction probabilities $\left<\epsilon^{\mathrm{c}}\right>$ in arbitrary snapshots of the model atmospheres, averaged over surfaces of constant vertical continuum optical depth $\tau^{\mathrm{c}}_{\lambda}$ at the three wavelength points indicated by vertical dotted lines in Fig.~\ref{fig:contflx}. For the $\mathrm{[Fe/H]}=-3.0$ model (upper left panel of Fig.~\ref{fig:epstau}), all photons that interact with the gas above optical depth $\tau^{\mathrm{c}}_{\lambda}\gtrsim10$ are absorbed and thermalize. At $3000$\,{\AA}, the atmosphere becomes slightly translucent before the optical surface ($\tau^{\mathrm{c}}_{\lambda}=1$) is reached, allowing photons to escape from larger depths than in the case where scattering is treated as absorption, while photons at 4000\,{\AA} and 5000\,{\AA} are still trapped. Higher up in the atmosphere, scattering becomes increasingly important at all wavelengths. Photon destruction probabilities grow with metallicity, but exhibit only weak variation at the optical surface for the models with $\mathrm{[Fe/H]}\le-1.0$. The slightly lower effective temperature of the $\mathrm{[Fe/H]}=-1.0$ model reduces $\epsilon^{\mathrm{c}}$ above the surface relative to the other model atmospheres due to the temperature-dependence of absorption opacity (see the discussion below). At $\mathrm{[Fe/H]}=0.0$, absorption dominates almost everywhere in the photosphere and beneath. The \ion{H}{I} $(n=2)$ bound-free absorption edge at 3647\,{\AA} (the so-called Balmer jump) causes a depression in the continuum flux on the blue side. Absorption through photoionization increases the photon destruction probability and moves the thermalization depth outward into cooler layers, causing a spike in the flux deviation $\Delta\left<F^{\mathrm{c}}\right>$ (upper panels of the plots in Fig.~\ref{fig:contflx}).

The granulation flow in 3D radiation-hydrodynamical simulations produces strong horizontal variations of the outward radiative intensities between the hot, bright granules and the cool, dark intergranular lanes. The left panel in Fig.~\ref{fig:surfint} shows continuum intensities in the stellar disk center at 3000\,{\AA}, computed with scattering as absorption for an arbitrary snapshot of the $\mathrm{[Fe/H]}=-3.0$ model. The shapes of granules and intergranular lanes are similar to solar granulation, but extend to much larger spatial scales \citep{Colletetal:2007}. Note that radiative emission in the dark intergranular lanes is comparatively small but nonzero, which is important for the formation of line profiles (see the discussion in Sect.~\ref{sec:3Dline}). The right panel in Fig.~\ref{fig:surfint} shows disk center continuum intensities computed with scattering; both panels in Fig.~\ref{fig:surfint} are normalized to the same average intensity. While the overall morphology of the granulation pattern is almost identical, the surface intensities appear slightly brighter due to the larger thermalization depth of the coherent scattering case.

The left panel of Fig.~\ref{fig:contcorr} quantifies the spatially resolved intensity ratio of the scattering and absorption calculations, showing the correlation with continuum surface intensities; the darker intergranular lanes gain more brightness in proportion to their intensity than the brighter granules. The reason for this variation is the temperature-dependence of the continuum photon destruction probabilities $\epsilon^{\mathrm{c}}$, visible in their correlation with continuum intensity at local optical surfaces ($\tau^{\mathrm{c}}=1$) in each column of the model atmosphere (right panel of Fig.~\ref{fig:contcorr}): the temperature-dependence of Rayleigh scattering opacity is weak if the gas is cool enough that hydrogen is dominantly neutral, $\sigma$ is therefore very similar at continuum optical surfaces in granules and lanes. Contrary to that, thermalizing opacity varies strongly with temperature. Absorption is thus dominant in hot granules, while scattering opacity is important in lanes. The result is a larger thermalization depth with respect to the optical surfaces in the lanes compared to granules and a relatively stronger intensity gain through scattering. The overall contrast in the continuum surface intensities reduces through scattering: while the horizontal average continuum intensity $\left<I^{\mathrm{c}}\right>$ at 3000\,{\AA} of the snapshot shown in Fig.~\ref{fig:surfint} increases by $\approx27$\,\% with respect to treating scattering as absorption, the relative rms variation at this wavelength decreases from $I^{\mathrm{c}}_{\mathrm{rms}}/\left<I^{\mathrm{c}}\right>=0.51$ to $I^{\mathrm{c}}_{\mathrm{rms}}/\left<I^{\mathrm{c}}\right>=0.43$ when coherent scattering is included in the radiative transfer computation.

\section{The effects of scattering on spectral lines in 3D}\label{sec:3Dline}

\begin{figure*}[htbp]
\centering
\includegraphics[width=7cm]{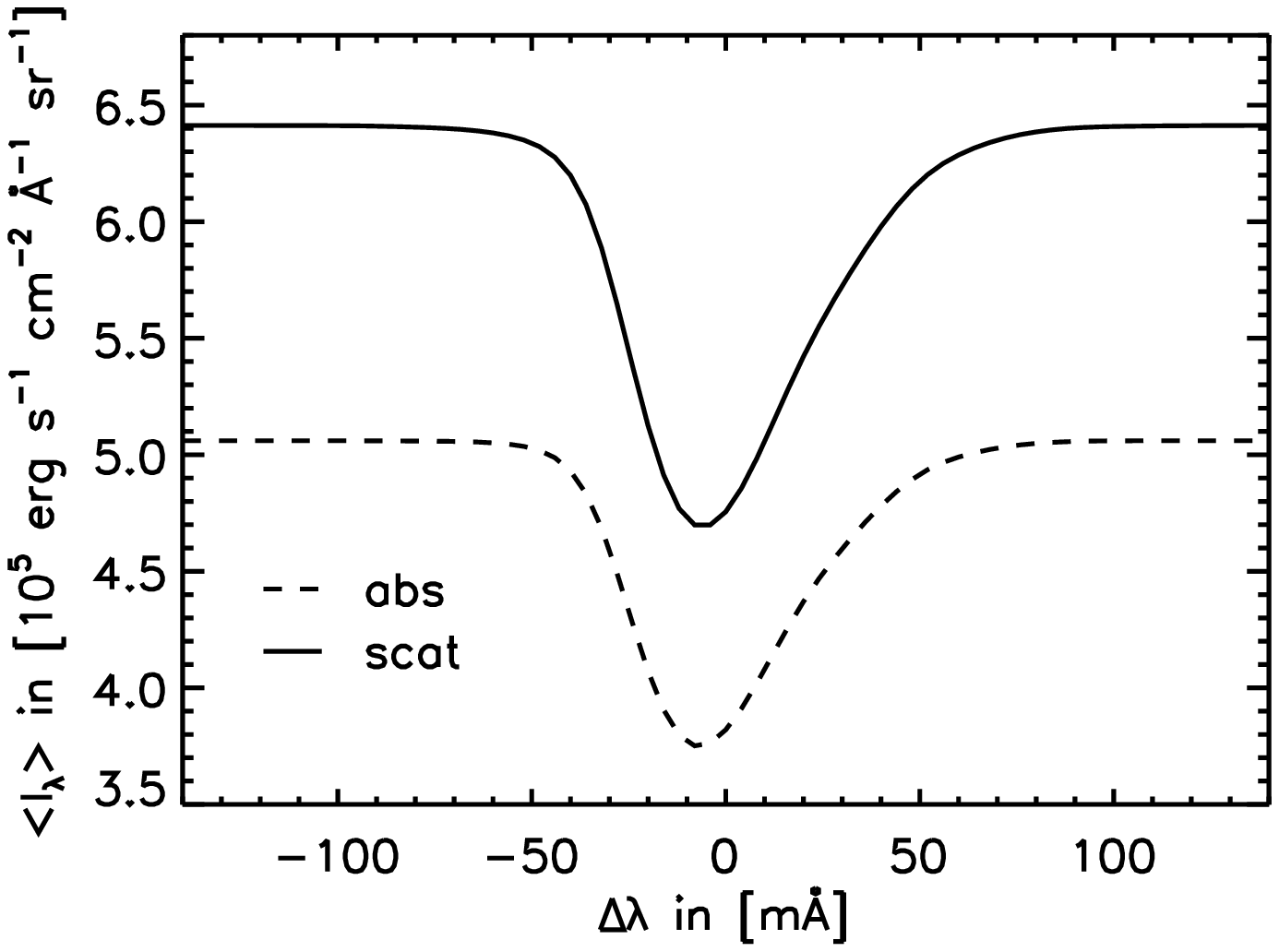}
\includegraphics[width=7cm]{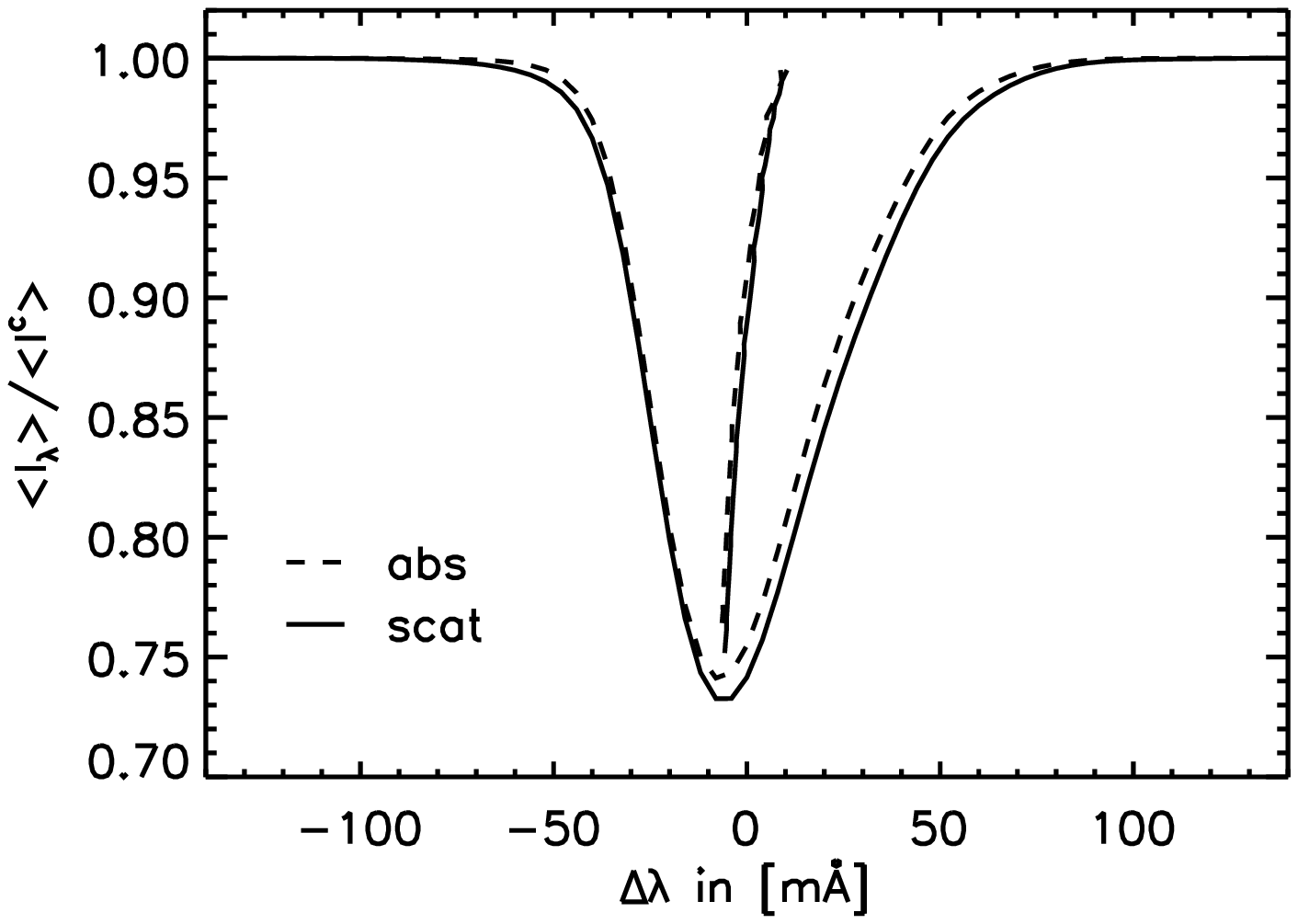}
\caption{Spatial averages of disk center intensity profiles (left panel) and normalized disk center intensity profiles with bisectors (right panel) of a fictitious low-excitation ($\chi=0$\,eV) \ion{Fe}{I} line plotted as functions of wavelength shift $\Delta\lambda$ in m{\AA} with $\lambda_{0}=3000$\,{\AA}, computed for an arbitrary snapshot of the 3D model with $\mathrm{[Fe/H]}=-3.0$ including coherent continuum scattering (solid lines) and treating scattering as absorption (dashed lines).}
\label{fig:FeIabsintbisec}
\end{figure*}

\begin{figure*}[htbp]
\centering
\includegraphics[width=7cm]{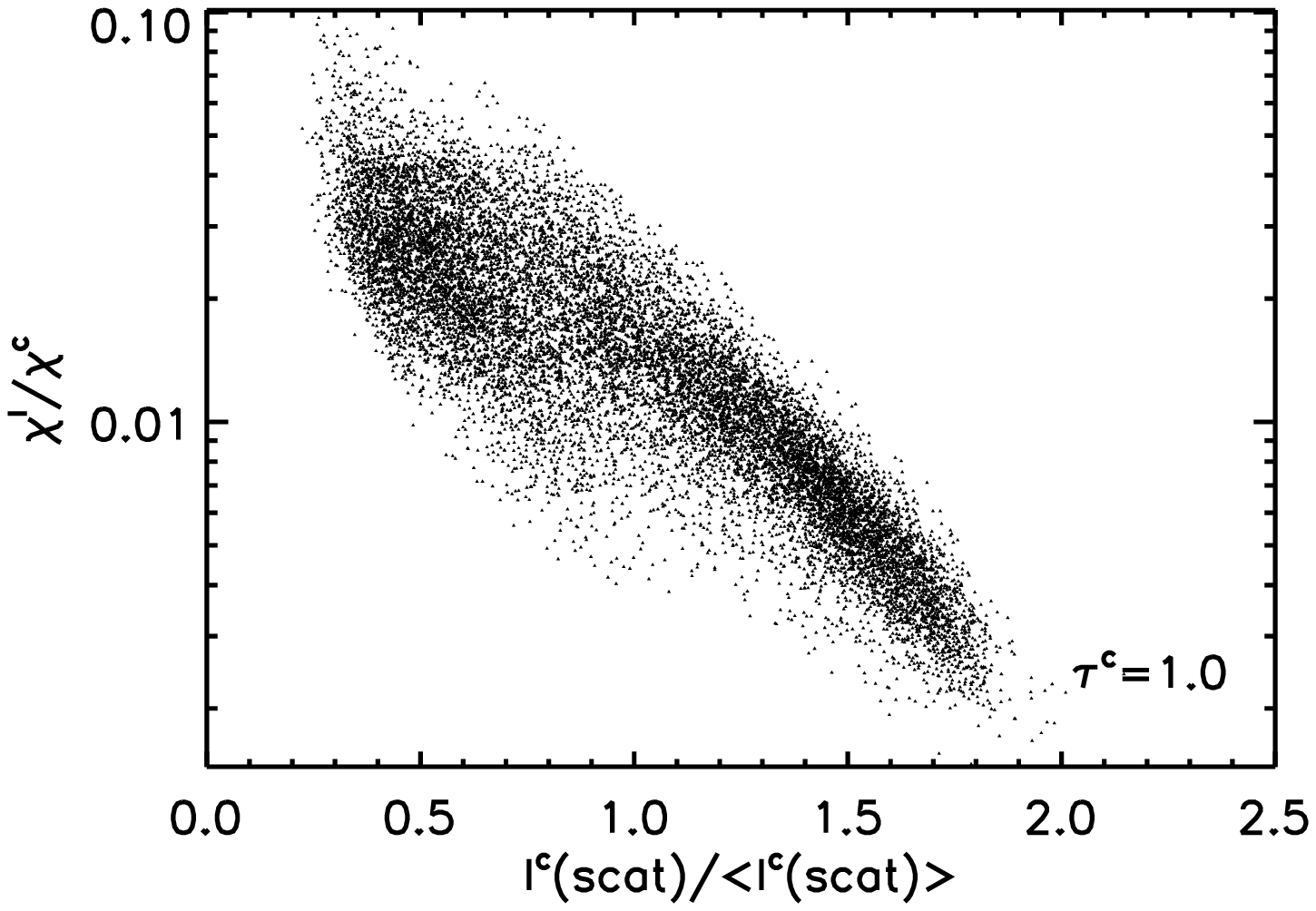}
\includegraphics[width=7cm]{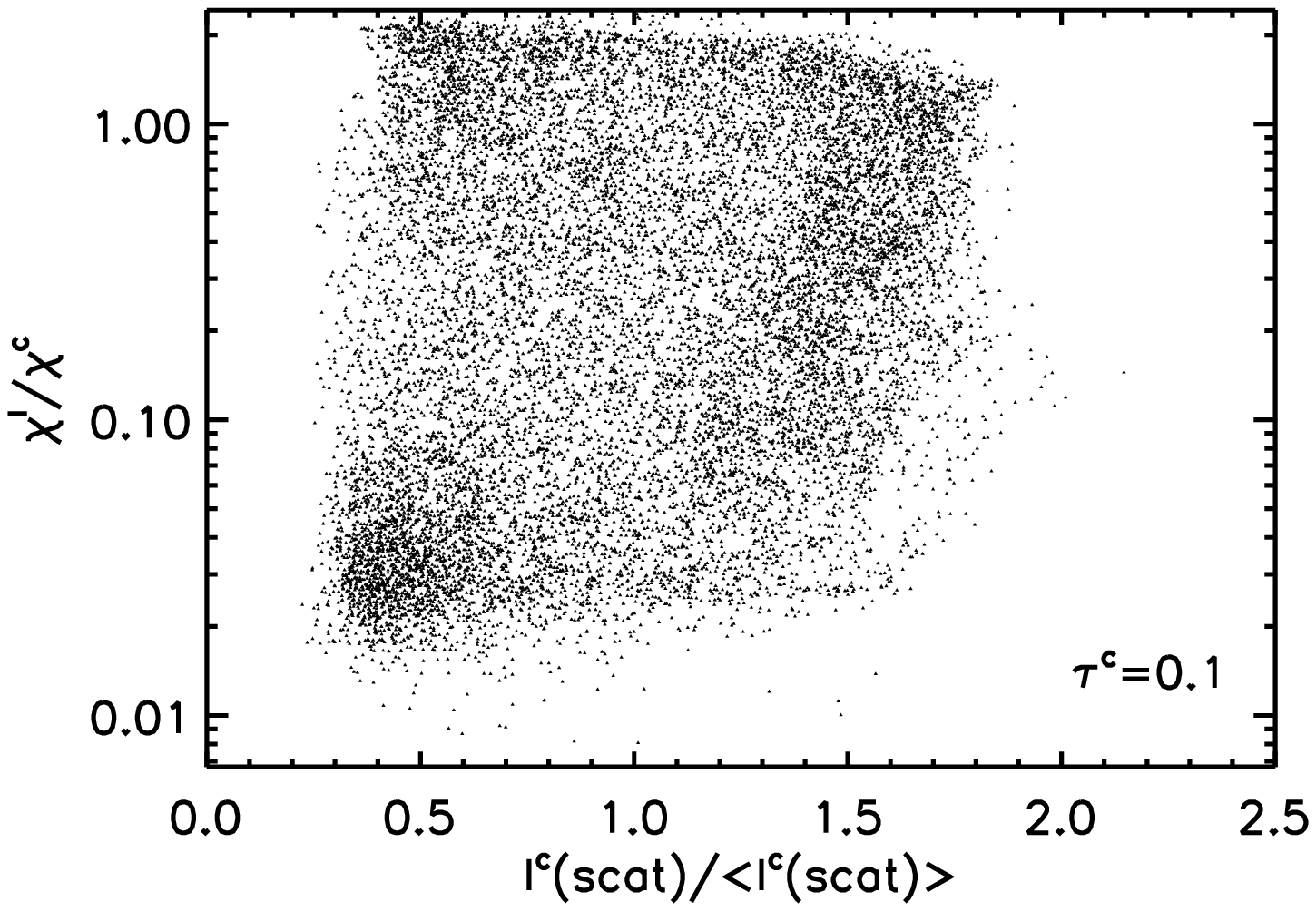}
\caption{Correlation between wavelength-integrated line-to-continuum opacity ratios $\chi^{\mathrm{l}}/\chi^{\mathrm{c}}$ and normalized continuum intensities $I^{\mathbf{c}}(\mathrm{scat})/\left<I^{\mathbf{c}}(\mathrm{scat})\right>$ at the continuum optical surface $\tau^{\mathrm{c}}=1$ (left panel), and at continuum optical depth $\tau^{\mathrm{c}}=0.1$ (right panel) for the \ion{Fe}{I} line of Fig.~\ref{fig:FeIabsintbisec} (3D model with $\mathrm{[Fe/H]}=-3.0$).}
\label{fig:lineopcorr}
\end{figure*}

\begin{figure*}[htbp]
\centering
\includegraphics[width=7cm]{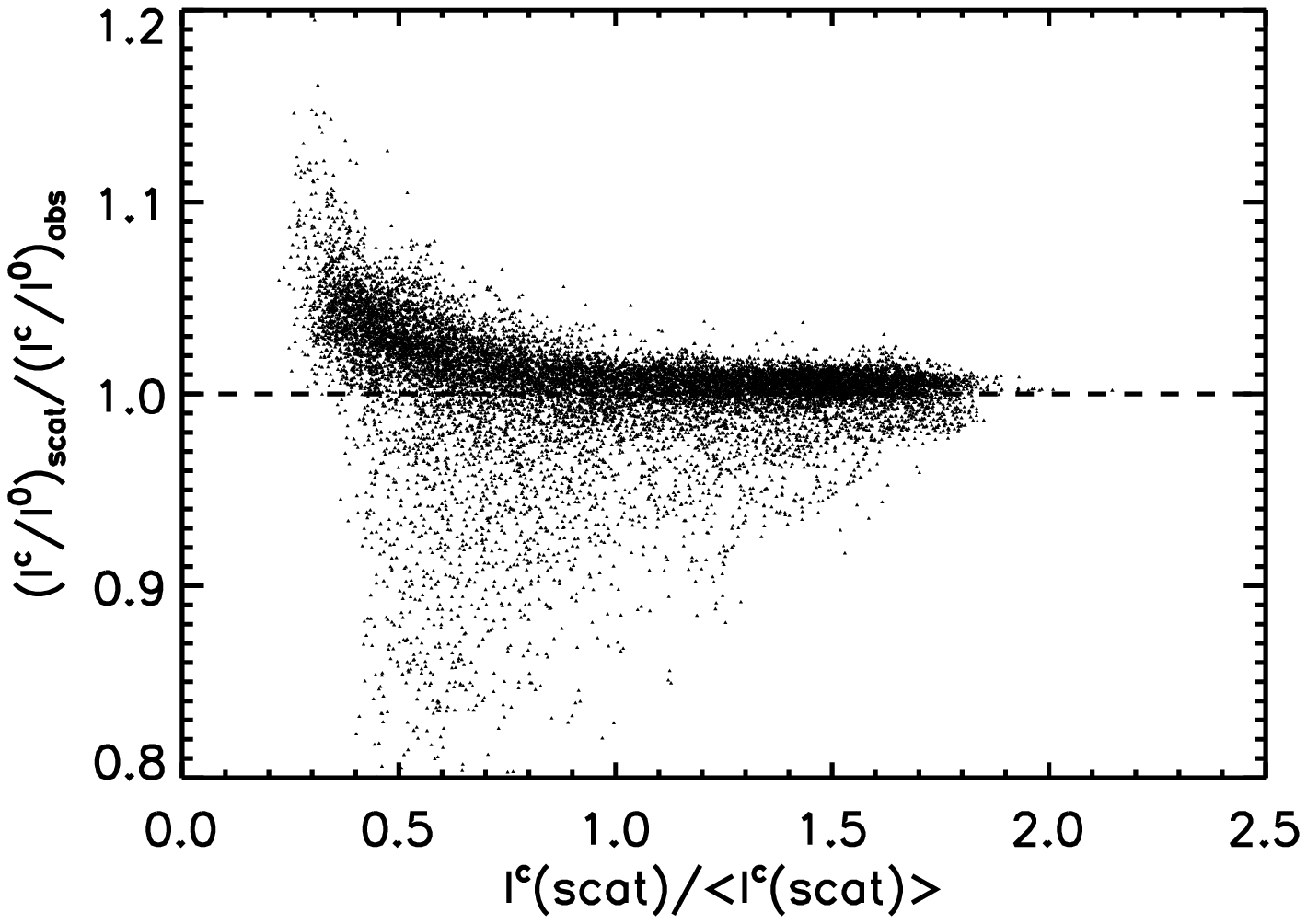}
\includegraphics[width=7cm]{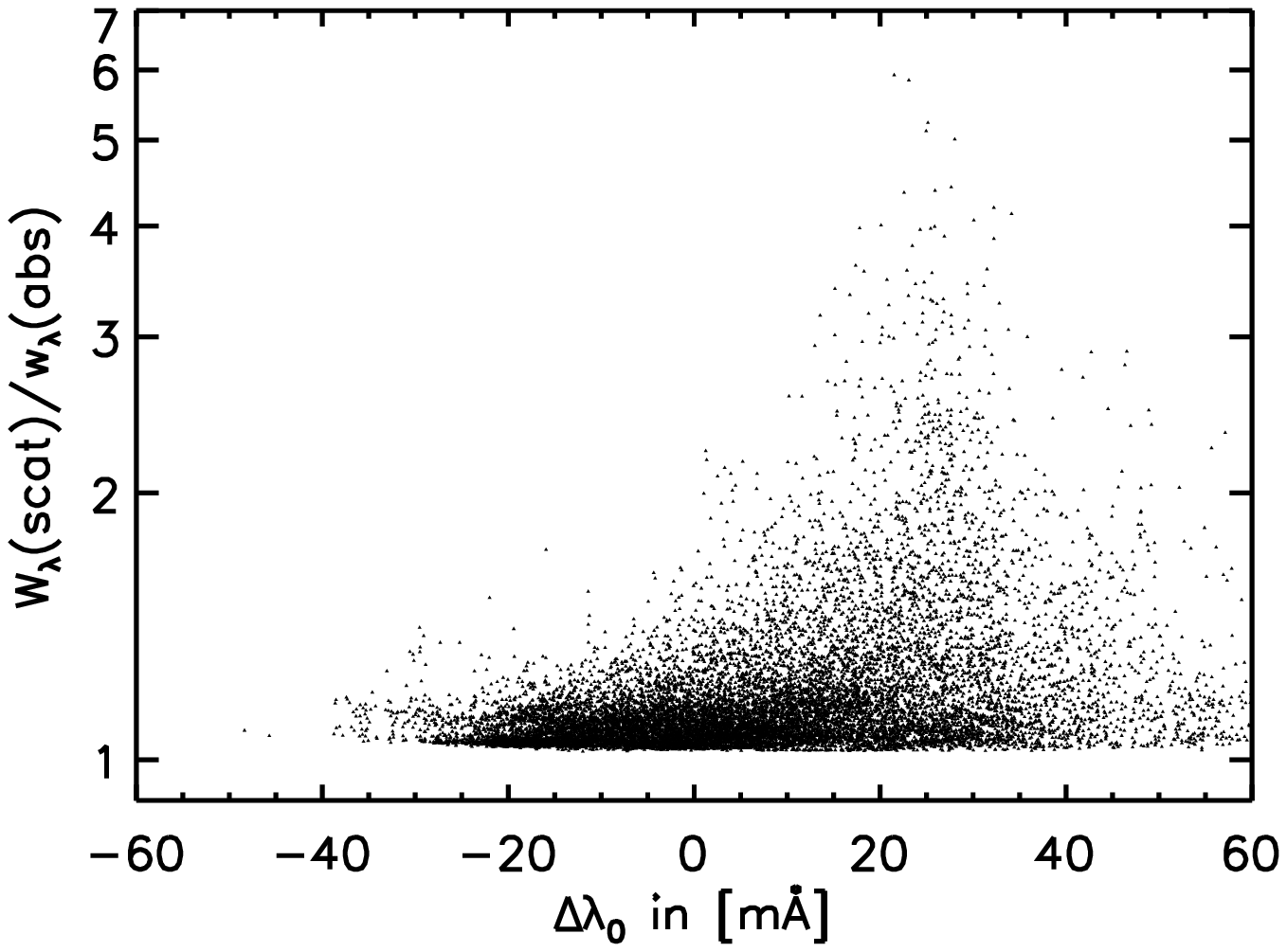}
\caption{Correlation between the growth of continuum-to-line core intensity ratios $(I^{\mathrm{c}}/I^0)_{\mathrm{scat}}/(I^{\mathrm{c}}/I^0)_{\mathrm{abs}}$ through scattering and normalized continuum intensities $I^{\mathbf{c}}(\mathrm{scat})/\left<I^{\mathbf{c}}(\mathrm{scat})\right>$ (left panel), as well as between the growth of equivalent widths $W_{\lambda}(\mathrm{scat})/W_{\lambda}(\mathrm{abs})$ through scattering and the line shift $\Delta\lambda_{0}$ (right panel) for the \ion{Fe}{I} line of Fig.~\ref{fig:FeIabsintbisec} (3D model with $\mathrm{[Fe/H]}=-3.0$).}
\label{fig:linecorr}
\end{figure*}

\begin{figure*}[htbp]
\centering
\includegraphics[width=7cm]{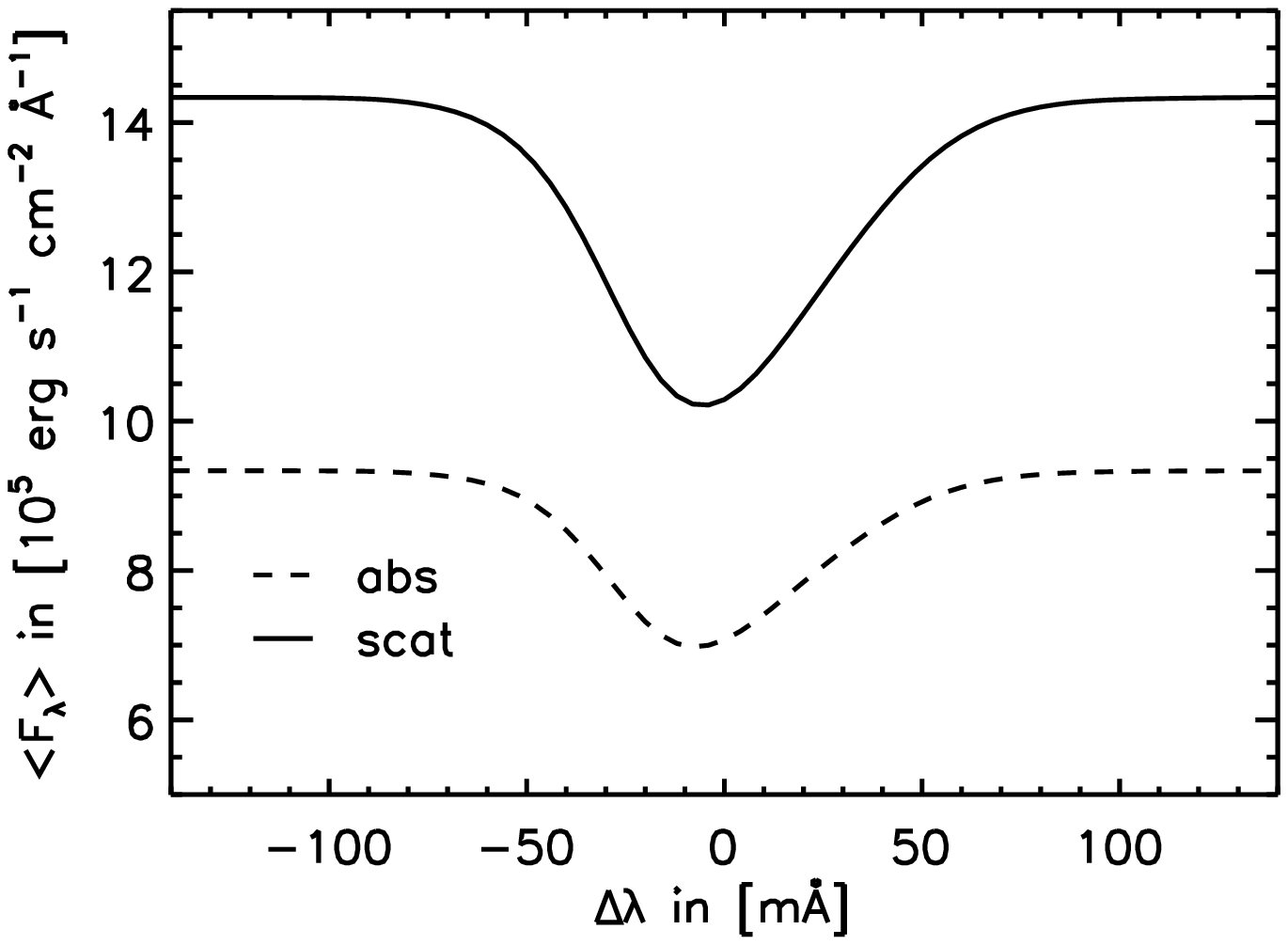}
\includegraphics[width=7cm]{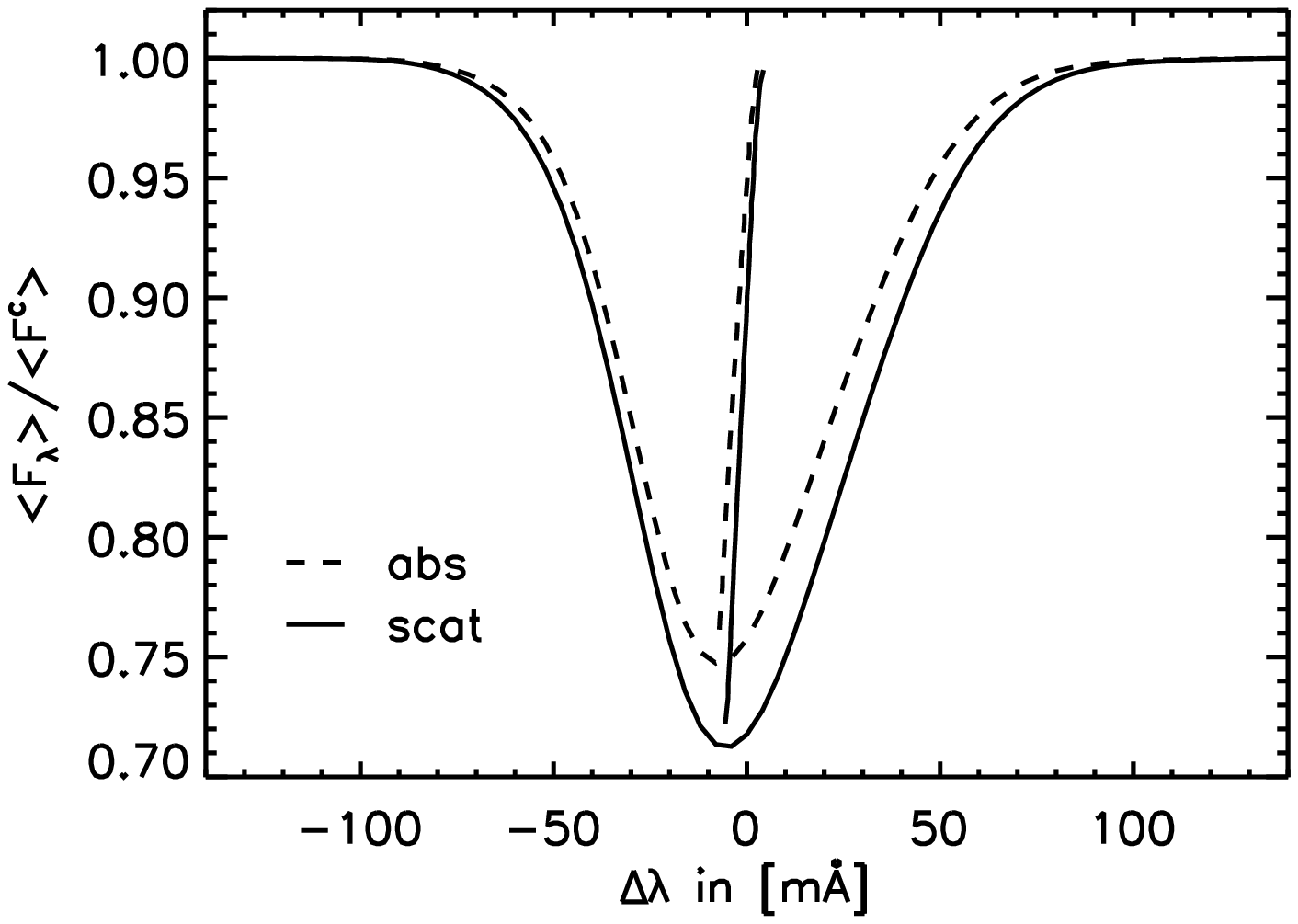}
\caption{Spatial averages of flux profiles (left panel) and normalized flux profiles with bisectors (right panel) of the \ion{Fe}{I} line of Fig.~\ref{fig:FeIabsintbisec} (3D model with $\mathrm{[Fe/H]}=-3.0$).}
\label{fig:fluxbisec}
\end{figure*}

\begin{figure*}[htbp]
\centering
\includegraphics[width=7cm]{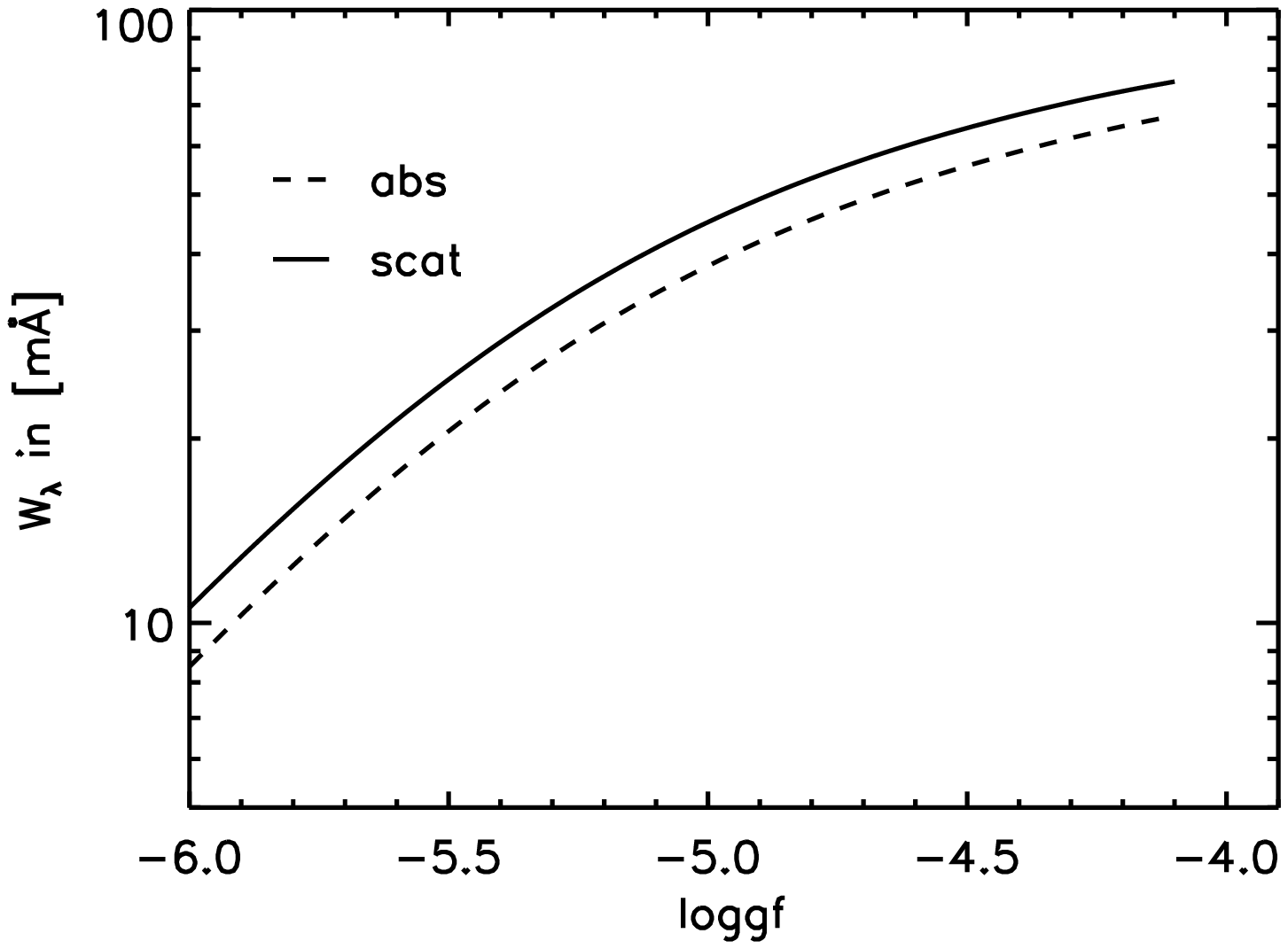}
\includegraphics[width=7cm]{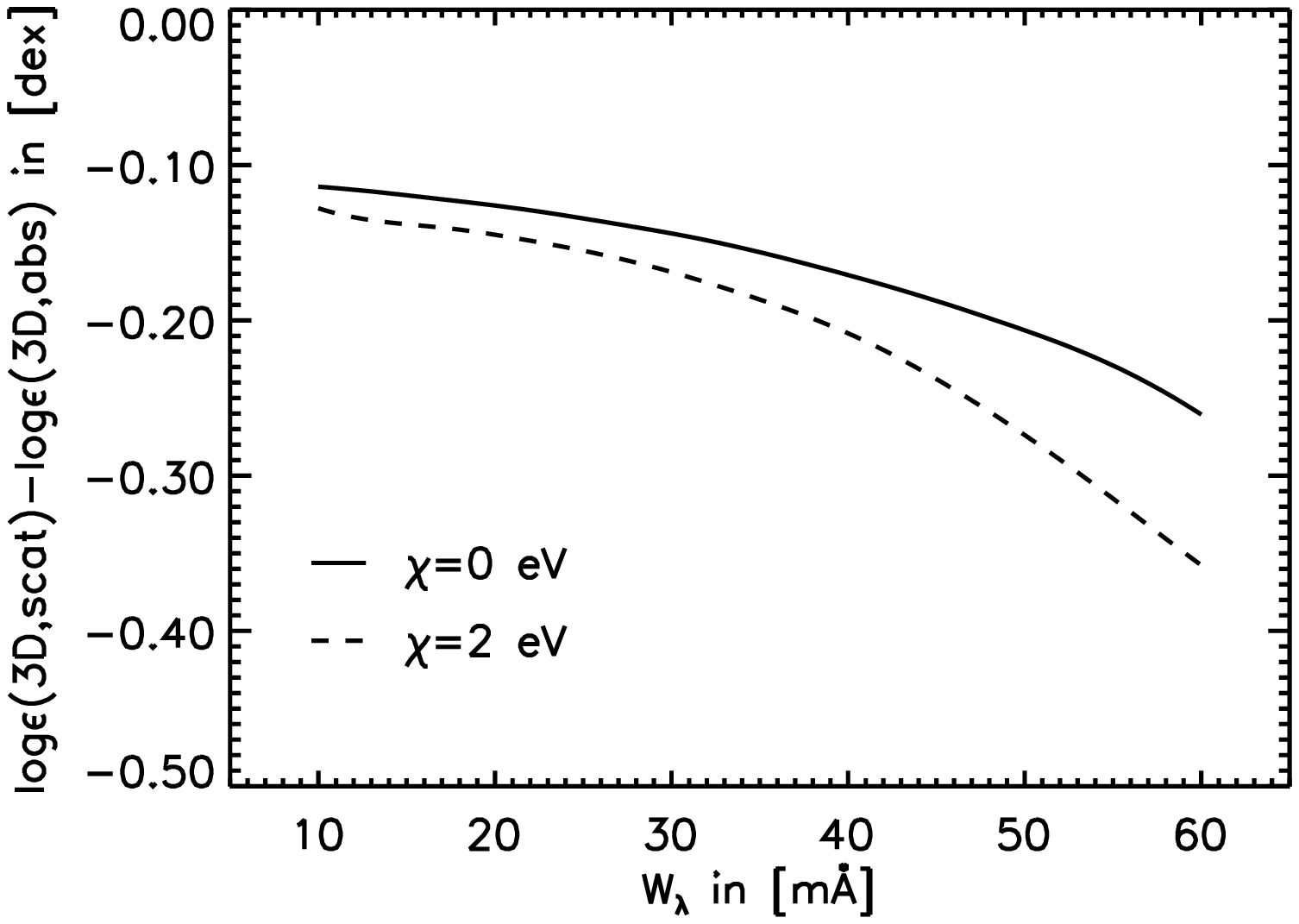}
\caption{\textit{Left:} Curves of growth for the \ion{Fe}{I} line of Fig.~\ref{fig:fluxbisec} (3D model with $\mathrm{[Fe/H]}=-3.0$) as a function of oscillator strength $\log gf$, computed with scattering as absorption (dashed line) and as coherent scattering (solid line). \textit{Right:} Corresponding 3D$-$3D scattering abundance corrections as a function of equivalent width $W_{\lambda}$ for the curves of growth shown in the left panel where $\chi=0$\,eV (solid line), and for a \ion{Fe}{I} line with $\chi=2$\,eV (dashed line).}
\label{fig:FeI3000cog}
\end{figure*}

We now add spectral lines from different atomic and molecular species to the continuum opacities. It is sufficient to restrict the wavelength range to $3000$\,{\AA}\,$\le\lambda\le5000$\,{\AA}, in which radiative flux is accessible for measurements with current instruments and in which continuum scattering is important, as it was demonstrated in Sect.~\ref{sec:3Dcont}. The profile broadening in all Fe line formation computations assumes the \citet{Unsoeld:1955} approximation for collisions with neutral hydrogen for simplicity, although the \texttt{SCATE} code includes tabulated collisional cross-sections based on quantum-mechanical calculations for many species and transitions (see Appendix~\ref{sec:code}). No collisional broadening was applied to molecular line profiles.

The effects of a background scattering continuum are demonstrated by analyzing the formation of fictitious \ion{Fe}{I} lines at 3000\,{\AA} in the atmosphere of the 3D model with $\mathrm{[Fe/H]}=-3.0$. The left panel of Fig.~\ref{fig:FeIabsintbisec} shows spatial averages of disk-center intensity profiles of a low-excitation ($\chi=0$\,eV) transition, where scattering opacity is treated as absorption (dashed line) and as coherent scattering (solid line). Line core brightness increases along with the continuum in the spatial average profile. The right panel of Fig.~\ref{fig:FeIabsintbisec} shows the same profiles normalized to their individual average continuum intensity, as well as the line bisectors in the profile centers. Line depth increases when scattering is included, and the profile bisector shifts to slightly longer wavelengths. Note the characteristic C-shape of the bisector, a consequence of spatial dominance of the upflowing gas in the bright granules, which causes blueshifts in the emitted light \citep[e.g.,][]{Dravins:1982}.

The depression of line core intensity in the normalized profile through continuum scattering stems from thermalization through line absorption. If line opacity is significant in continuum-forming layers, the thermalization depth of the radiation field moves outward into cooler parts of the atmosphere, since the joint photon destruction probability $\epsilon_{\nu}$ of continuum and line opacity is larger than the continuum photon destruction probability $\epsilon^{\mathrm{c}}$:
\begin{equation}
\epsilon_{\nu}=\frac{\chi^{\mathrm{l}}_{\nu}}{\chi^{\mathrm{l}}_{\nu}+\chi^{\mathrm{c}}}+\frac{\chi^{\mathrm{c}}}{\chi^{\mathrm{l}}_{\nu}+\chi^{\mathrm{c}}}\epsilon^{\mathrm{c}}=\frac{\chi^{\mathrm{l}}_{\nu}+\kappa^{\mathrm{c}}}{\chi^{\mathrm{l}}_{\nu}+\chi^{\mathrm{c}}}\ge\epsilon^{\mathrm{c}};
\end{equation}
line opacity $\chi^{\mathrm{l}}_{\nu}$ is treated as absorption in our computations. The radiation temperature in the deeper parts of the line contribution function decreases towards the lower local gas temperature, and the line core gains less brightness through scattering than the continuum. Normalization turns this disproportional growth into a deeper line profile. If line opacity is not significant in continuum-forming layers, line core brightness increases along with the continuum through the contribution from scattered photons, and the intensity gain is mostly divided out when the line profile is normalized.

Both magnitude and height-dependence of the line opacity thus determine the importance of continuum scattering for line formation: the ability of lines to thermalize the radiation field grows with $\chi^{\mathrm{l}}_{\nu}$; the opacity contribution of low-excitation lines of neutral atoms such as the \ion{Fe}{I} line in Fig.~\ref{fig:FeIabsintbisec} is biased towards higher atmospheric layers (compare the distribution of line-to-continuum opacity ratios $\chi^{\mathrm{l}}/\chi^{\mathrm{c}}$ at the continuum optical surface $\tau^{\mathrm{c}}=1.0$ and at $\tau^{\mathrm{c}}=0.1$ in Fig.~\ref{fig:lineopcorr}), such species therefore have a weaker effect on the thermalization depth in continuum-forming layers. High-excitation lines exert larger influence, as the exponential temperature-dependence of the Boltzmann excitation equilibrium moves their opacity contribution closer to $\tau^{\mathrm{c}}=1.0$.

The growth of continuum-to-line core intensity ratios through scattering correlates with normalized continuum intensity across the stellar surface (left panel in Fig.~\ref{fig:linecorr}): in the hot granules, lines gain less strength since line opacity is relatively small and continuum absorption relatively large in continuum-forming layers (left panel of Fig.~\ref{fig:lineopcorr} and right panel of Fig.~\ref{fig:contcorr}), and both continua and cores increase in brightness (left panels of Fig.~\ref{fig:contcorr} and Fig.~\ref{fig:linecorr}). In the cool intergranular lanes, continua gain more intensity than the cores since line opacity plays a more important role in thermalizing the radiation field.

The spatial dependence of line growth is translated into wavelength space as Doppler-shifts change sign between the upflowing gas in granules and downflowing gas in intergranular lanes. The right panel of Fig.~\ref{fig:linecorr} shows the correlation between the growth of equivalent widths $W_{\lambda}$ through scattering and wavelength shifts $\Delta\lambda_{0}$ of the individual line profiles in each column. We define $\Delta\lambda_{0}$ through the deviation of the intensity profile bisector at half-depth from the center wavelength $\lambda_{0}$,
\begin{equation}
\Delta\lambda_{0}=\frac{1}{2}\left(\lambda_{\mathrm{hd}}^{\mathrm{blue}}+\lambda_{\mathrm{hd}}^{\mathrm{red}}\right)-\lambda_{0},
\end{equation}
since the 3D velocity field may produce blended profiles with multiple minima in some columns. The stronger gains in line strength in the intergranular lanes are shifted towards larger wavelengths, causing the red wing of the normalized profile to appear relatively darker.

The flux profiles in Fig.~\ref{fig:fluxbisec} demonstrate that the effects of scattering also become stronger towards the limb: continuum optical surfaces move outward into cooler layers with smaller continuum photon destruction probability (see Fig.~\ref{fig:epstau}), which increases the brightness gain in the continuum and therefore produces stronger lines (right panel of Fig.~\ref{fig:fluxbisec}) through the absorption mechanism that was discussed above.

We quantify the impact of scattering on flux profiles by computing synthetic curves of growth, which are sampled with a set of typically 9 oscillator strengths and interpolated using cubic splines, treating scattering as absorption (dashed line in the left panel of Fig.~\ref{fig:FeI3000cog}) and as coherent scattering (solid line). The deeper profiles of the scattering case lead to larger equivalent widths $W_{\lambda}$. The distance between the two curves at each $W_{\lambda}$ defines a logarithmic 3D$-$3D abundance correction\footnote{In spectroscopic notation, $\log\epsilon(\mathrm{A})\equiv\log_{10}(N_{\mathrm{A}}/N_{\mathrm{H}})+12.0$, where $N_{\mathrm{A}}$ and $N_{\mathrm{H}}$ are the number densities of species A and hydrogen}
\begin{equation}
\Delta\log\epsilon\equiv\log\epsilon(\mathrm{3D,scat})-\log\epsilon(\mathrm{3D,abs});
\end{equation}
the solid line in the right panel of Fig.~\ref{fig:FeI3000cog} shows the result. As larger thermalization of the radiation field near line core wavelengths produces increasingly deeper flux profiles, saturation in the curve of growth is delayed and stronger lines exhibit larger abundance corrections. $\Delta\log\epsilon$ also grows with excitation level $\chi$ of the transition (dashed line in the right panel of Fig.~\ref{fig:FeI3000cog}): larger line opacity in continuum-forming layers leads to abundance corrections of up to $-0.36$\,dex at $W_{\lambda}=60$\,m{\AA}. The effect of continuum scattering on abundance measurements can easily be verified using an order-of-magnitude estimate, assuming a rectangular spectral line with sufficient absorption opacity that the core flux $F^{0}$ is insensitive to continuum emission. Scattering then increases only the continuum flux $F^{\mathrm{c}}$, and the ratio of equivalent widths $W_{\lambda}(\mathrm{scat})/W_{\lambda}(\mathrm{abs})$ is independent of line width. For a linear curve of growth and an initial profile depth $d\equiv(1-F^{0}/F^{\mathrm{c}}(\mathrm{abs}))=0.5$, a continuum flux gain $F^{\mathrm{c}}(\mathrm{scat})/F^{\mathrm{c}}(\mathrm{abs})=1.5$ through scattering then translates into an abundance correction of the order of $\sim-0.1$\,dex.

\section{3D$-$3D scattering abundance corrections}\label{sec:scatcorr3D}

\subsection{Curves of growth for Fe I and Fe II lines}

We compute scattering abundance corrections for fictitious \ion{Fe}{I} and \ion{Fe}{II} lines with excitation potentials $\chi=0$\,eV and $\chi=2$\,eV at 3000\,{\AA}, 4000\,{\AA} and 5000\,{\AA} and for all 3D models to investigate the dependence on ionization stage, excitation potential, wavelength and metallicity. Figure~\ref{fig:m3cog} shows the results for the $\mathrm{[Fe/H]}=-3.0$ model. As expected from the discussion in Sect.~\ref{sec:3Dline}, we find the largest $\Delta\log\epsilon$ for the strongest high-excitation \ion{Fe}{II} lines at 3000\,{\AA}, where thermalizing opacity becomes most important in continuum-forming layers near line core frequencies and desaturation has the strongest effect. Abundance corrections reach $\Delta\log\epsilon\approx-0.5$\,dex at $W_{\lambda}=60$\,m{\AA} (dot-dashed line in the upper left panel of Fig.~\ref{fig:m3cog}). The corrections become less severe for weaker lines, neutral species and lower excitation levels. At 4000\,{\AA} (upper right panel), scattering effects are still significant; strong high-excitation \ion{Fe}{II} lines reach $\Delta\log\epsilon\approx-0.09$\,dex at $W_{\lambda}=60$\,m{\AA}. At 5000\,{\AA} (lower left panel), scattering is negligible with $\Delta\log\epsilon\approx-0.02$\,dex at $W_{\lambda}=60$\,m{\AA}, which is smaller than typical abundance measurement errors.

The results for the $\mathrm{[Fe/H]}=-2.0$ model and the $\mathrm{[Fe/H]}=-1.0$ model (Fig.~\ref{fig:m2cog} and Fig.~\ref{fig:m1cog}) exhibit similar behavior as scattering opacity is important for continuum formation (Fig.~\ref{fig:contflx} and Fig.~\ref{fig:epstau}). Abundance corrections are again largest for both metallicities at 3000\,{\AA} and for the strongest high-excitation \ion{Fe}{II} lines, reaching $\Delta\log\epsilon\approx-0.5$\,dex ($\mathrm{[Fe/H]}=-2.0$) and $\Delta\log\epsilon\approx-0.4$\,dex ($\mathrm{[Fe/H]}=-1.0$) at $W_{\lambda}=60$\,m{\AA}. The smaller effective temperature of the $\mathrm{[Fe/H]}=-1.0$ leads to relatively larger scattering effects compared to the other models (see Sect.~\ref{sec:3Dcont}).

The situation changes at solar metallicity ($\mathrm{[Fe/H]}=0.0$, Fig.~\ref{fig:m0cog}): scattering is weak around the optical surface at all wavelengths (lower right panel of Fig.~\ref{fig:epstau}), leading to generally smaller abundance corrections. At 3000\,{\AA}, $\Delta\log\epsilon$ reaches only $-0.1$\,dex for the strongest high-excitation \ion{Fe}{II} lines and only $-0.03$\,dex at small equivalent widths. At 4000\,{\AA} and 5000\,{\AA}, scattering can be neglected.

\subsection{Curves of growth for molecular lines}\label{sec:cogmolec3D}

Spectral lines from molecules are important tools for abundance measurements of carbon, nitrogen and oxygen in metal-poor stars. Computing synthetic line profiles and curves of growth requires solving a set of equilibrium equations to obtain population numbers of the different atoms and molecules (see Appendix~\ref{sec:code}). In late-type stellar atmospheres, simple molecules such as carbon monoxide (CO), the hydrides CH, NH and OH and cyanide (CN) form in sufficiently cool layers. CH, OH and NH molecules have observable transitions in the blue and UV wavelength regions. We follow \citet{Colletetal:2007} and compute synthetic curves of growth for fictitious molecular lines at 3150\,{\AA} (OH), 3360\,{\AA} (NH) and 4360\,{\AA} (CH).

The resulting abundance corrections for the $\mathrm{[Fe/H]}=-3.0$ model are shown in the lower right panel of Fig.~\ref{fig:m3cog}. We find the largest scattering effect for the OH line due to its short wavelength, reaching $-0.13$\,dex for the strongest lines (dashed line in the lower right panel of Fig.~\ref{fig:m3cog}). NH lines at 3360\,{\AA} exhibit a slightly smaller abundance correction (dotted line) with $\Delta\log\epsilon=-0.10$\,dex at $W_{\lambda}=60$\,m{\AA}; CH lines at 4360\,{\AA} experience only weak influence from continuum scattering (solid line). The magnitude of $\Delta\log\epsilon$ for molecular lines is very similar to \ion{Fe}{I} lines, as molecules form mostly in higher, cooler layers of the atmosphere.

The abundance corrections of molecular lines at $\mathrm{[Fe/H]}=-2.0$ are similarly large as for the most metal-poor model, but they become less severe at $\mathrm{[Fe/H]}=-1.0$ where the OH correction reduces to $-0.09$\,dex at $W_{\lambda}=60$\,{\AA}. $\Delta\log\epsilon$ is almost negligible at $\mathrm{[Fe/H]}=0.0$, the strongest OH features reach a correction of $-0.03$\,dex.

Note that we vary oscillator strengths instead of elemental abundances to calculate the curve of growth. The resulting correction $\Delta\log\epsilon$ is therefore an approximation, as shifts in the molecular equilibria are not taken into account. Decreasing $\log\epsilon(\mathrm{O})$ by $0.13$\,dex in the $\mathrm{[Fe/H]}=-3.0$ model and increasing $\log gf$ by the same amount leads to a deviation of the resulting equivalent widths of $\lesssim2$\,\%, the systematic errors in the abundance corrections of Fig.~\ref{fig:m3cog} through Fig.~\ref{fig:m0cog} are therefore small.

\begin{figure}[!t]
\centering
\includegraphics[width=\linewidth]{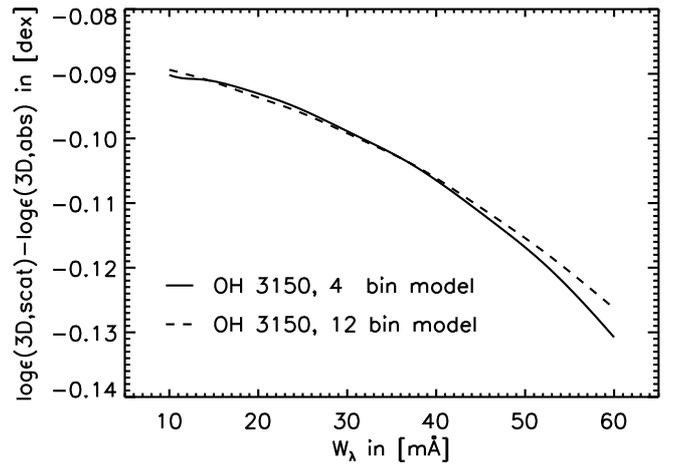}
\caption{3D$-$3D scattering abundance corrections as a function of equivalent width $W_{\lambda}$ for a fictitious OH line at 3150\,{\AA} computed for 3D models with $\mathrm{[Fe/H]}=-3.0$ using 4 opacity bins (solid line) and using 12 opacity bins (dashed line).}
\label{fig:OH12bin4bincog}
\end{figure}

\begin{figure*}[p]
\centering
\mbox{
\includegraphics[width=6.5cm]{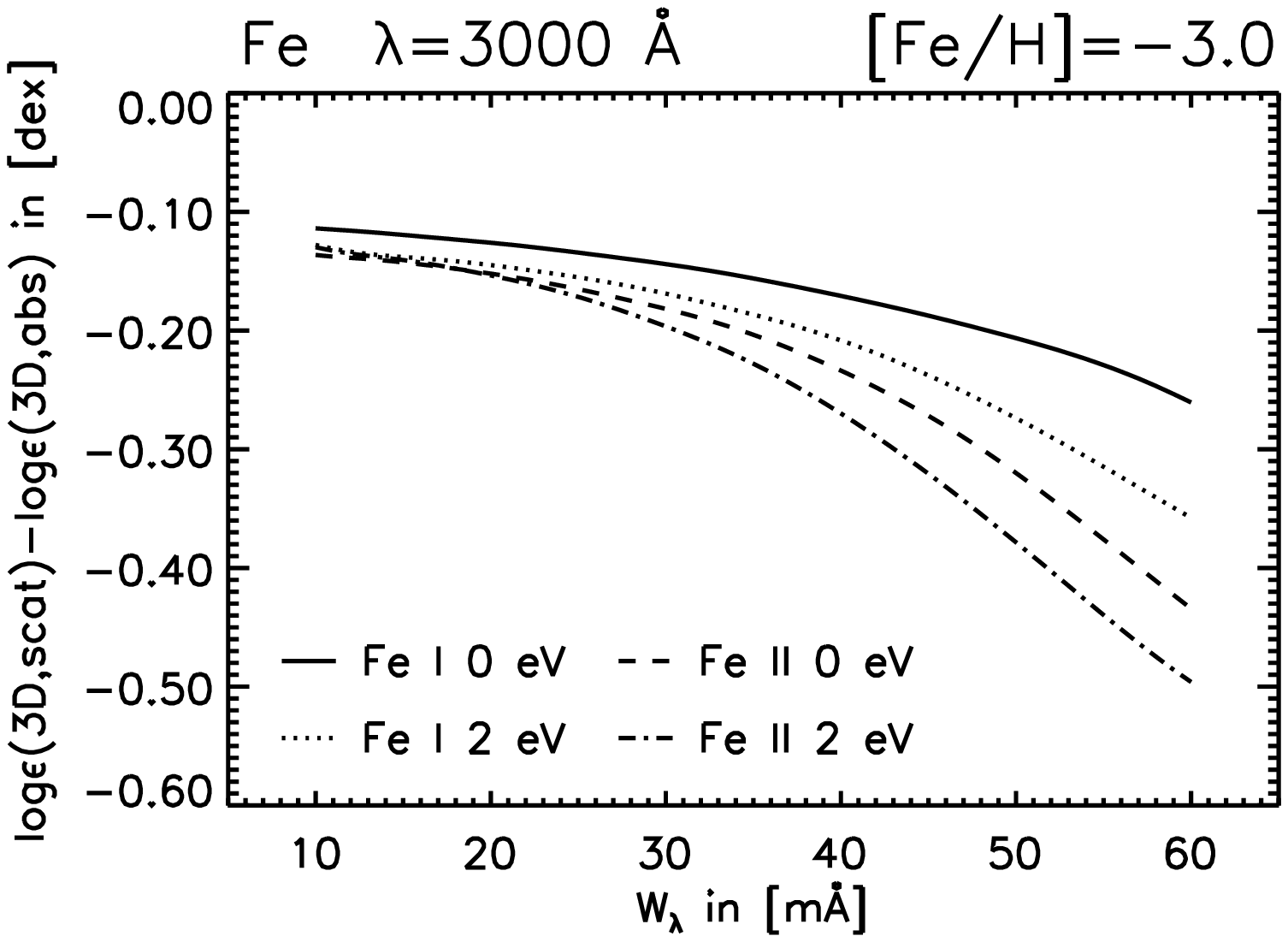}
\includegraphics[width=6.5cm]{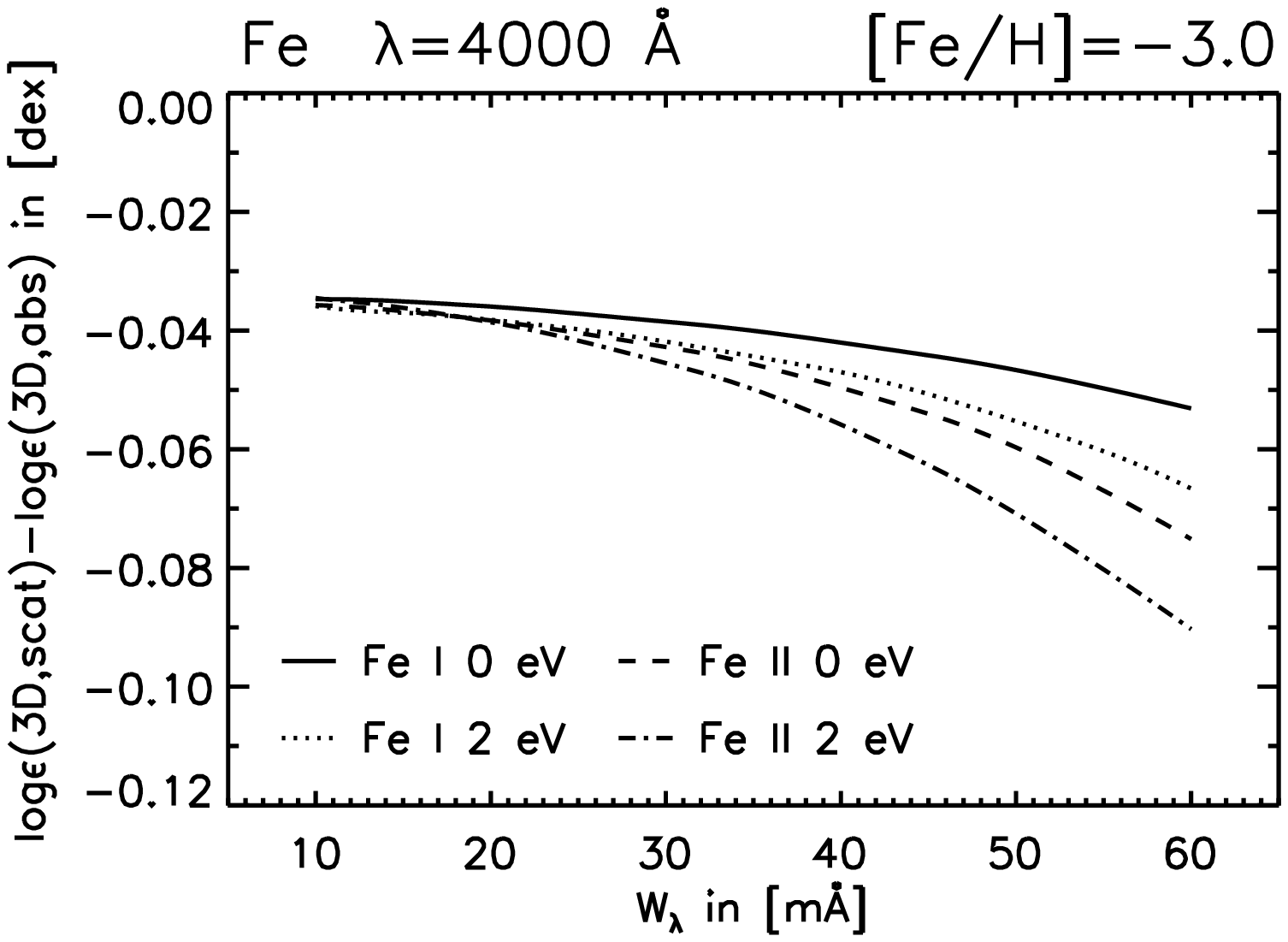}
}
\mbox{
\includegraphics[width=6.5cm]{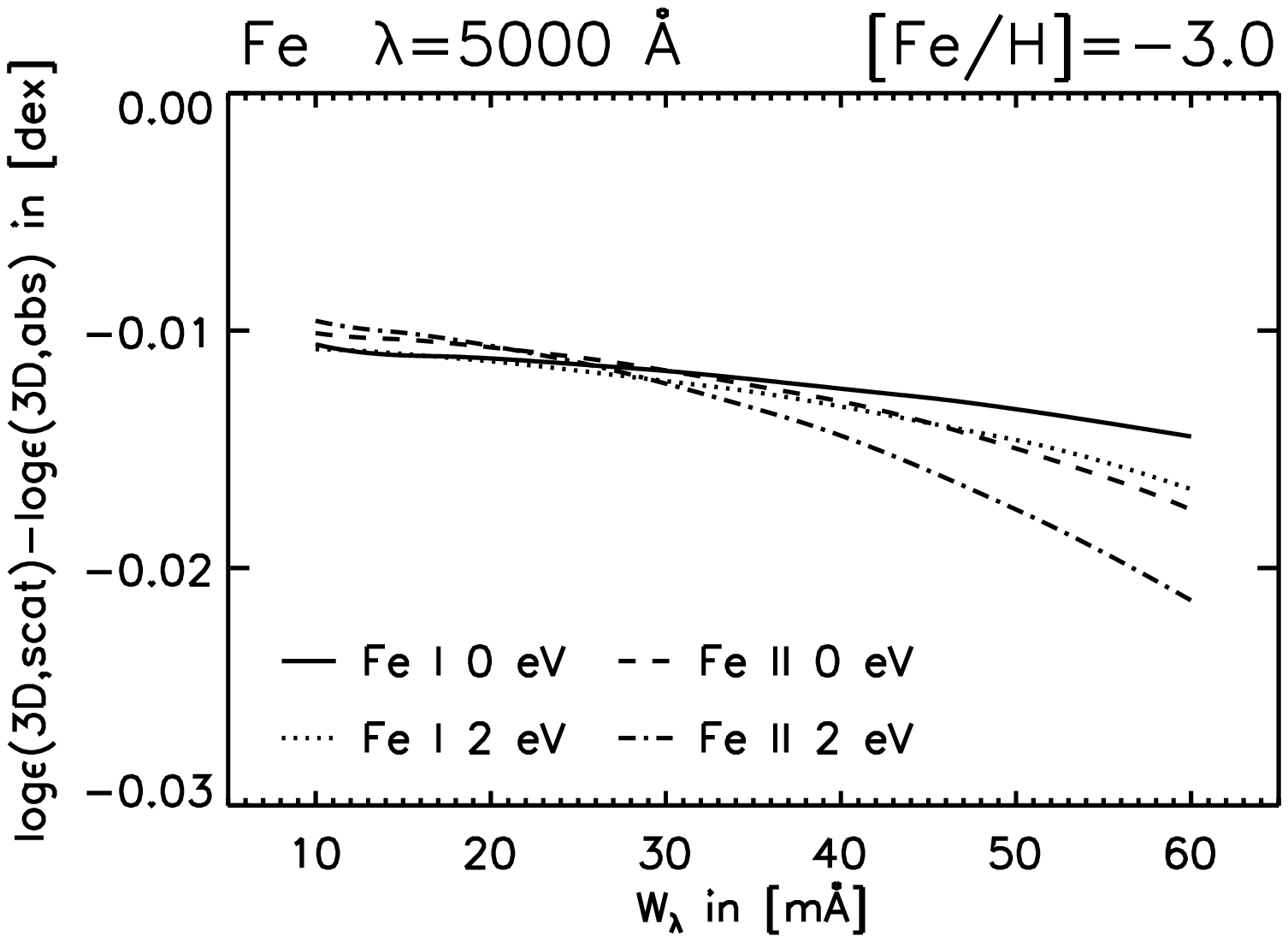}
\includegraphics[width=6.5cm]{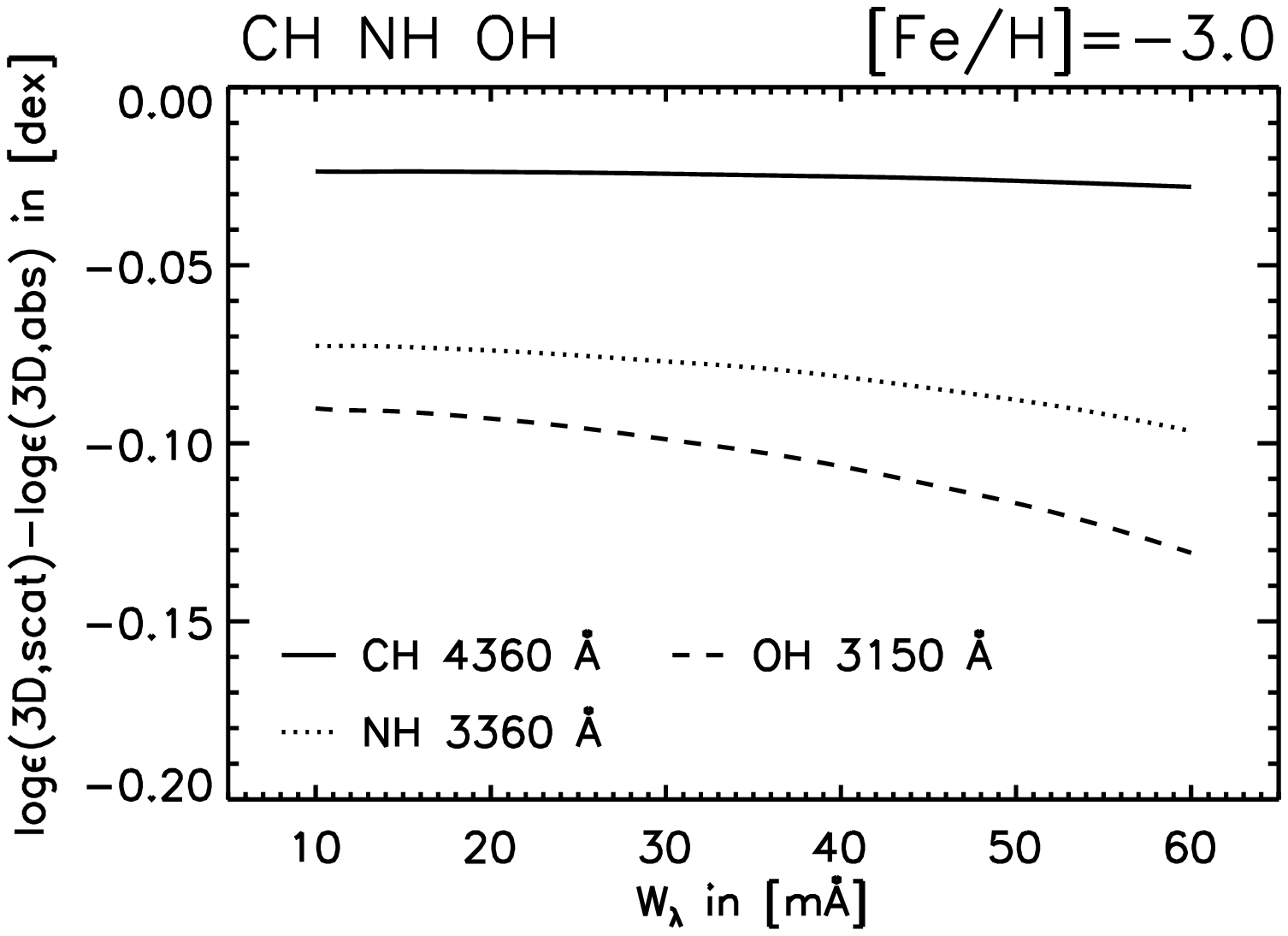}
}
\caption{\textit{Upper left panel to lower left panel:} 3D$-$3D scattering abundance corrections for fictitious \ion{Fe}{I} lines and \ion{Fe}{II} lines with excitation potential $\chi=0$\,eV and $\chi=2$\,eV at 3000\,{\AA}, 4000\,{\AA} and 5000\,{\AA}, computed for the 3D model with $\mathrm{[Fe/H]}=-3.0$. \textit{Lower right panel:} 3D$-$3D scattering abundance corrections for typical CH, NH and OH lines for the same model.}
\label{fig:m3cog}
\end{figure*}

\begin{figure*}[p]
\centering
\mbox{
\includegraphics[width=6.5cm]{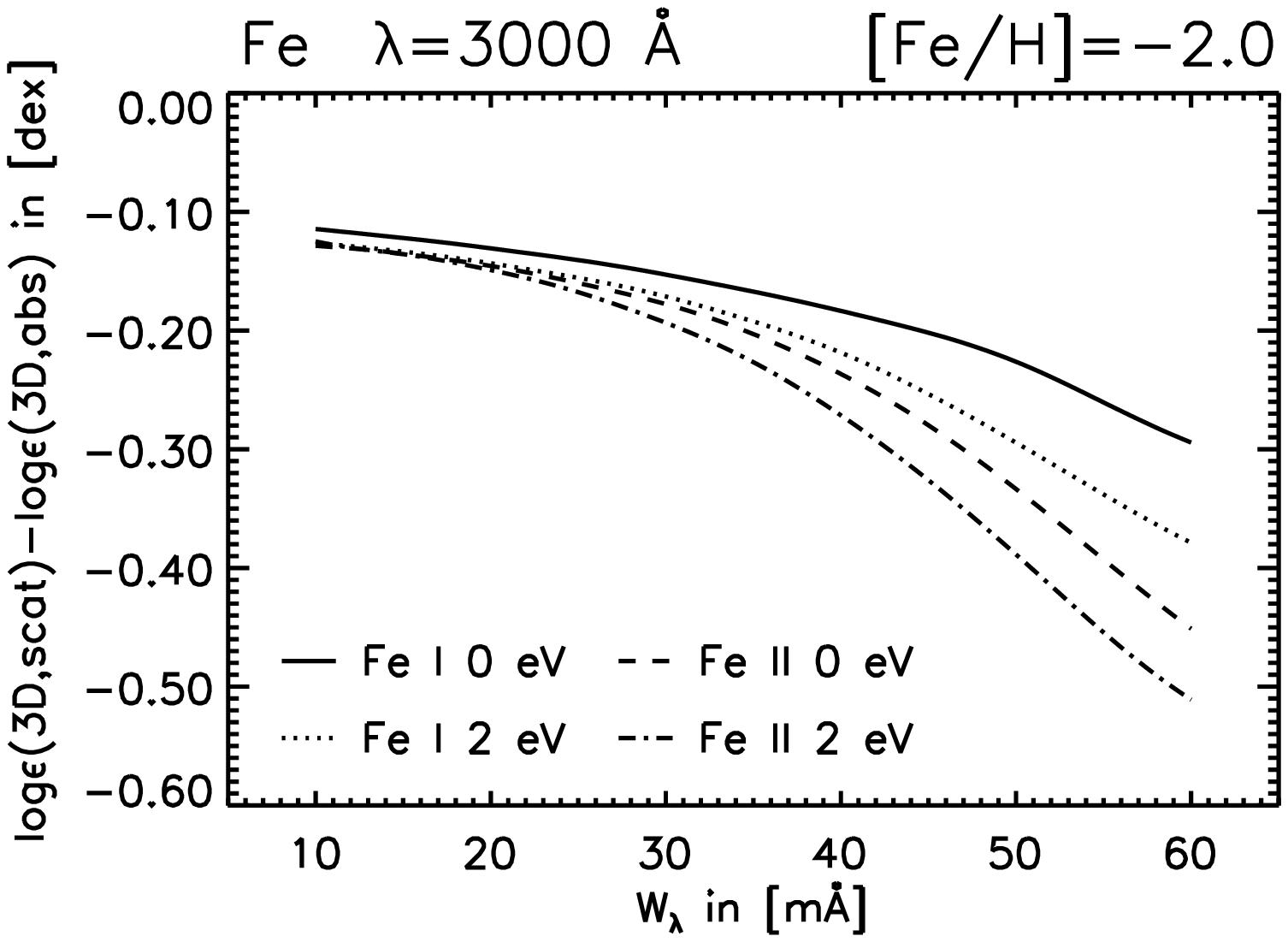}
\includegraphics[width=6.5cm]{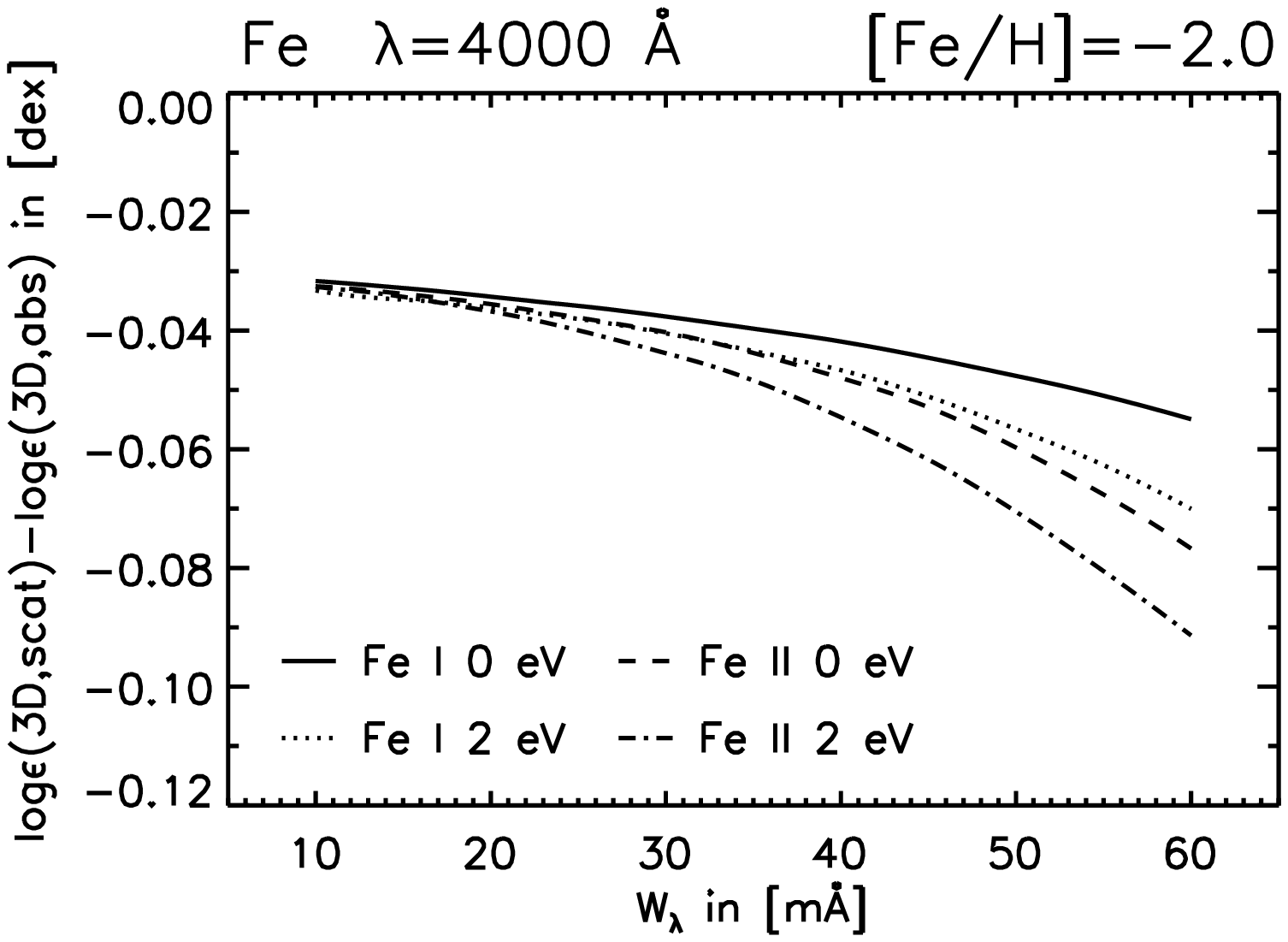}
}
\mbox{
\includegraphics[width=6.5cm]{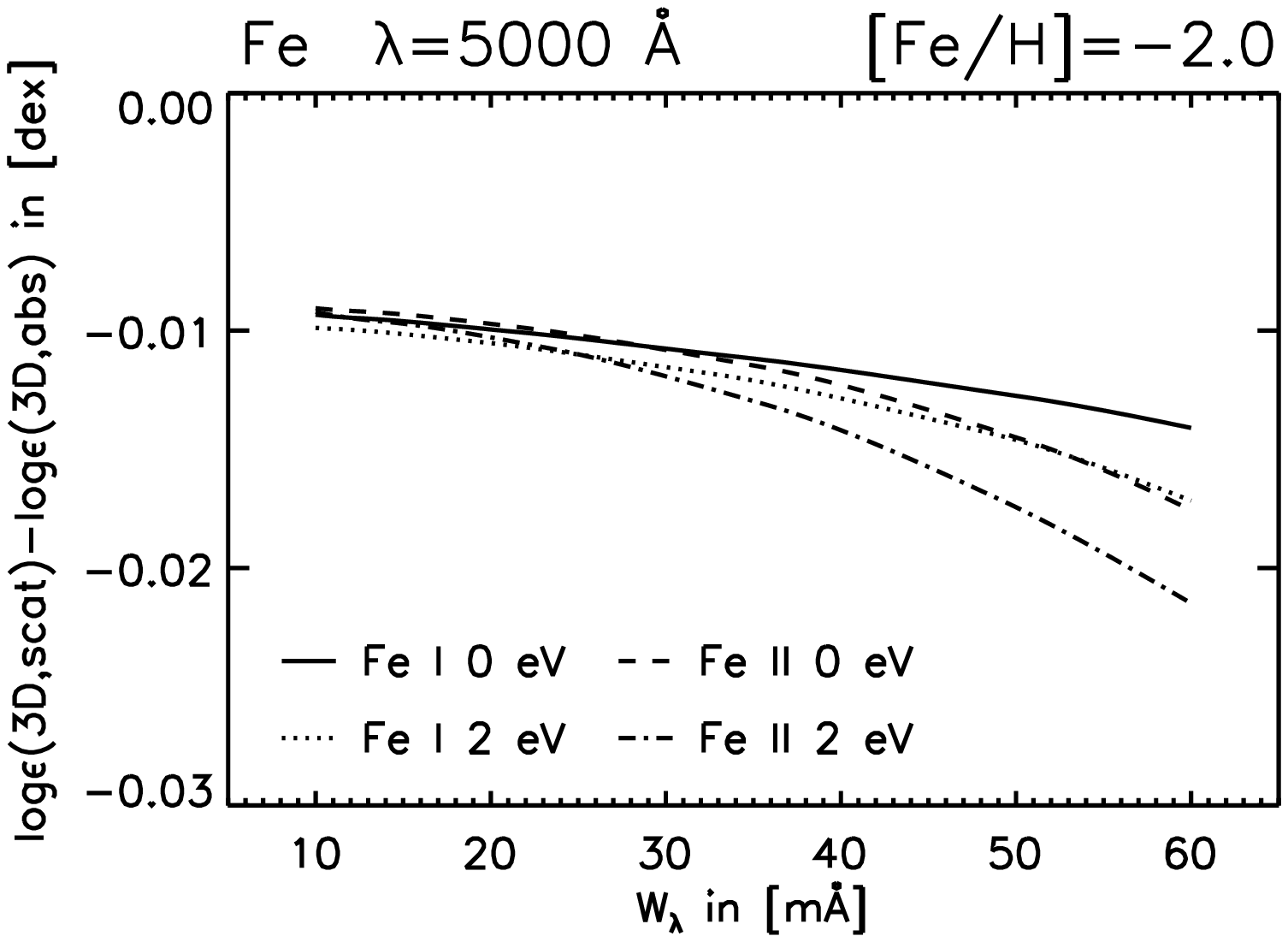}
\includegraphics[width=6.5cm]{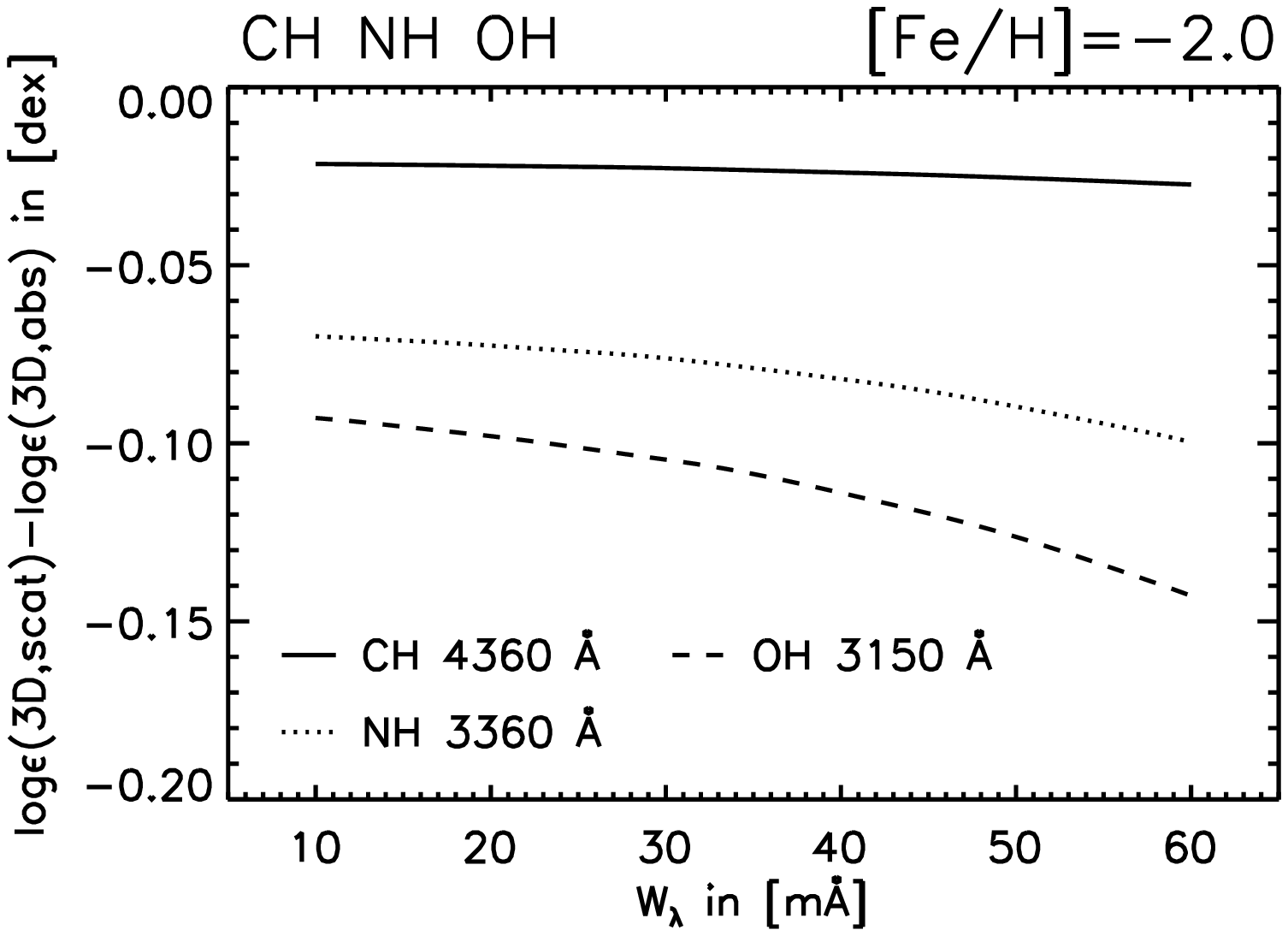}
}
\caption{Same as Fig.~\ref{fig:m3cog}, but computed for the 3D model with $\mathrm{[Fe/H]}=-2.0$.}
\label{fig:m2cog}
\end{figure*}

\begin{figure*}[p]
\centering
\mbox{
\includegraphics[width=6.5cm]{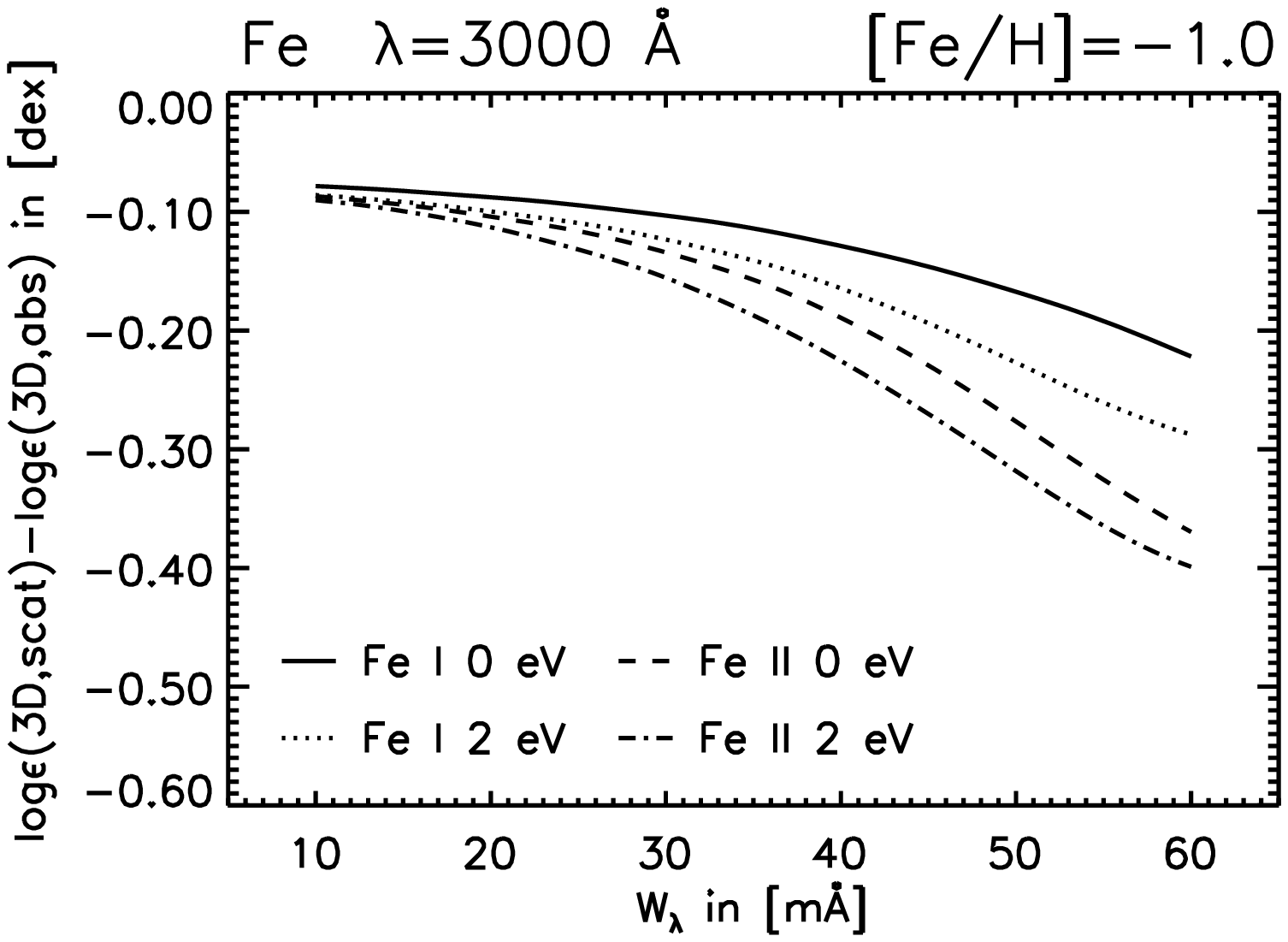}
\includegraphics[width=6.5cm]{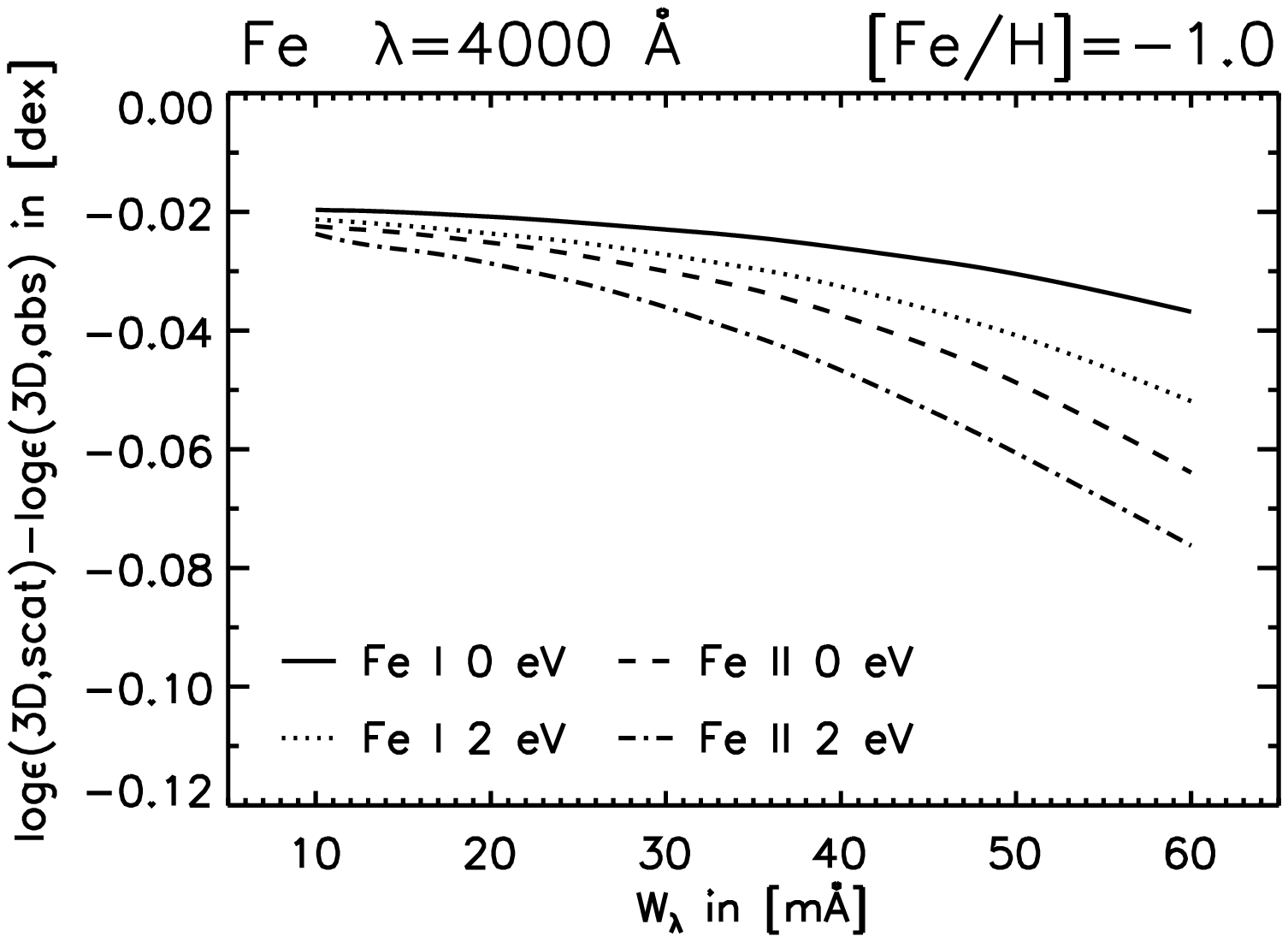}
}
\mbox{
\includegraphics[width=6.5cm]{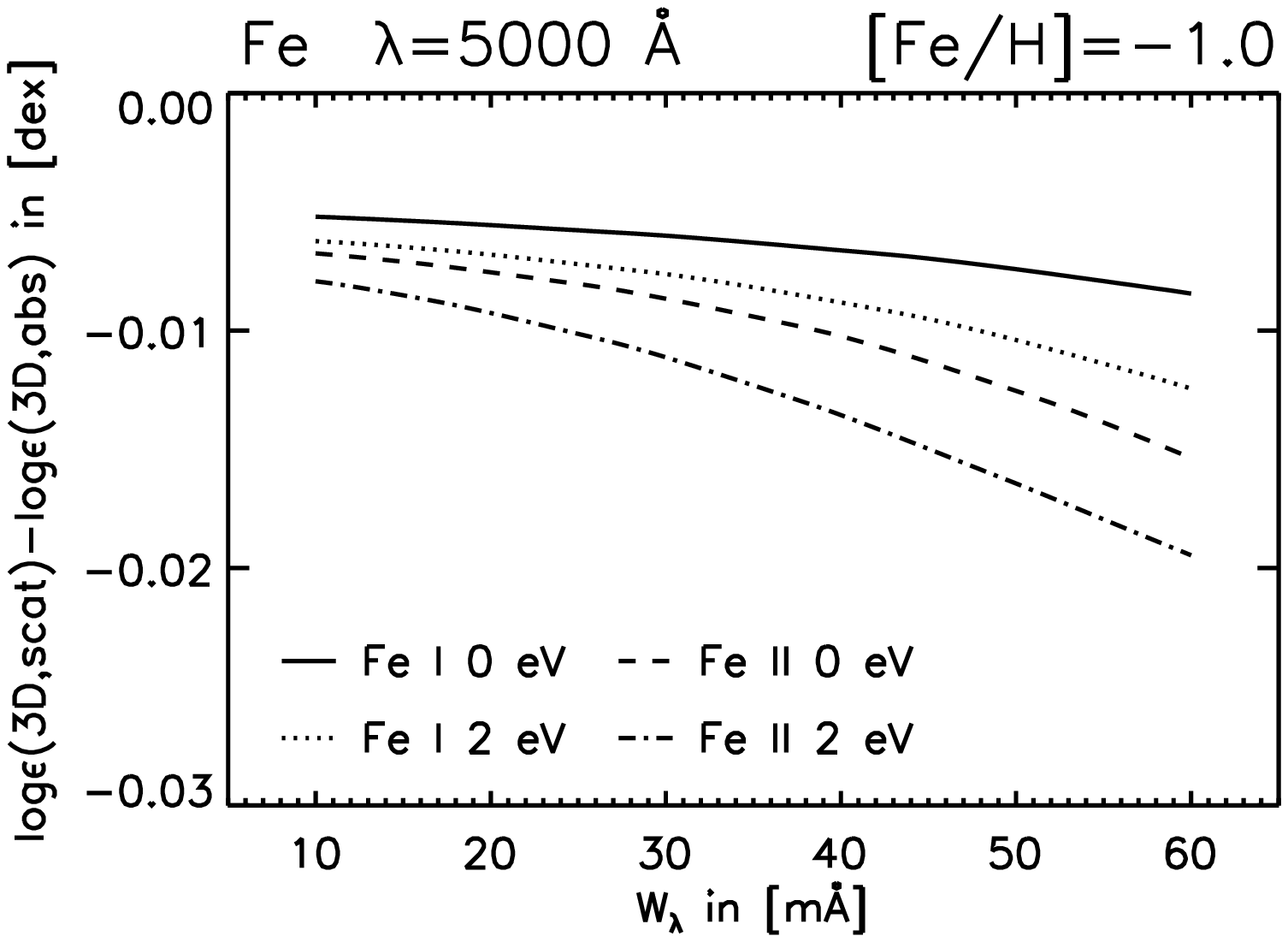}
\includegraphics[width=6.5cm]{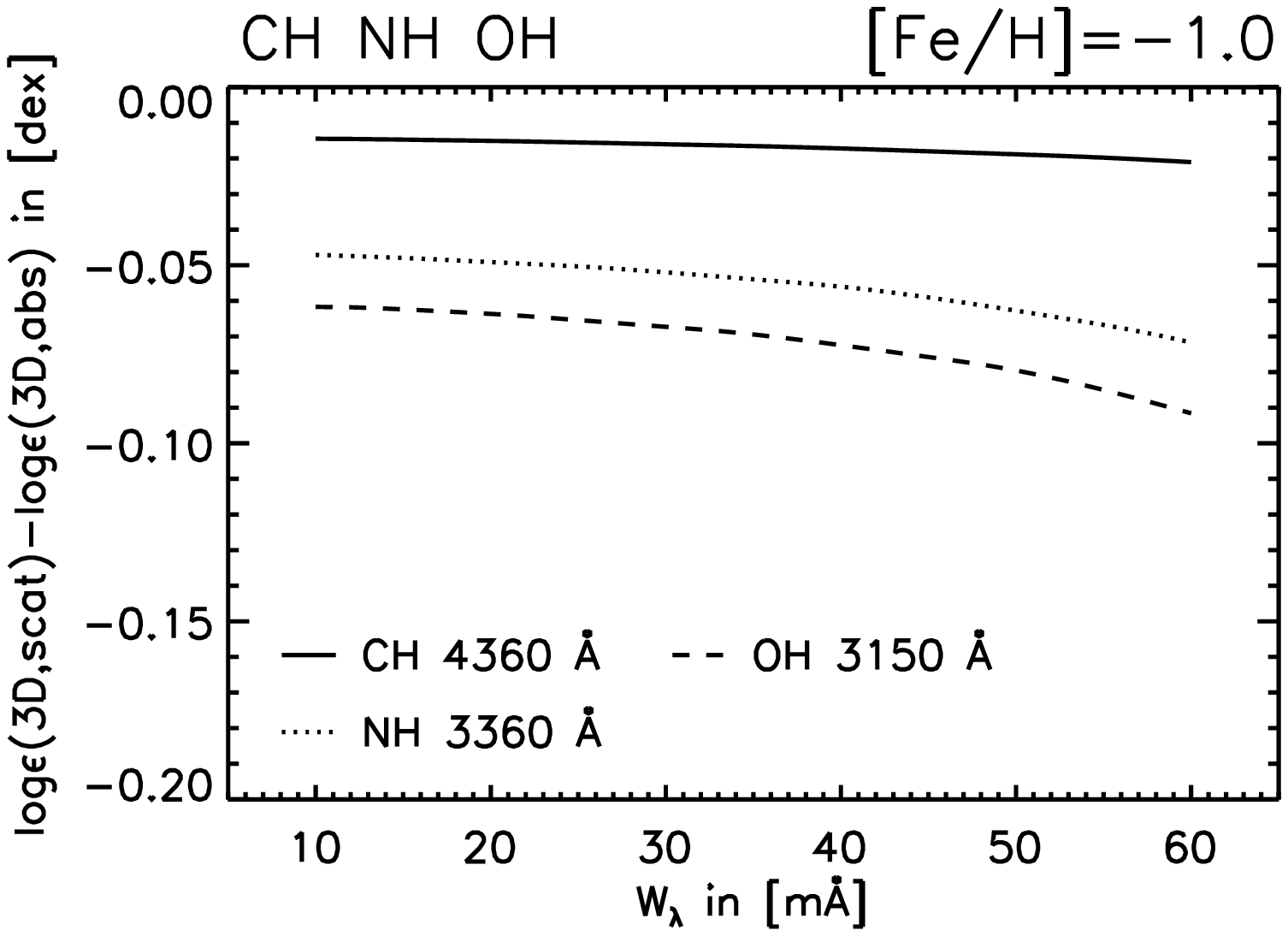}
}
\caption{Same as Fig.~\ref{fig:m3cog}, but computed for the 3D model with $\mathrm{[Fe/H]}=-1.0$.}
\label{fig:m1cog}
\end{figure*}

\begin{figure*}[p]
\centering
\mbox{
\includegraphics[width=6.5cm]{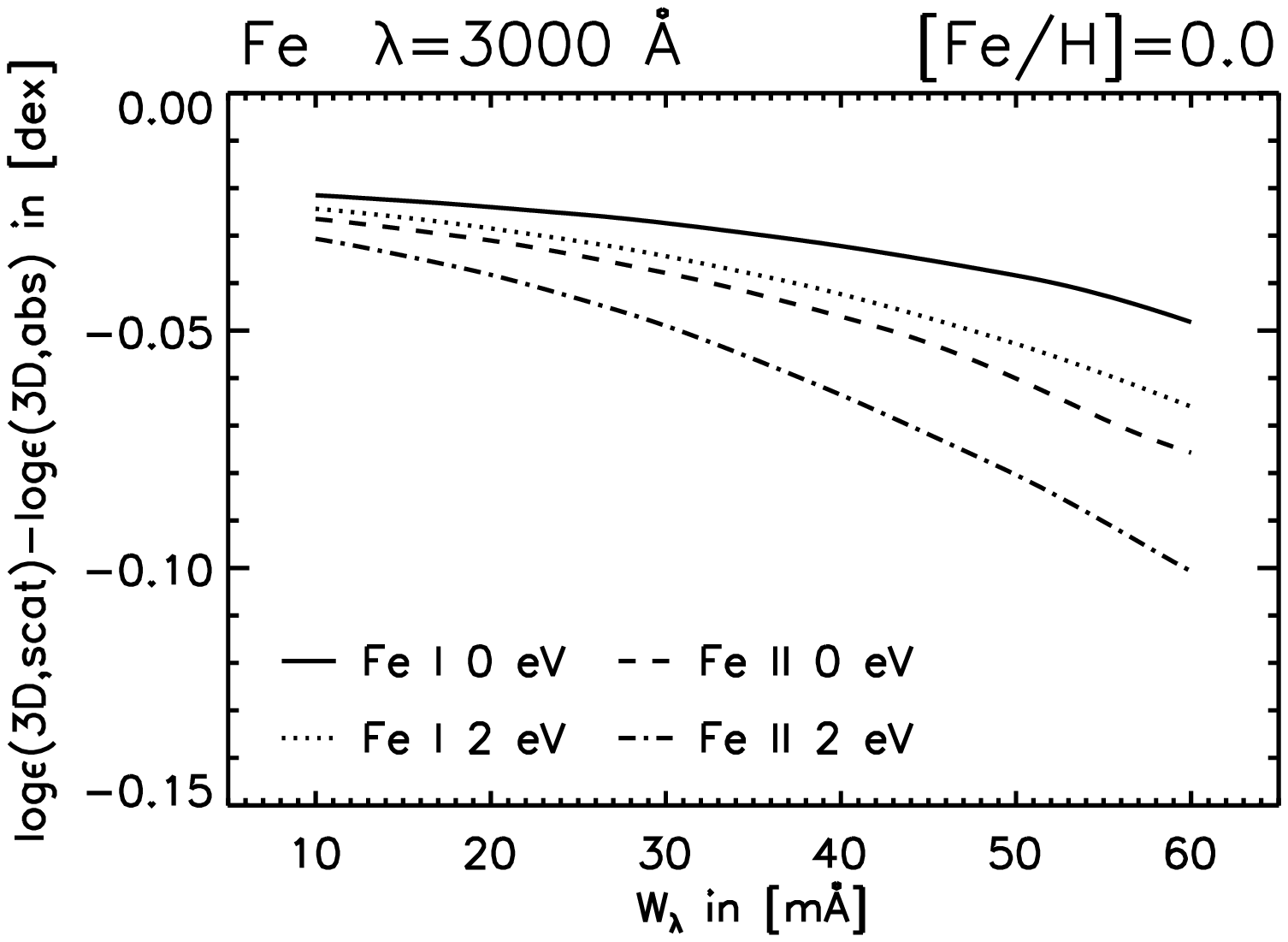}
\includegraphics[width=6.5cm]{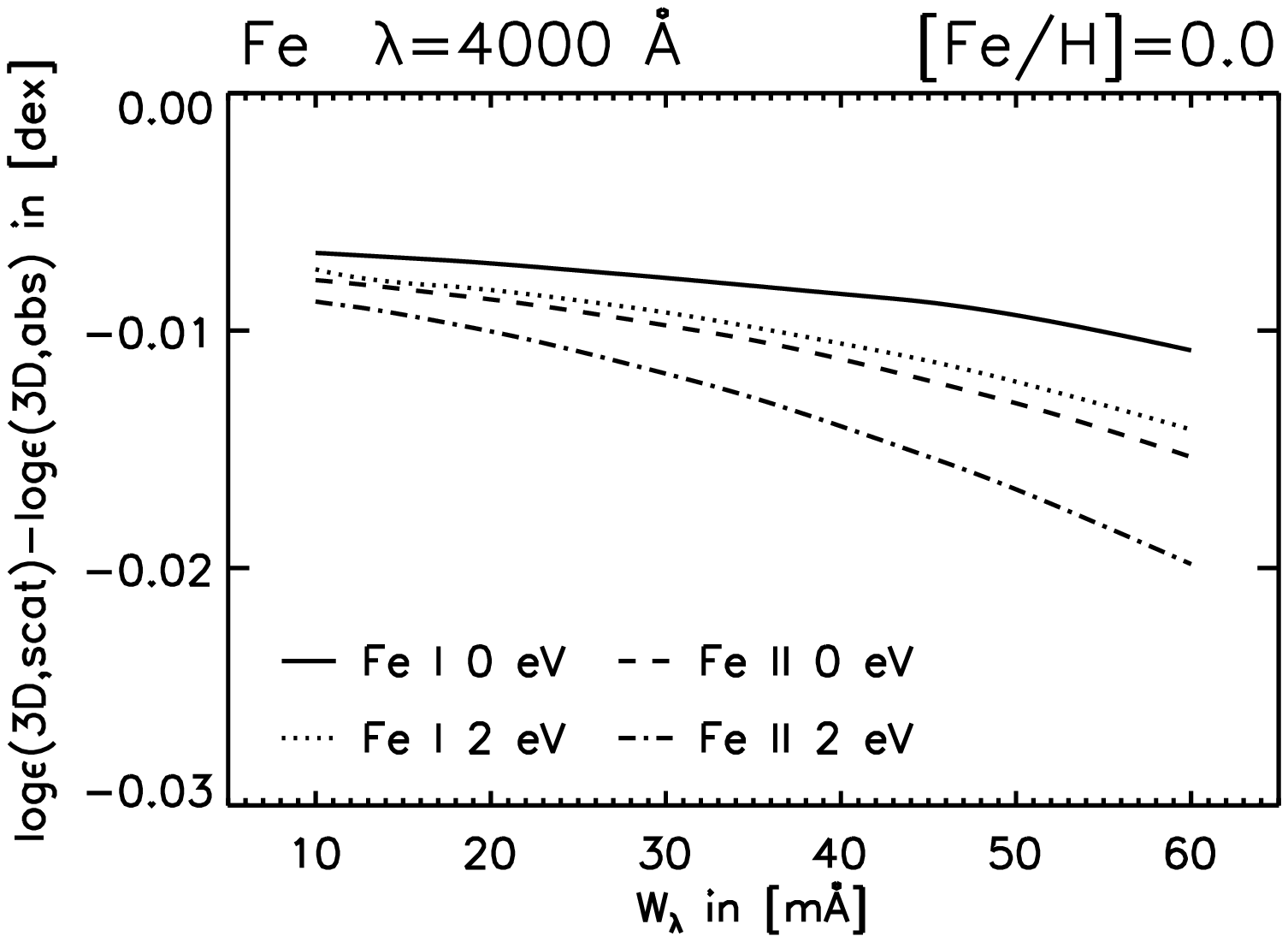}
}
\mbox{
\includegraphics[width=6.5cm]{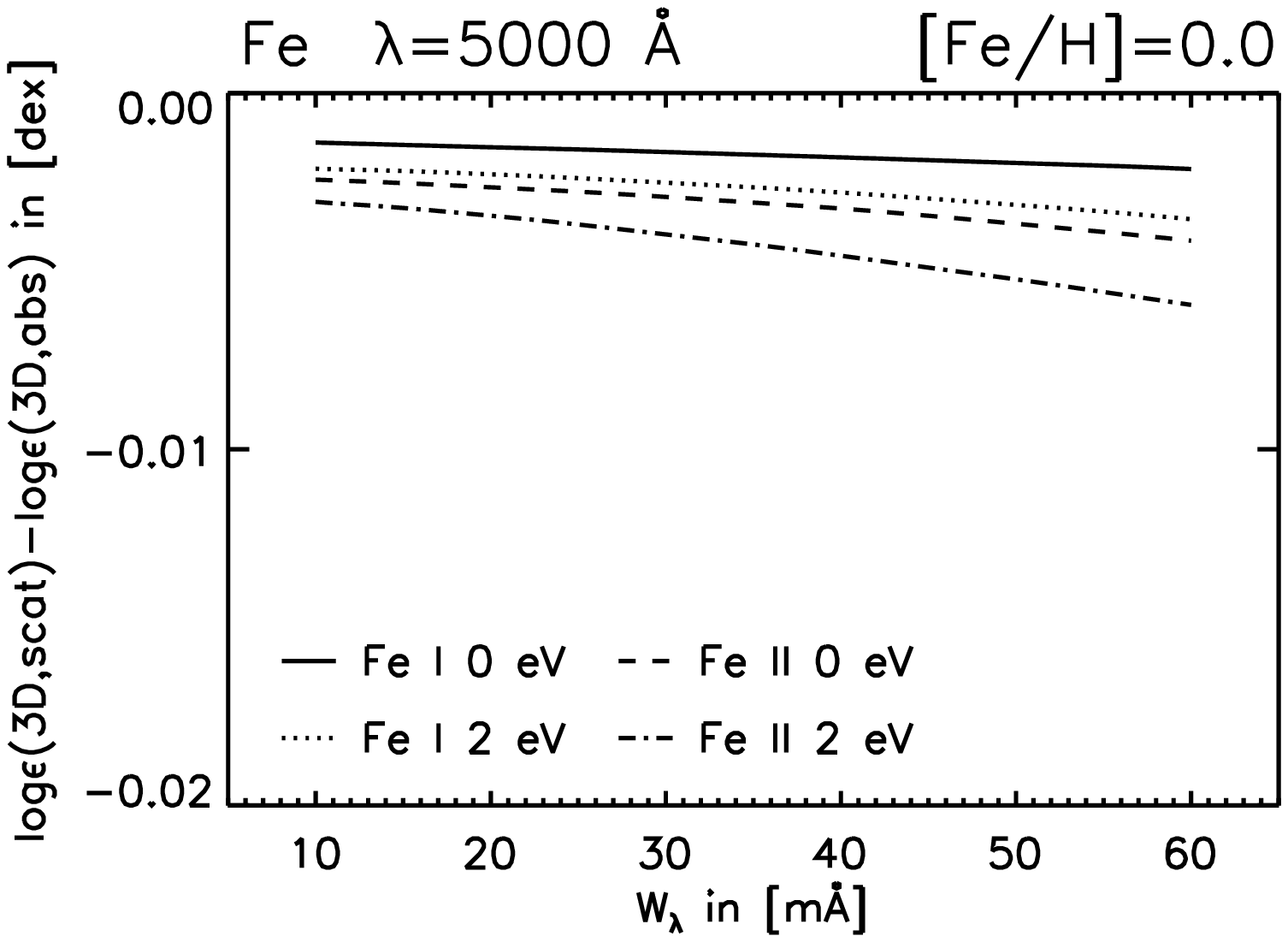}
\includegraphics[width=6.5cm]{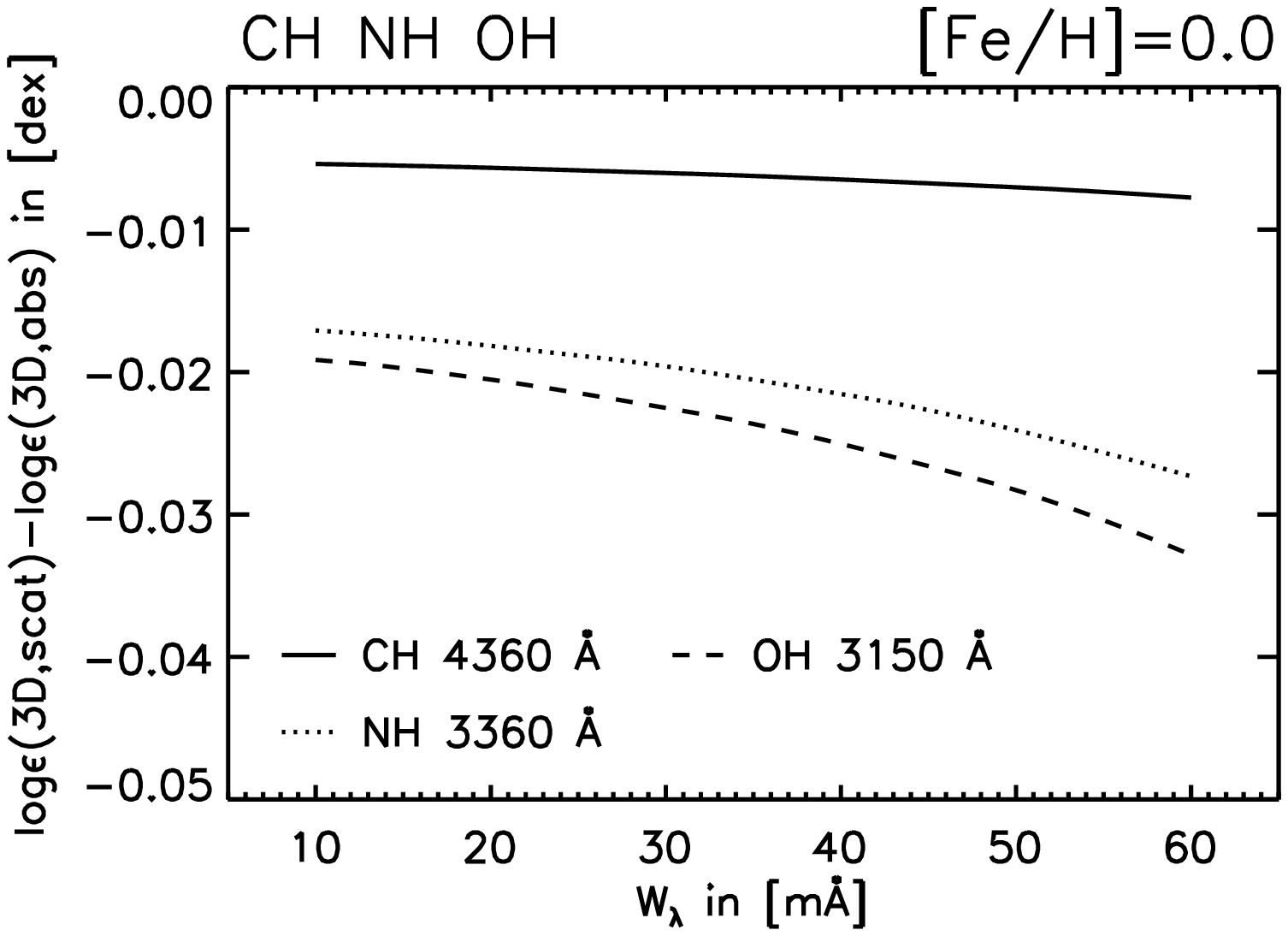}
}
\caption{Same as Fig.~\ref{fig:m3cog}, but computed for the 3D model with $\mathrm{[Fe/H]}=0.0$. Note the different scaling of the vertical axes.}
\label{fig:m0cog}
\end{figure*}

Computing line profiles for the OH molecule, \citet{GonzalezHernandezetal:2010} recently reported on a dependence of their 3D oxygen abundances on the opacity treatment in their 3D model atmospheres of metal-poor dwarf stars. The shallower temperature gradient of their model with 12 opacity bins significantly reduces the 3D$-$1D abundance corrections due to the strong temperature-dependence of molecular equilibrium populations. We therefore investigate the importance of the opacity treatment for the 3D$-$3D scattering effects of the OH 3150 line at $\mathrm{[Fe/H]}=-3.0$. The resulting abundance corrections as a function of equivalent width for the 4 bin 3D model and the 12 bin 3D model (see Sect.~\ref{sec:3Datmo}) are shown in Fig.~\ref{fig:OH12bin4bincog}. The overall dependence of $\Delta\log\epsilon$ on the opacity treatment is very small, as the temperature-decreased OH population of the 12 bin model is compensated by increased abundance (or oscillator strength in our case) on the curve of growth to reach a given equivalent width. The differences in the atmospheric stratification become slightly more noticeable for stronger lines, which sample a wider height range, but the deviation between the predictions of $\Delta\log\epsilon$ remains below $0.01$\,dex. The \emph{relative} effects of scattering on abundance measurements should therefore be insensitive to the opacity treatment used for constructing the model atmosphere when comparing 3D line formation computations.

\afterpage{\clearpage}

\section{3D$-$1D abundance corrections}\label{sec:scatcorr3D1D}

Abundance analyses with 1D hydrostatic model atmospheres are still commonplace in astrophysical research. We therefore compare the curves of growth derived from the 3D models with the results obtained for 1D hydrostatic \texttt{MARCS} models.

\begin{figure*}[htbp]
\centering
\includegraphics[width=11cm]{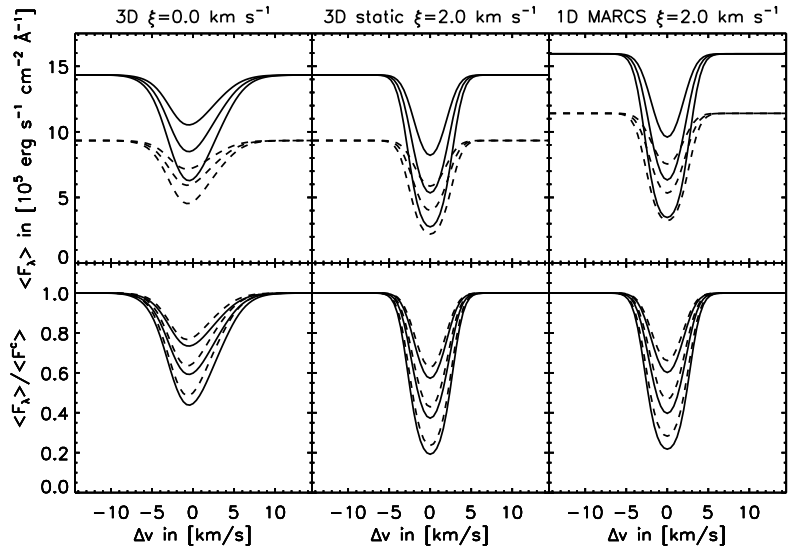}\vspace{0.8cm}
\caption{Spatial averages of flux profiles as functions of Doppler shift $\Delta v$ from the line center for the \ion{Fe}{I} line of Fig.~\ref{fig:fluxbisec}, computed treating scattering as absorption (dashed lines) and as coherent scattering (solid lines) using the 3D model with $\mathrm{[Fe/H]}=-3.0$ (left column), the 3D model with $\mathrm{[Fe/H]}=-3.0$ and with velocity fields set to zero (3D static model, center column) and the 1D hydrostatic \texttt{MARCS} model with $\mathrm{[Fe/H]}=-3.0$ (right column). The bottom row shows normalized profiles. The 3D static model and the 1D model assume microturbulent broadening with $\xi=2.0$\,km\,s$^{-1}$; line strengths of the 1D calculations needed to be increased by $+0.7$\,dex to obtain similar equivalent widths (see text).}
\label{fig:FeIprofiles}
\end{figure*}

\begin{figure*}[htbp]
\centering
\includegraphics[width=7cm]{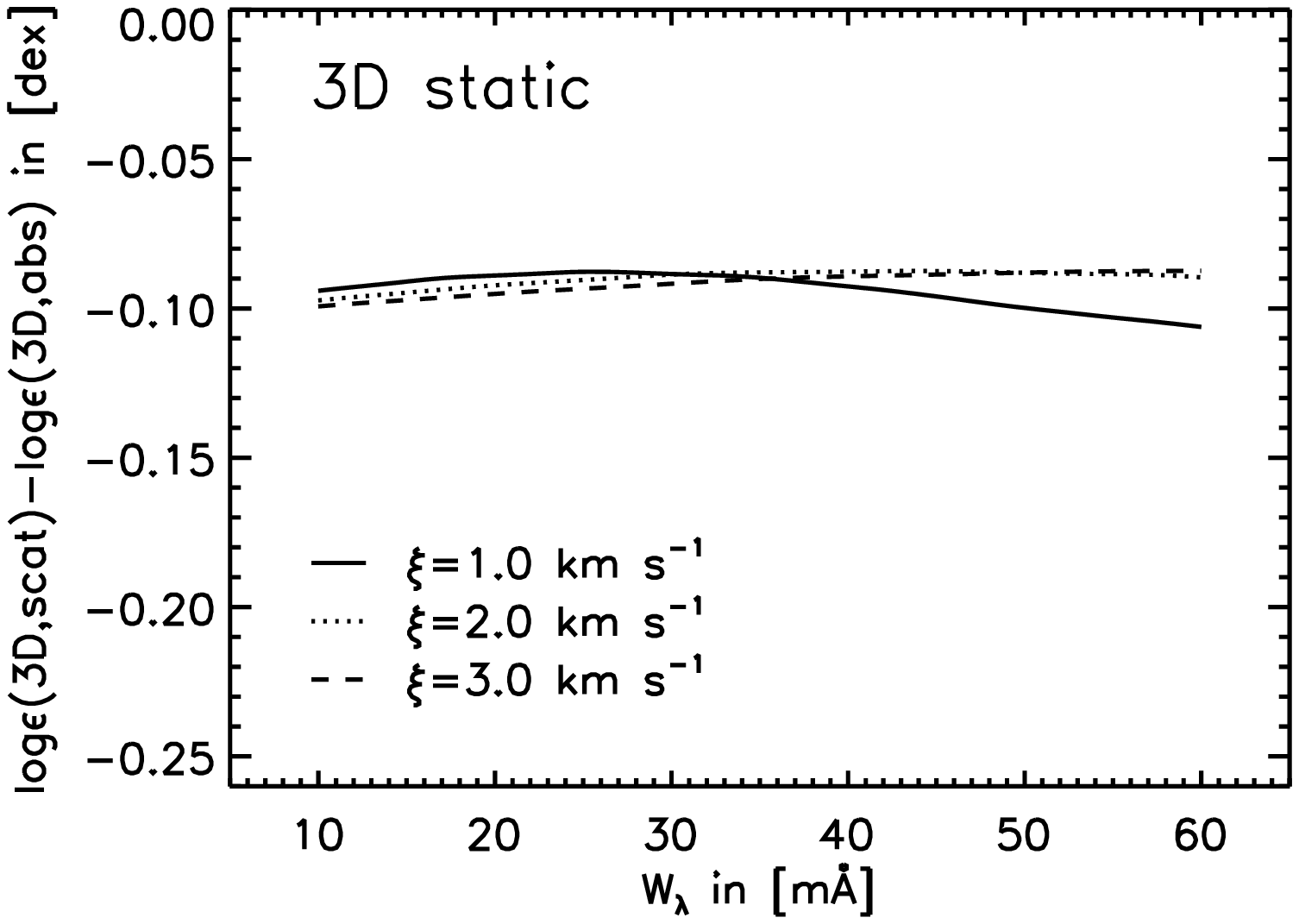}
\includegraphics[width=7cm]{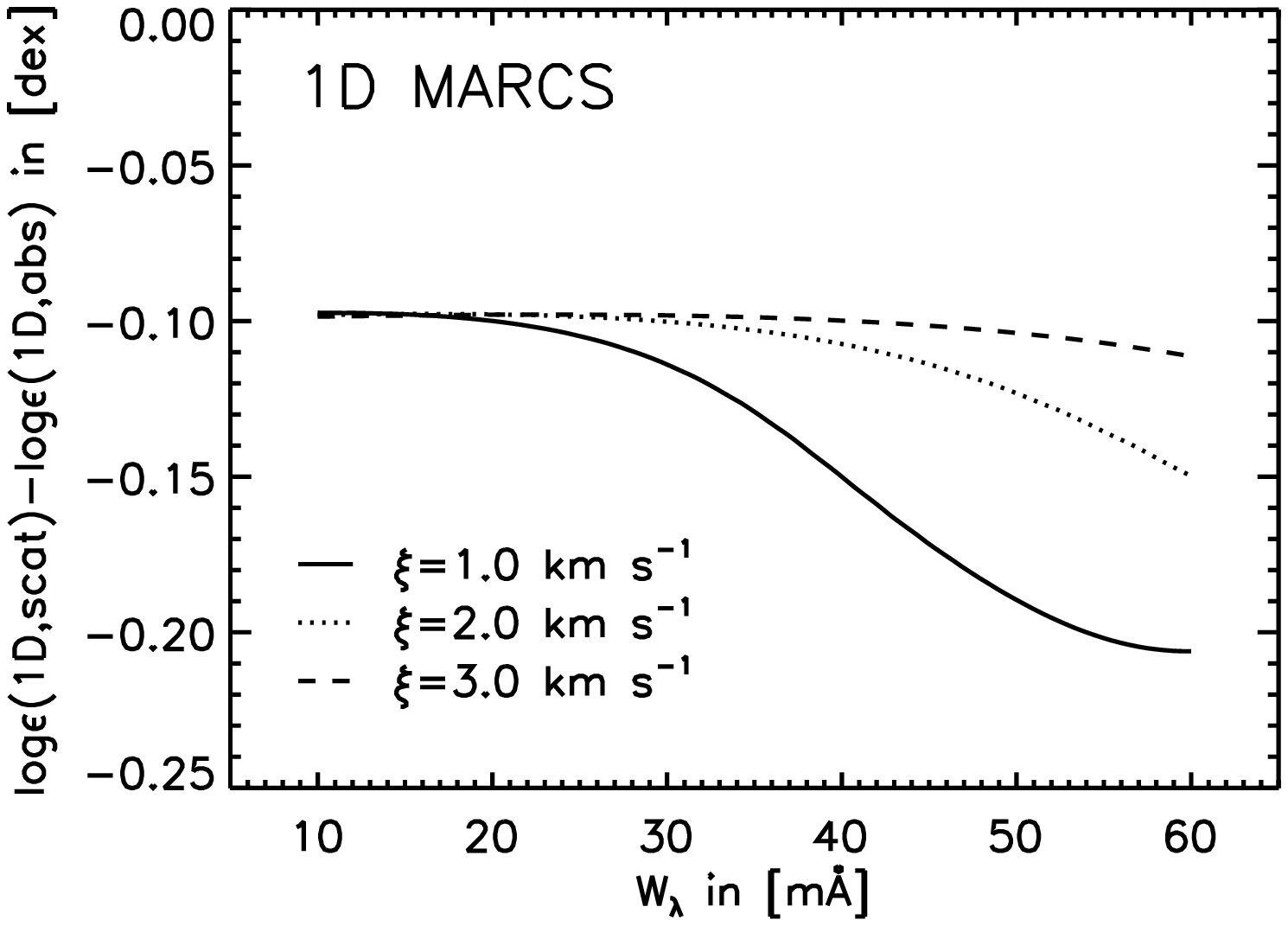}
\caption{Abundance corrections for the same \ion{Fe}{I} line as shown in Fig.~\ref{fig:fluxbisec}, but computed for the 3D model with $\mathrm{[Fe/H]}=-3.0$ and with velocity fields set to zero (left panel) and for the corresponding 1D hydrostatic \texttt{MARCS} model with $\mathrm{[Fe/H]}=-3.0$ (right panel), assuming microturbulent broadening with $\xi=1.0$\,km\,s$^{-1}$ (solid lines), $\xi=2.0$\,km\,s$^{-1}$ (dotted lines), and $\xi=3.0$\,km\,s$^{-1}$ (dashed lines).}
\label{fig:FeI30003Dstatic1D}
\end{figure*}

In order to establish the importance of the 3D structure and velocity fields, we first determine the scattering effects on a \ion{Fe}{I} line with $\chi=0$\,eV that forms in a 3D atmosphere with zero gas velocities and in a 1D hydrostatic \texttt{MARCS} model atmosphere with $\mathrm{[Fe/H]}=-3.0$. Missing broadening through gas motion is replaced with microturbulent broadening in these two cases, choosing $\xi=1.0$\,km\,s$^{-1}$, $\xi=2.0$\,km\,s$^{-1}$ and $\xi=3.0$\,km\,s$^{-1}$. Oscillator strengths for the calculations with the 1D model need to be increased by $0.7$\,dex to obtain approximately the same range of equivalent widths as in the 3D case. The adjustment compensates for lower absorber population numbers, which again stem from the much shallower temperature gradient above the surface of the 1D model compared to the 3D atmosphere; low-excitation transitions at low metallicity are particularly affected \citep[see][]{Colletetal:2007}. In the absence of Doppler-shifts, profiles are exactly symmetric (center column and right column of Fig.~\ref{fig:FeIprofiles}). Resulting abundance corrections are shown in Fig.~\ref{fig:FeI30003Dstatic1D}. In the 3D static case, emission from granules and intergranular lanes is no longer separated in wavelength. At low microturbulence ($\xi=1.0$\,km\,s$^{-1}$), the line profiles saturate earlier compared to the 3D case with velocity field. Desaturation through scattering effects is reduced as the more weakly scattering granules dominate the radiative flux, resulting in less dependence of $\Delta\log\epsilon$ on line strength (solid line in the left panel of Fig.~\ref{fig:FeI30003Dstatic1D}). Increasing microturbulent broadening delays saturation of line profiles in the scattering and absorption cases, extending the linear part of the curve of growth,  and reduces thermalization as line opacity is distributed over a wider profile. $\Delta\log\epsilon$ is almost constant (dotted line and dashed line in the left panel of Fig.~\ref{fig:FeI30003Dstatic1D}). The weakest lines at $W_{\lambda}=10$\,m{\AA} are independent from microturbulent broadening and yield approximately the same abundance corrections for all values of $\xi$.

The 1D \texttt{MARCS} model exhibits similar scattering abundance corrections at the lowest equivalent widths with $\Delta\log\epsilon=-0.1$\,dex. They are more severe for stronger lines compared to the 3D static case, reaching $-0.2$\,dex at $W_{\lambda}=60$\,m{\AA} for $\xi=1.0$\,km\,s$^{-1}$ (right panel of Fig.~\ref{fig:FeI30003Dstatic1D}). Increasing microturbulence again desaturates both the absorption and scattering line profiles and reduces thermalization, resulting in a weaker dependence of $\Delta\log\epsilon$ on line strength.

We repeat the profile computation for the same fictitious lines as in Sect.~\ref{sec:scatcorr3D} using the 1D \texttt{MARCS} models and compare 3D$-$1D abundance corrections with coherent scattering and treating scattering as absorption. Microturbulent broadening was fixed at $\xi=2.0$\,km\,s$^{-1}$ to simplify the discussion. Contrary to the differential comparison of 3D curves of growth in the previous section, the \emph{absolute} 3D$-$1D abundance corrections in each case are more sensitive to the stratification of the model atmospheres. However, a test using our 12 bin model with $\mathrm{[Fe/H]}=-3.0$ for the OH line showed that the deviation between the 3D$-$1D abundance corrections with scattering and treating scattering as absorption was insensitive to the temperature profile of the 3D model. The following discussion thus focusses on the relative effect of scattering.

\begin{figure*}[p]
\centering
\mbox{
\includegraphics[width=6.5cm]{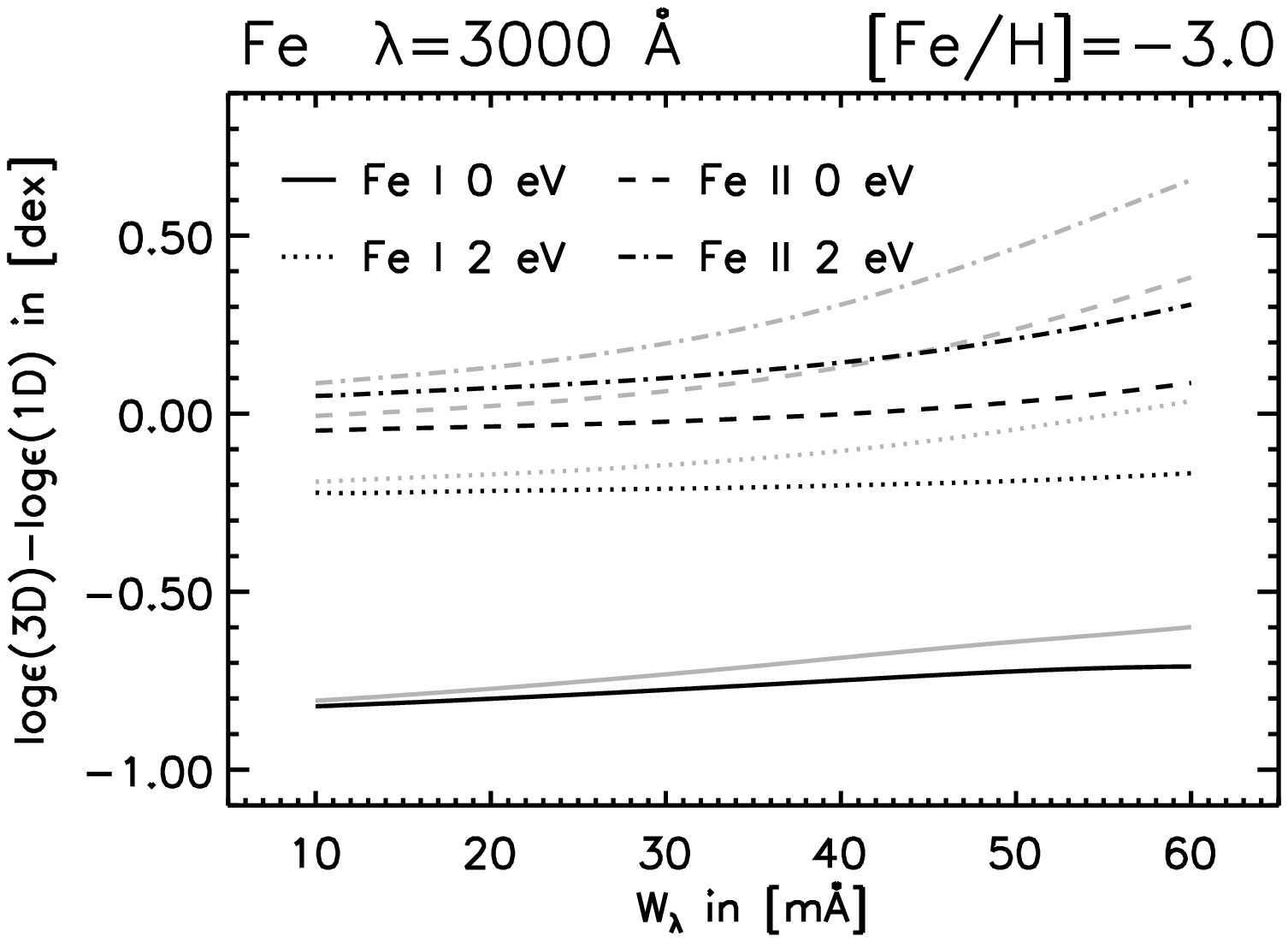}
\includegraphics[width=6.5cm]{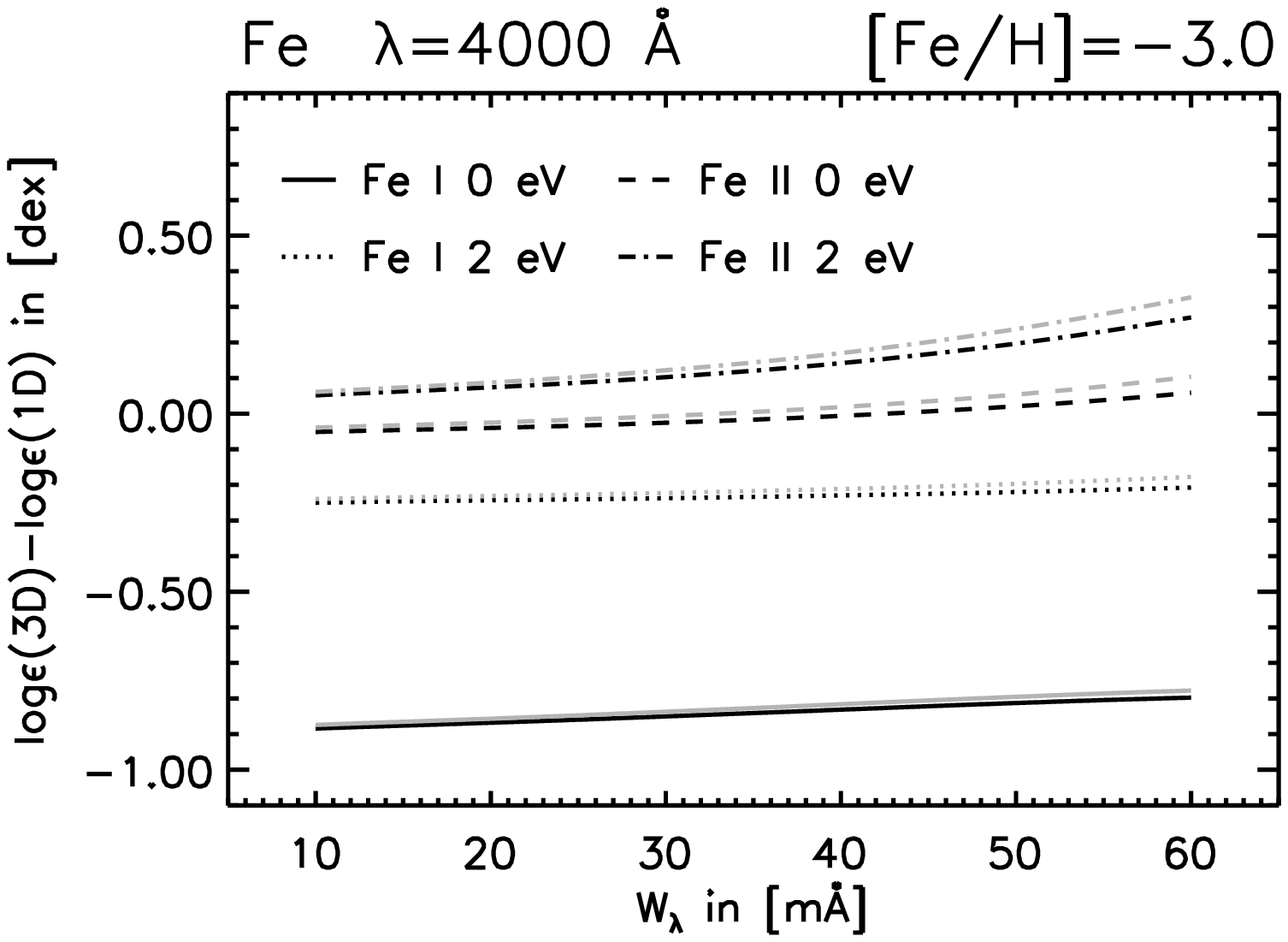}
}
\mbox{
\includegraphics[width=6.5cm]{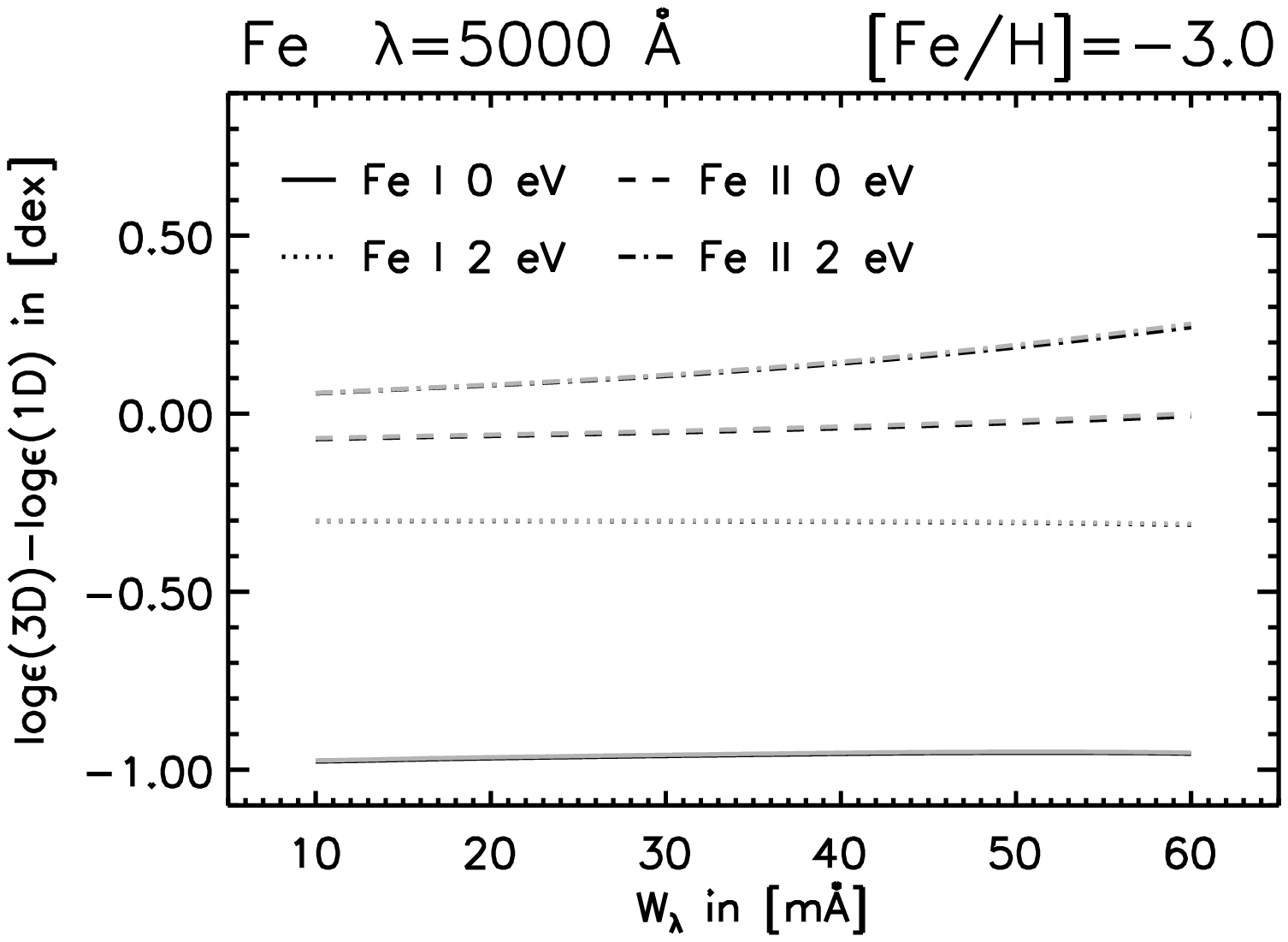}
\includegraphics[width=6.5cm]{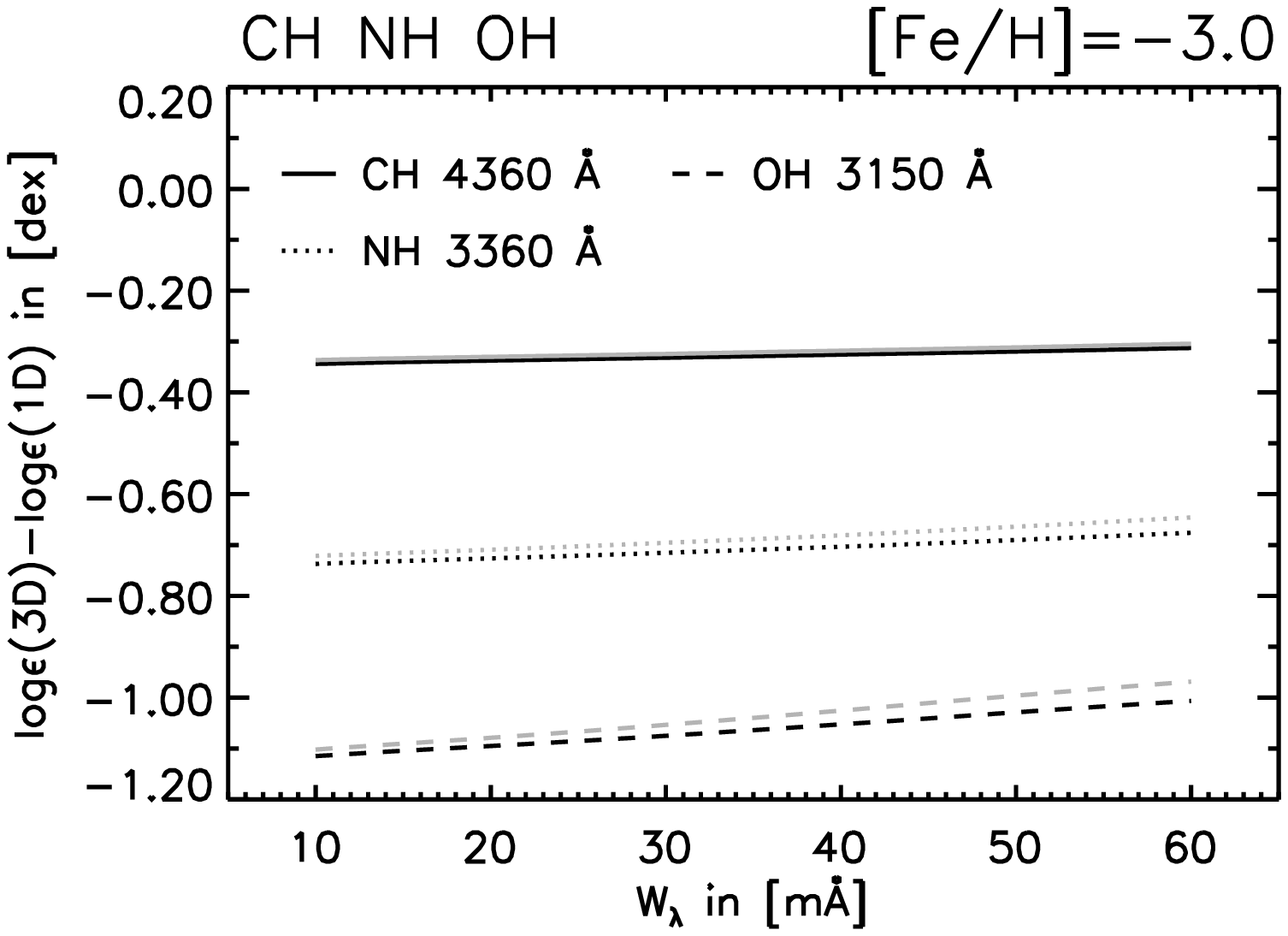}
}
\caption{\textit{Upper left panel to lower left panel:} 3D$-$1D abundance corrections for fictitious \ion{Fe}{I} lines and \ion{Fe}{II} lines with excitation potential $\chi=0$\,eV and $\chi=2$\,eV at 3000\,{\AA}, 4000\,{\AA} and 5000\,{\AA}, computed for the 3D and 1D models with $\mathrm{[Fe/H]}=-3.0$, including coherent scattering (black) and treating scattering as absorption (gray). The 1D calculations assume microturbulent broadening with $\xi=2.0$\,km\,s$^{-1}$. \textit{Lower right panel:} Abundance corrections for typical CH, NH and OH lines for the same models. Note that feedback of the 3D abundance corrections on the molecular equilibrium is \emph{not} taken into account (see text).}
\label{fig:m33D1Dcog}
\end{figure*}

The upper left panel of Fig.~\ref{fig:m33D1Dcog} shows the 3D$-$1D abundance corrections for the \ion{Fe}{I} and \ion{Fe}{II} lines at 3000\,{\AA} for the models with $\mathrm{[Fe/H]}=-3.0$, computed with coherent scattering (black lines) and treating scattering as absorption (gray lines). The overall behavior of the scattering-as-absorption calculations follows the results of \citet{Colletetal:2007}, who used model atmospheres with comparable stellar parameters in their analysis, and will therefore not be further discussed. The 3D$-$1D abundance corrections with scattering are very similar to the 3D$-$1D absorption results for all weak lines as the 3D$-$3D and 1D$-$1D scattering effects reach similar strength ($\approx-0.1$\,dex in either case, see the right panel of Fig.~\ref{fig:FeI3000cog} and Fig.~\ref{fig:FeI30003Dstatic1D}). As $W_{\lambda}$ grows, the calculations predict stronger scattering corrections in 3D, which leads to a growing downward deviation from the scattering-as-absorption 3D$-$1D results for all species. High-excitation \ion{Fe}{II} lines exhibit the largest difference, requiring an adjustment of the 3D$-$1D scattering-as-absorption abundance corrections by up to $0.35$\,dex. It is important to keep in mind that the effect of scattering on abundance corrections in 1D depends on the choice of microturbulent broadening (right panel of Fig.~\ref{fig:FeI30003Dstatic1D}); choosing a smaller $\xi$ thus reduces the deviation between the 3D$-$1D abundance corrections. However, the general influence of microturbulence on 3D$-$1D corrections through the 1D curve of growth is often larger than that \citep[$>0.1$\,dex for low-excitation \ion{Fe}{I} lines with $W_{\lambda}=60$\,m{\AA} when $\xi$ is increased by $0.5$\,km\,s$^{-1}$; see Fig.~5 and Fig.~6 in][]{Colletetal:2007}.

At longer wavelengths, the importance of scattering decreases, and the 3D$-$1D abundance corrections with scattering come into closer agreement with the scattering-as-absorption calculations for all species and line strengths (upper right panel and lower left panel of Fig.~\ref{fig:m33D1Dcog}); the deviations reach $0.05$\,dex at $4000$\,{\AA} for the strongest \ion{Fe}{II} lines and almost vanish at $5000$\,{\AA}. The same is observed for CH, NH and OH molecular lines, with OH exhibiting the largest deviation of $0.05$\,dex between the 3D$-$1D corrections when scattering is included for the strongest lines, while the differences are negligible in the case of CH lines. The treatment of molecular lines in a 3D$-$1D comparison is more complex due to the strong non-linearities that stem from the temperature sensitivity of the equilibrium populations and their dependence on the chemical composition of the model, which may change dramatically at the lowest metallicities \citep{Colletetal:2007}. We fix the abundances of carbon, nitrogen and oxygen to the scaled solar composition; this leads to an offset of the \emph{absolute} 3D$-$1D corrections with respect to the results of \citet{Colletetal:2007}, but only weakly affects the differential comparison of scattering effects. The 3D$-$1D abundance corrections for molecules should therefore not be directly applied to 1D abundance analyses.

\begin{figure*}[p]
\centering
\mbox{
\includegraphics[width=6.5cm]{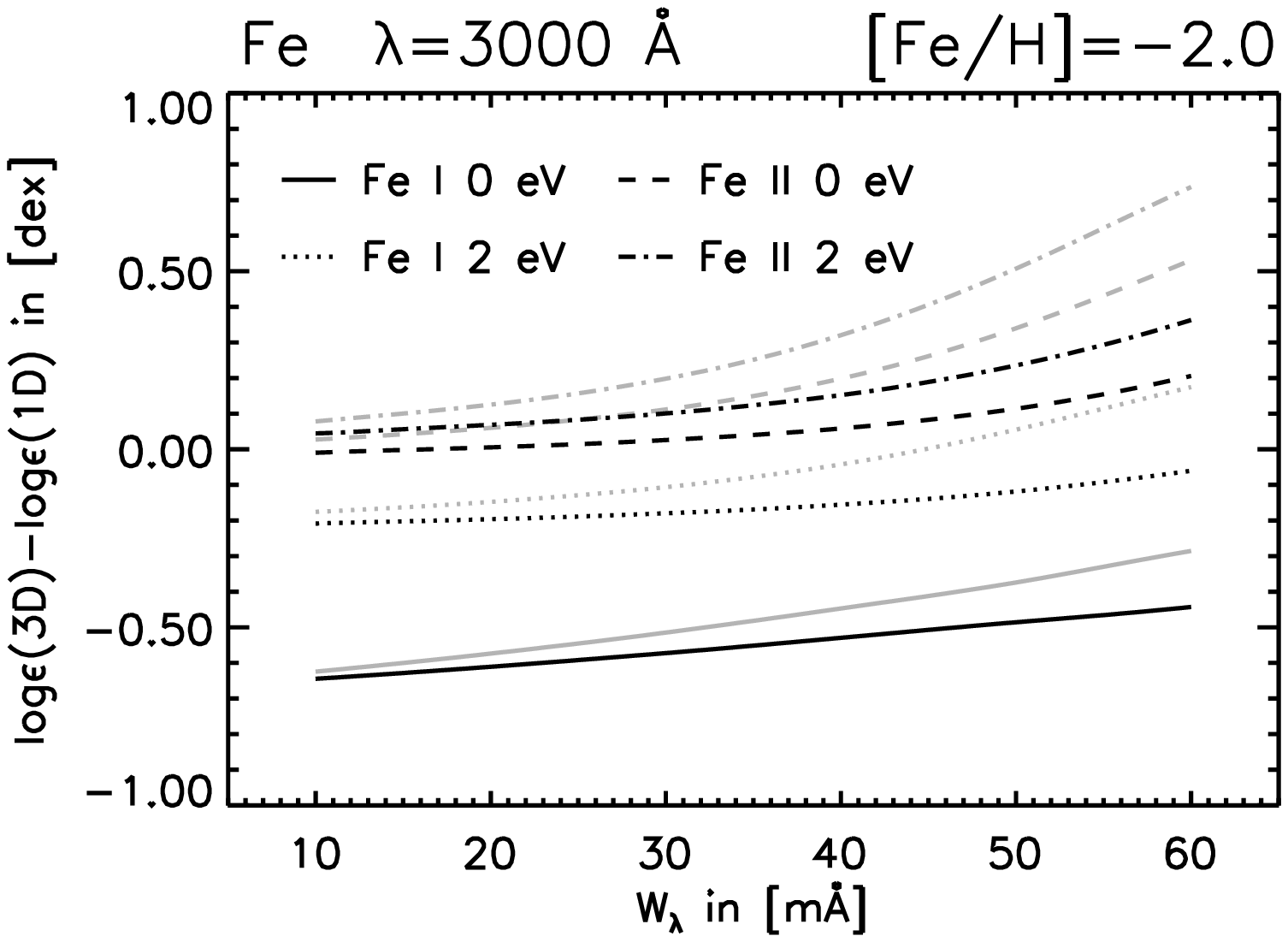}
\includegraphics[width=6.5cm]{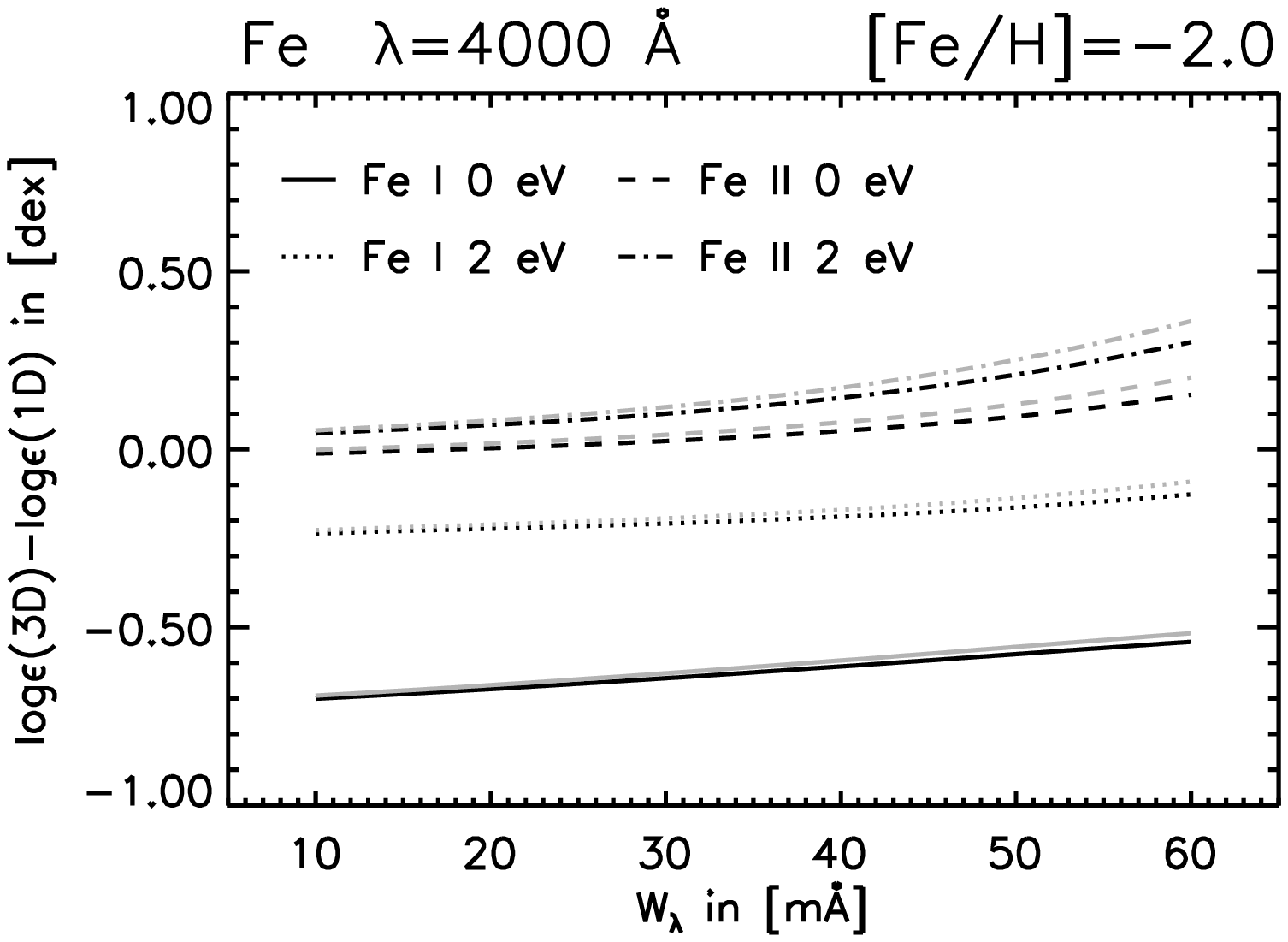}
}
\mbox{
\includegraphics[width=6.5cm]{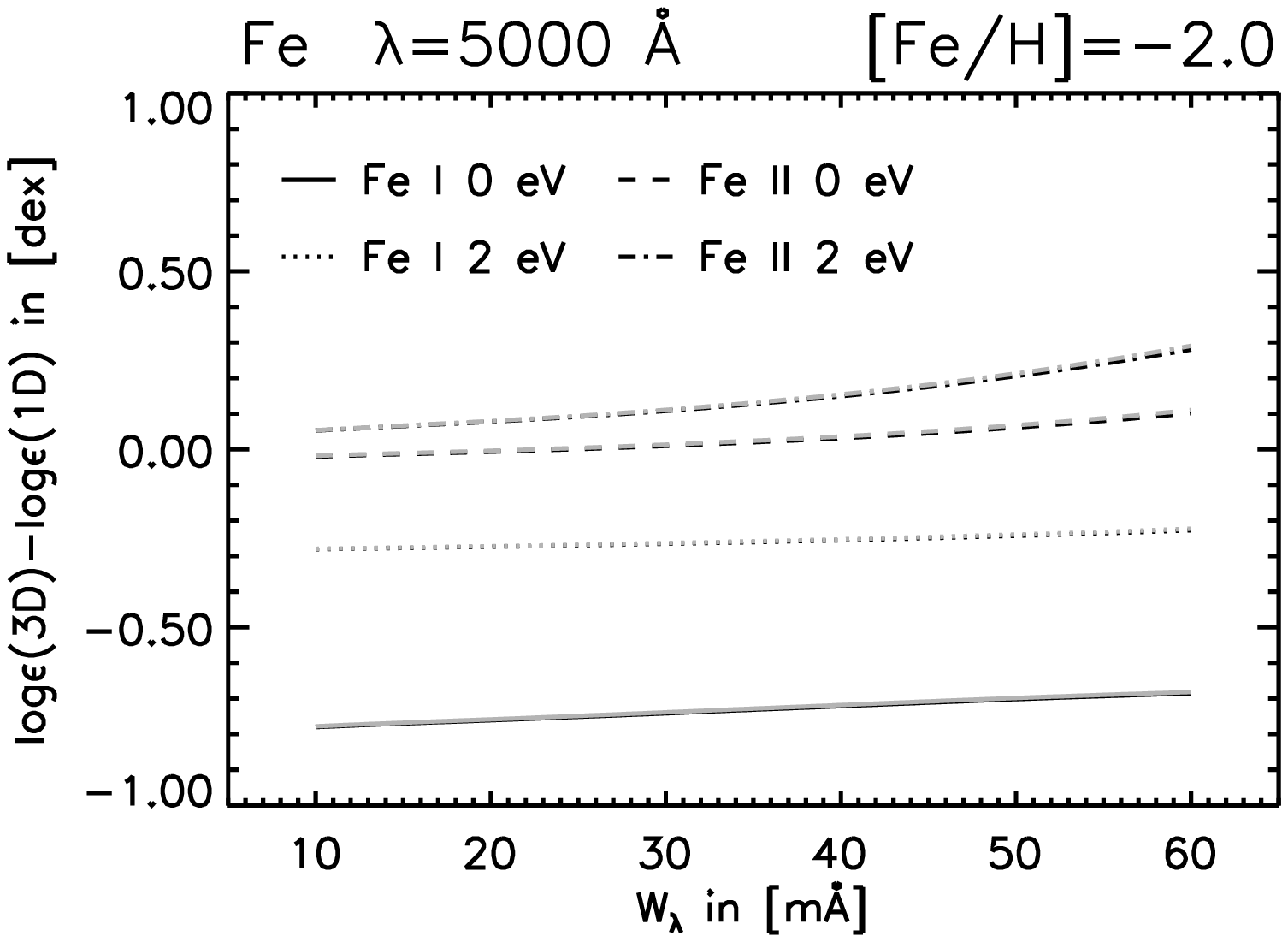}
\includegraphics[width=6.5cm]{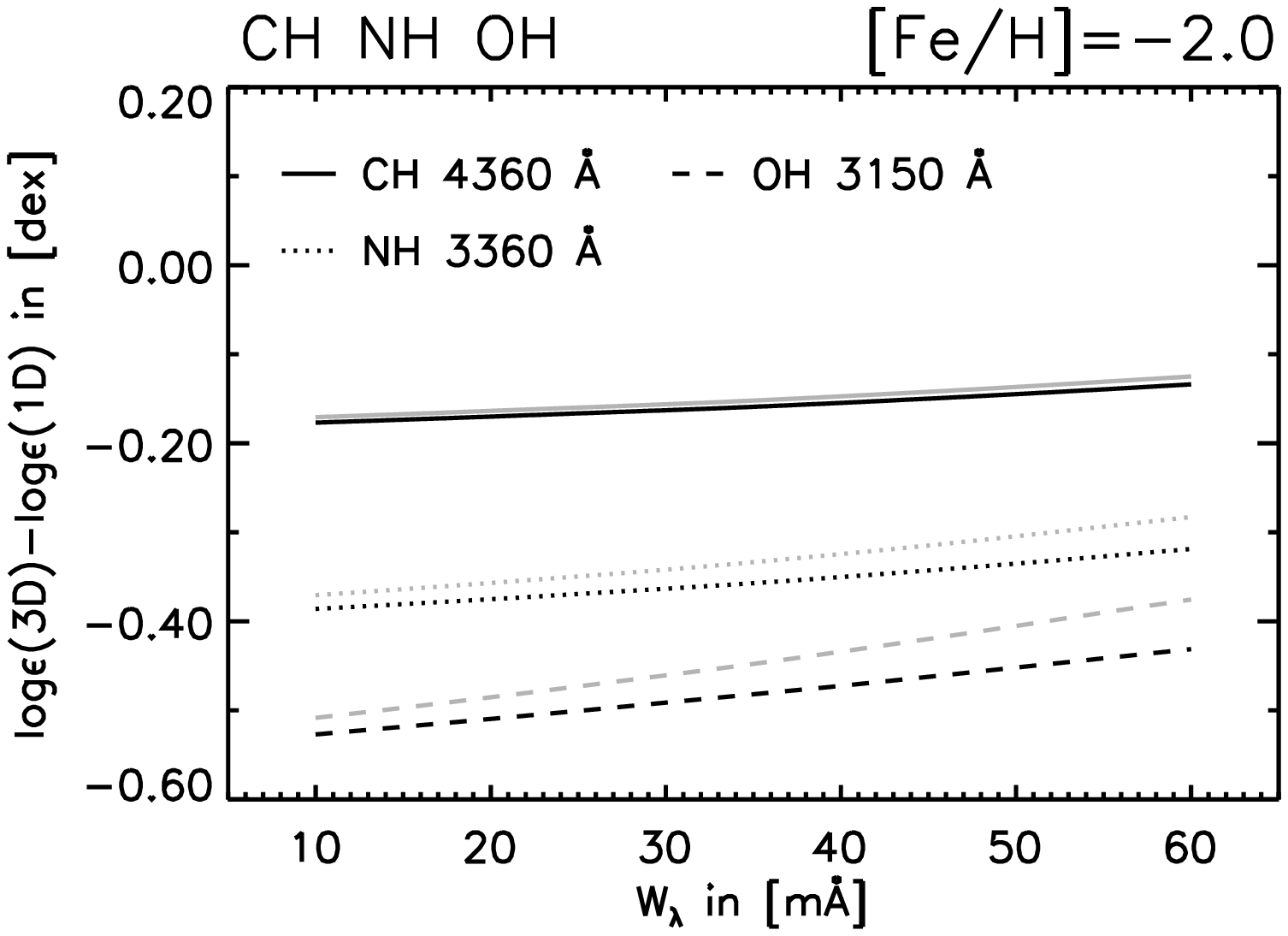}
}
\caption{Same as Fig.~\ref{fig:m33D1Dcog}, but computed for the models with $\mathrm{[Fe/H]}=-2.0$.}
\label{fig:m23D1Dcog}
\end{figure*}

\begin{figure*}[p]
\centering
\mbox{
\includegraphics[width=6.5cm]{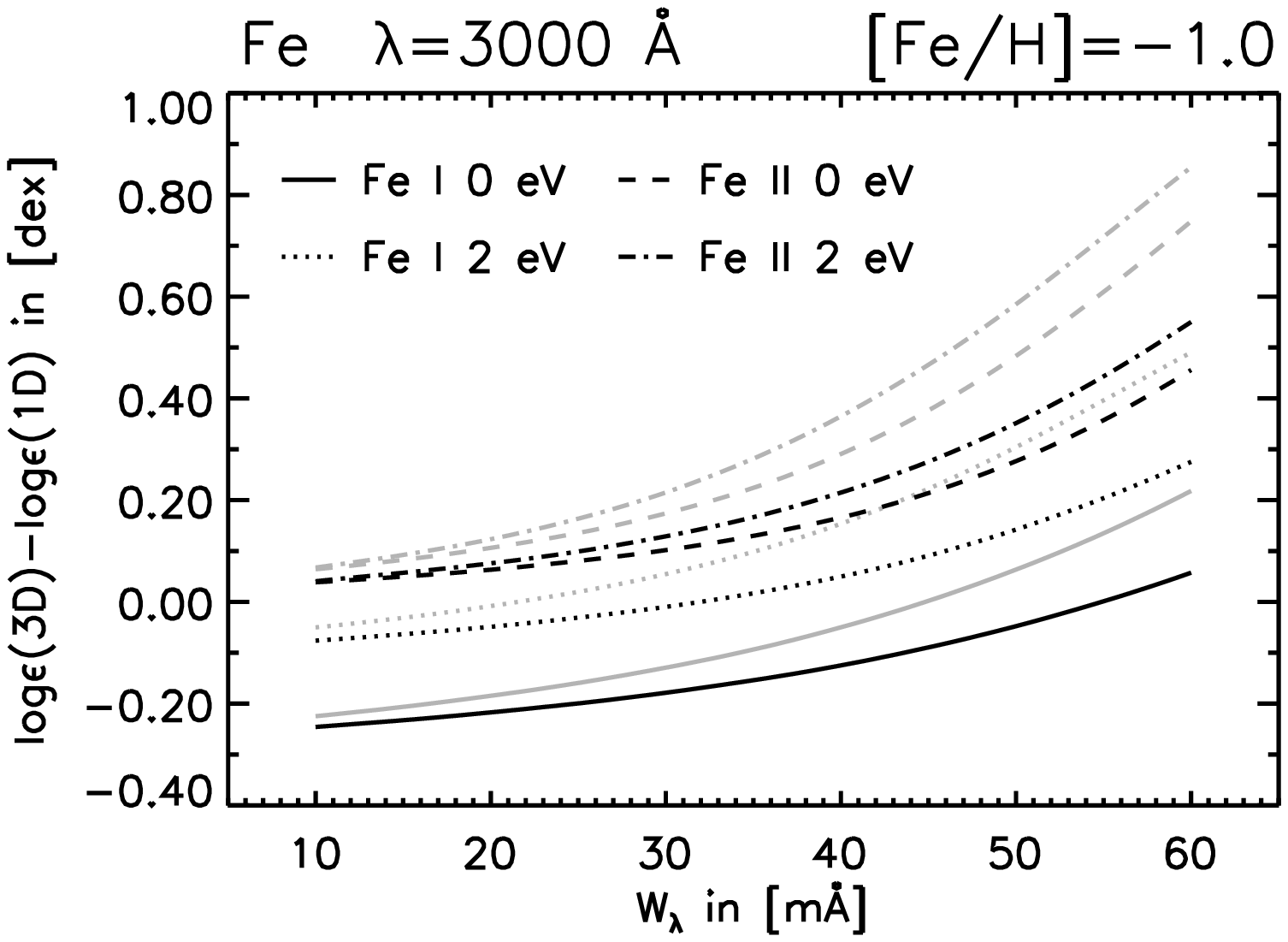}
\includegraphics[width=6.5cm]{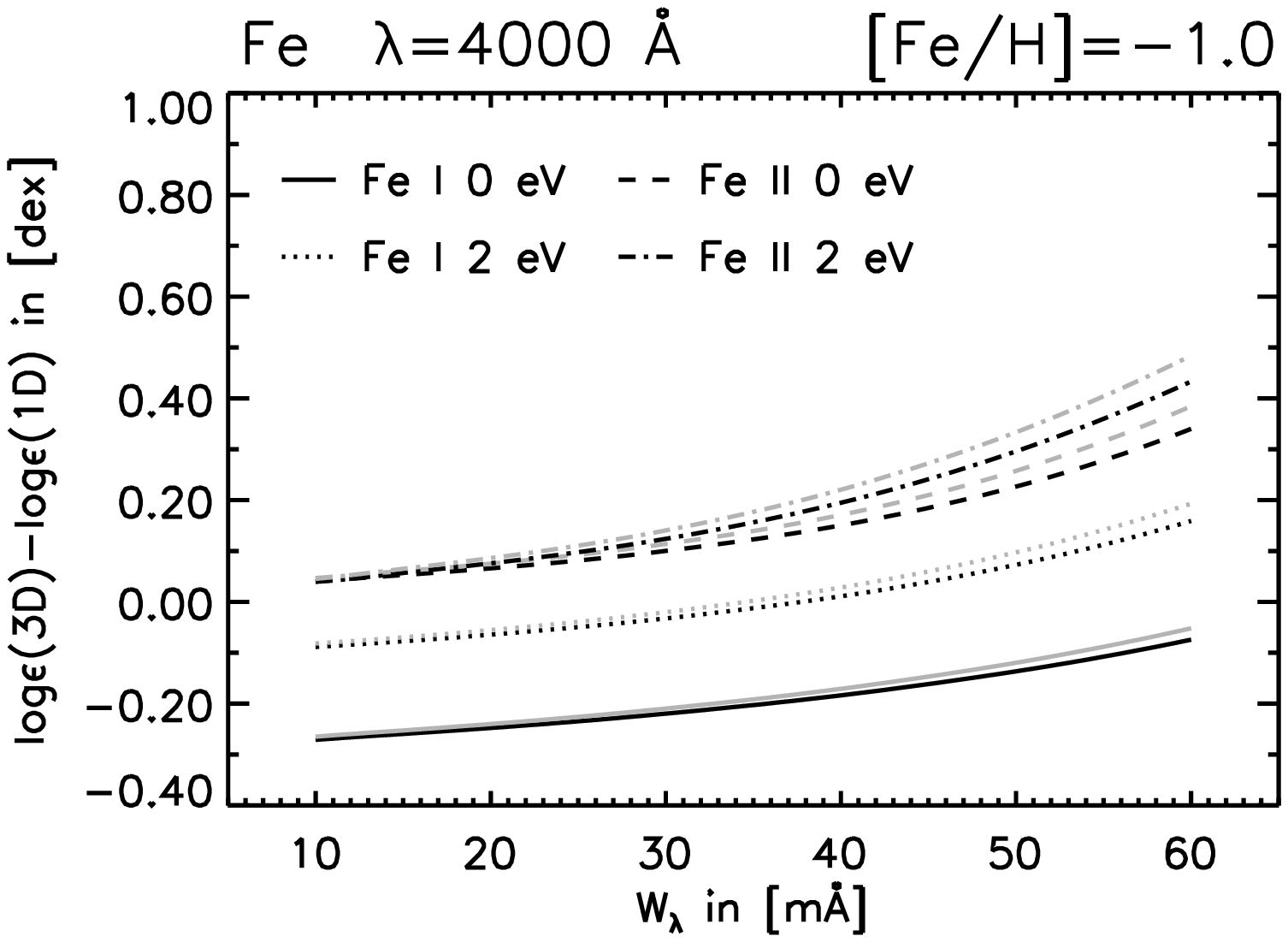}
}
\mbox{
\includegraphics[width=6.5cm]{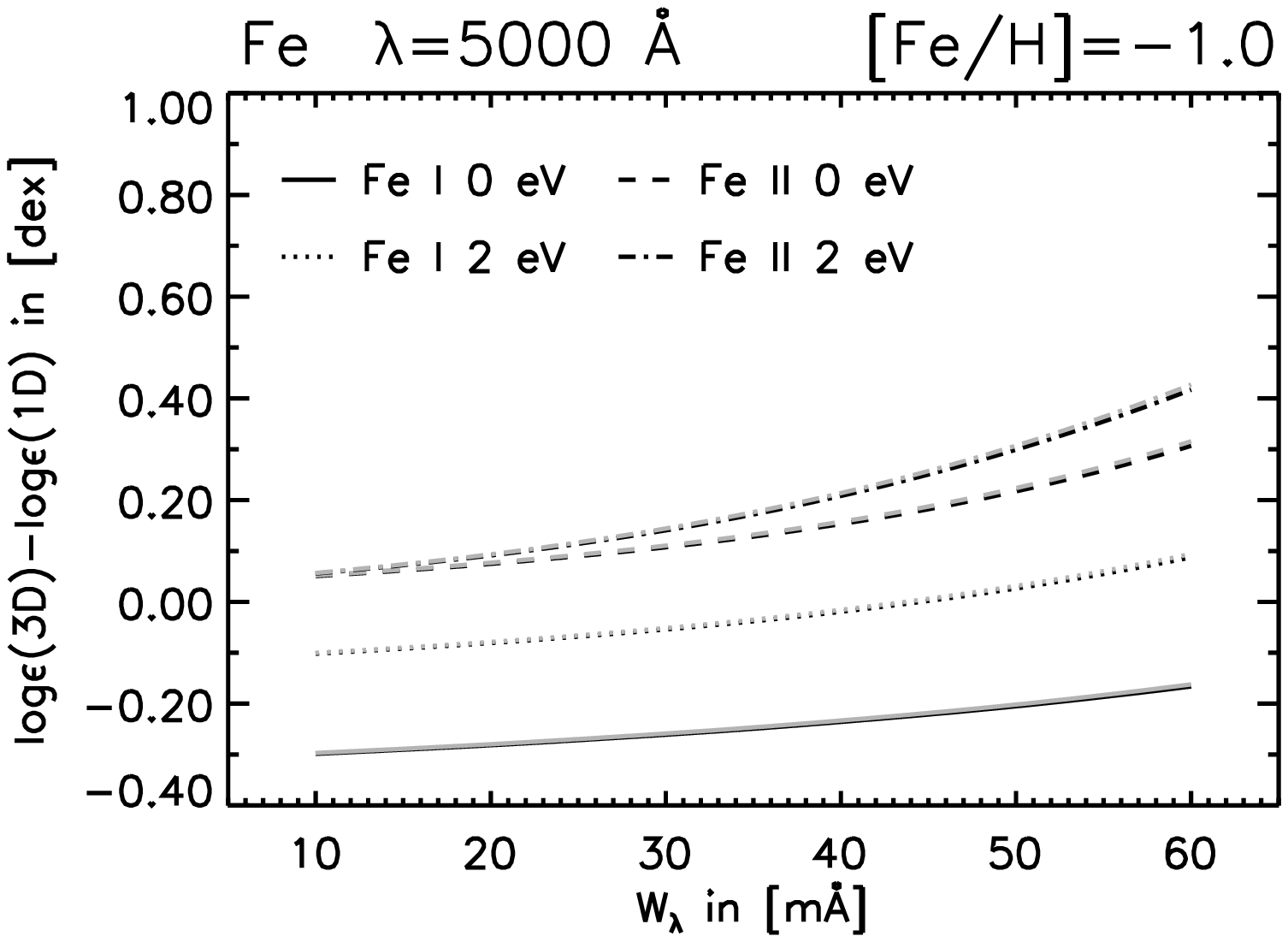}
\includegraphics[width=6.5cm]{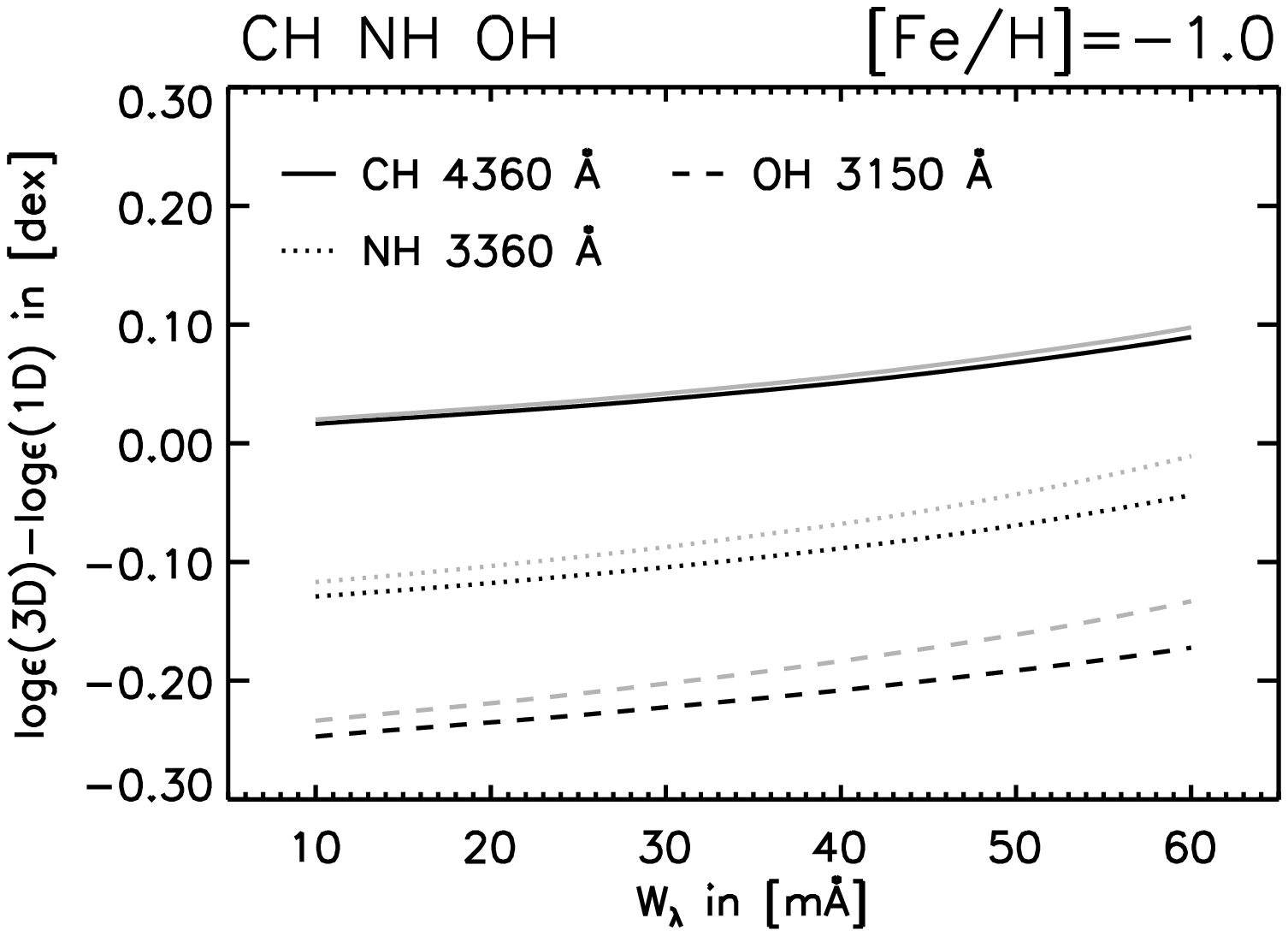}
}
\caption{Same as Fig.~\ref{fig:m33D1Dcog}, but computed for the models with $\mathrm{[Fe/H]}=-1.0$.}
\label{fig:m13D1Dcog}
\end{figure*}

\begin{figure*}[p]
\centering
\mbox{
\includegraphics[width=6.5cm]{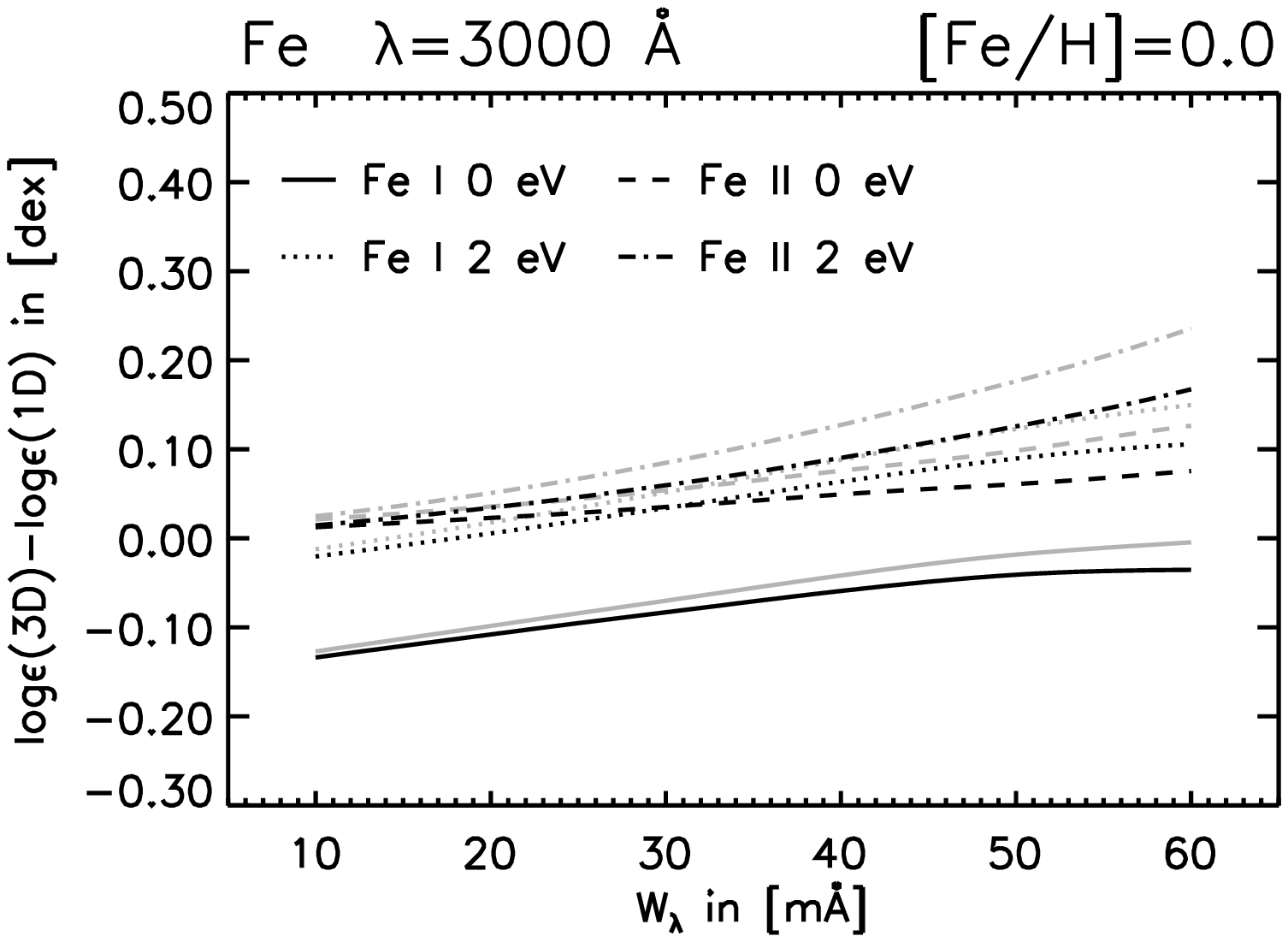}
\includegraphics[width=6.5cm]{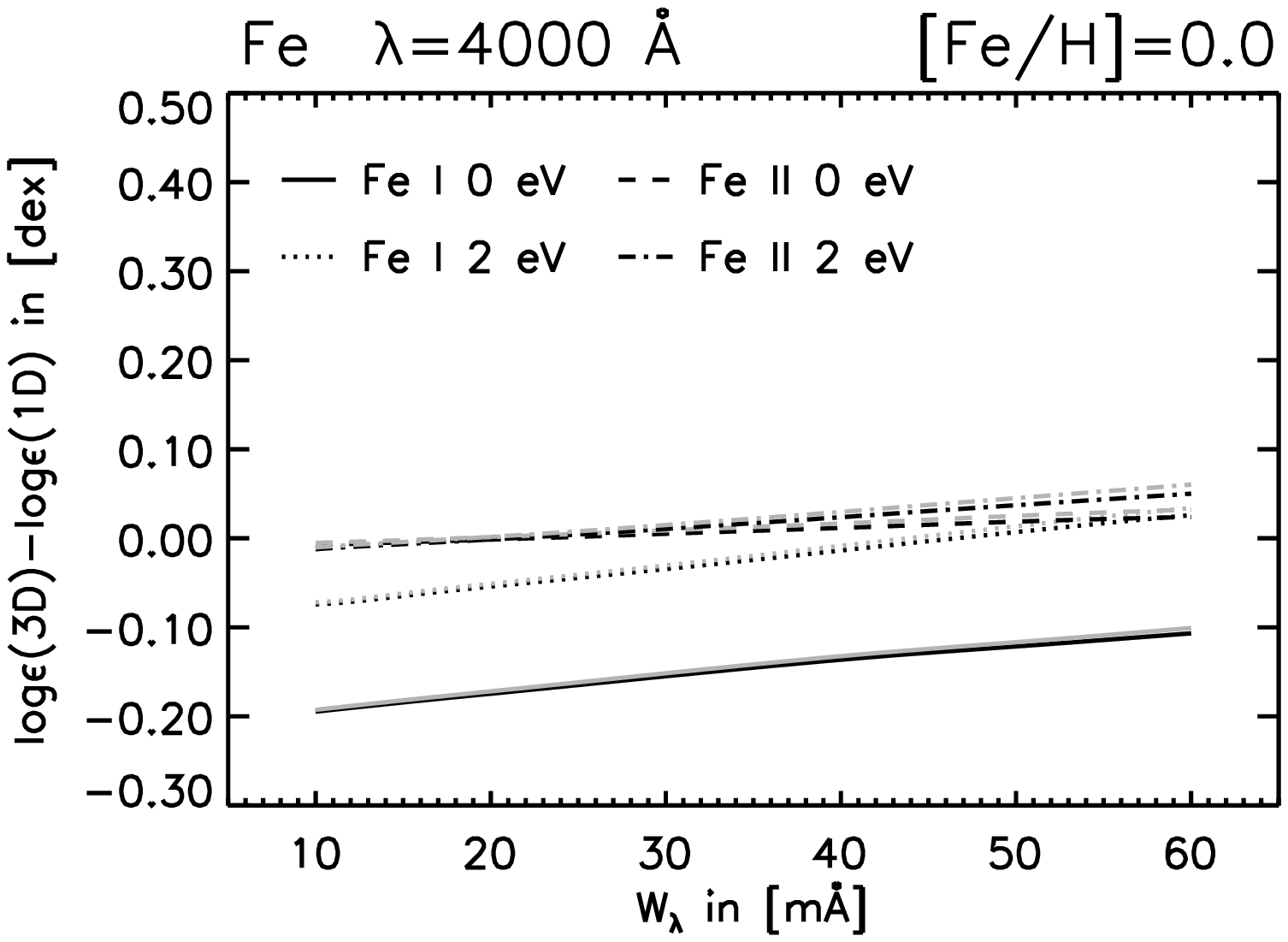}
}
\mbox{
\includegraphics[width=6.5cm]{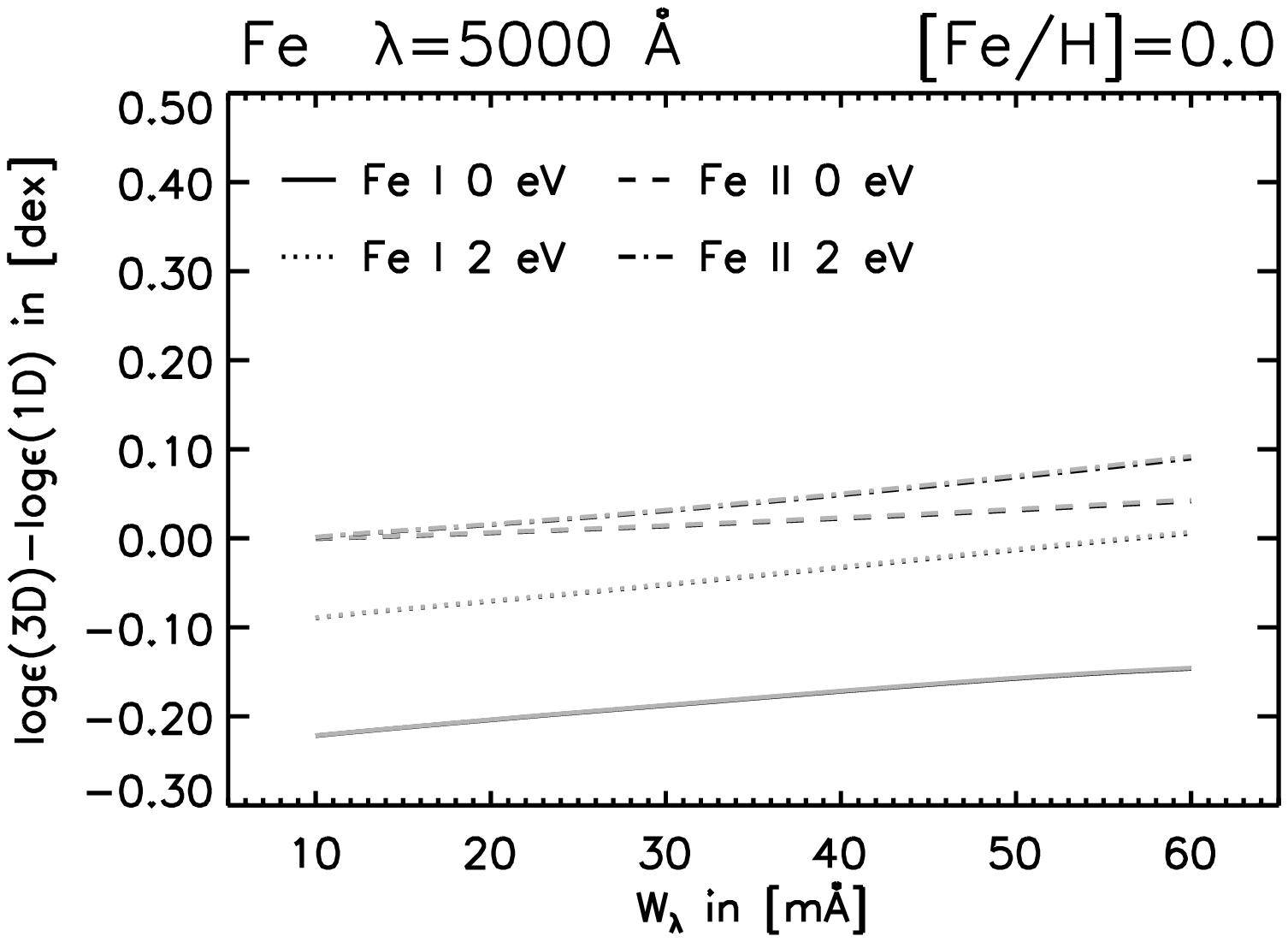}
\includegraphics[width=6.5cm]{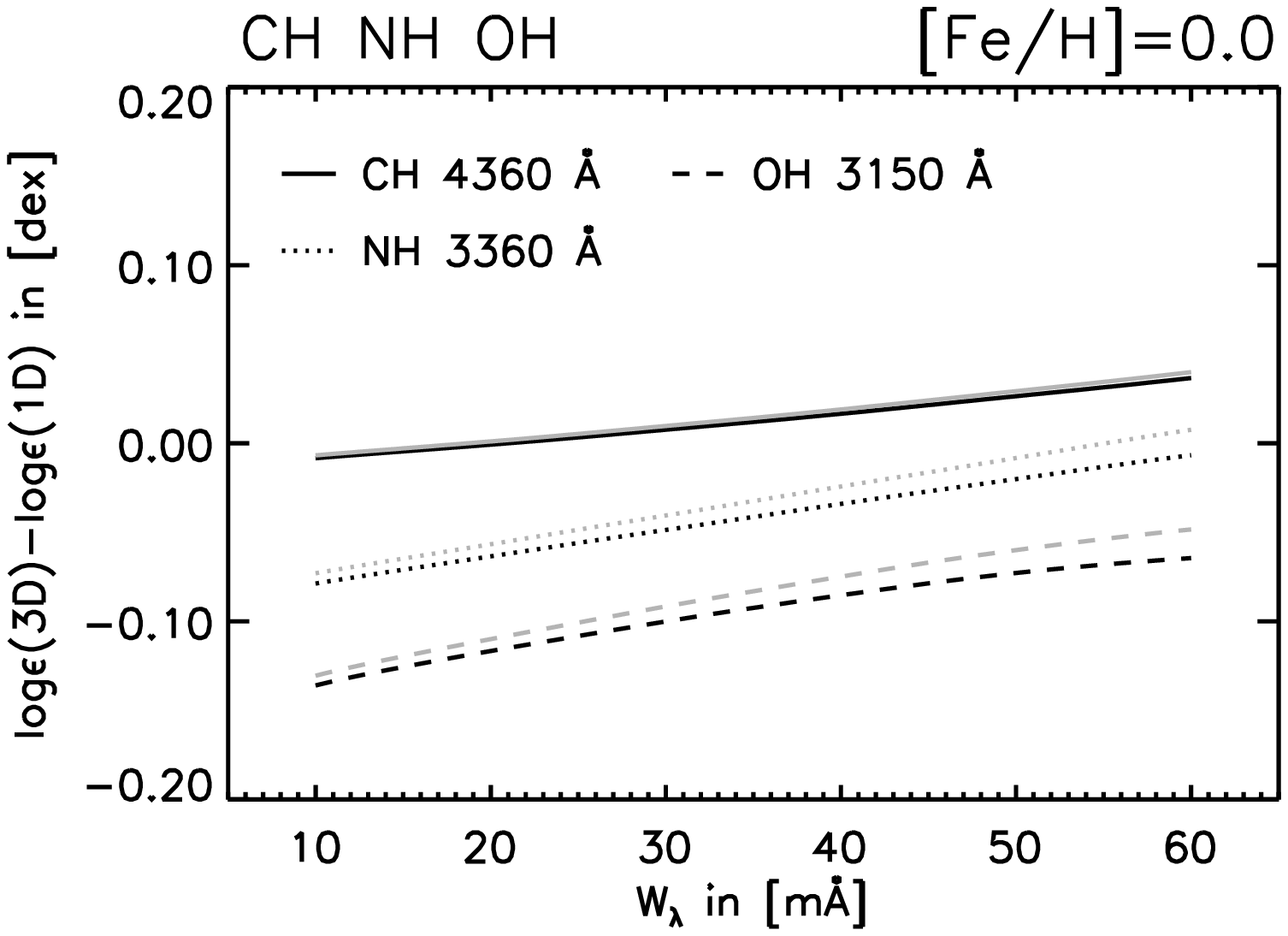}
}
\caption{Same as Fig.~\ref{fig:m33D1Dcog}, but computed for the models with $\mathrm{[Fe/H]}=0.0$.}
\label{fig:m03D1Dcog}
\end{figure*}

At metallicities $\mathrm{[Fe/H]}=-2.0$ and $\mathrm{[Fe/H]}=-1.0$, the predicted 3D$-$1D abundance corrections generally decrease due to the growing similarity between the temperature stratifications of the 3D models and the 1D models \citep[for further discussion see][]{Colletetal:2007}, but the deviations of the scattering cases are very similar to the calculations with the $\mathrm{[Fe/H]}=-3.0$ models (Fig.~\ref{fig:m23D1Dcog} and Fig.~\ref{fig:m13D1Dcog}), as scattering is still important at the shortest wavelengths and for the strongest lines (see Fig.~\ref{fig:m2cog} and Fig.~\ref{fig:m1cog}). At solar metallicity, the 3D$-$1D abundance corrections for the strongest high-excitation \ion{Fe}{II} lines at $3000$\,{\AA} deviate by less than 0.1\,dex when scattering is included (upper left panel of Fig.~\ref{fig:m03D1Dcog}), and they are negligible at longer wavelengths.

The effects of the 3D temperature stratification and 3D structure dominate the 3D$-$1D abundance corrections at all metallicities for all Fe lines with $\lambda\ge4000$\,{\AA} and for molecular lines, but scattering radiative transfer nevertheless leads to significant deviations for the strongest Fe lines at 3000\,{\AA}. It is not clear whether this removes or adds trends with wavelength in the 3D and 1D line-by-line abundance measurements for a given species, since the magnitude of the deviation depends on line strength and on the microturbulence parameter which is needed to broaden 1D line profiles; detailed stellar abundance measurements will be needed for further insight.

\section{Conclusions}\label{sec:conclusion}

We investigate the effects of a scattering background continuum on spectral line formation based on 3D hydrodynamical and 1D hydrostatic model atmospheres of red giant stars with different metallicity. Continuum scattering is treated in the coherent and isotropic approximation; opacities assume LTE. We compute radiative transfer using the \texttt{SCATE} line formation code to determine synthetic continuum flux levels and flux profiles of fictitious \ion{Fe}{I} and \ion{Fe}{II} lines between $3000$\,{\AA}\,$\le\lambda\le5000$\,{\AA}, as well as selected fictitious CH, NH and OH molecular transitions.

Rayleigh scattering contributes significant opacity in the blue and UV wavelength bands, allowing photons to escape from larger atmospheric depths compared to the case of treating scattering as absorption. As a consequence, the continuum surface flux increases above 2000\,{\AA} and below 5000\,{\AA}, while scattering is negligible in the infrared and at even shorter wavelengths, where photoionization processes thermalize the radiation field. In the 3D models, the strongest relative brightness gain is observed in intergranular lanes, where scattering opacity is more important and photon destruction probabilities near the optical surface are smaller compared with the hot granules.

Continuum scattering affects the strength of LTE lines, depending on ionization stage, excitation potential, wavelength and metallicity. If a line contributes significant absorption opacity in continuum-forming layers, the thermalization depth near line core frequencies moves outward into cooler layers and the core gains less brightness through scattering than the continuum. Normalization of the flux profiles translates this disproportional intensity gain into a deepening of the line. We find the largest effects at the lowest metallicity, the shortest wavelengths, for singly ionized lines and for higher excitation levels, where continuum scattering is strongest and line opacity is significant in continuum-forming layers. The temperature contrast across the surface of 3D model atmospheres results in differential line growth, as intergranular lanes experience a stronger scattering effect than granules. Doppler-shifts reverse sign between granules and intergranular lanes, translating spatial variation into wavelength space. The red wing of the lines appears thus more strongly deepened than the blue wing, and the line center shifts slightly towards longer wavelengths.

We quantify the importance of continuum scattering in 3D by comparing synthetic curves of growth using scattering radiative transfer and treating scattering as absorption. Increasing line strength through scattering desaturates line profiles, leading to a growing deviation between the curves of growth for stronger lines. Transformed into a 3D$-$3D abundance correction for given equivalent width, we find up to $\approx-0.5$\,dex deviation for \ion{Fe}{II} with $\chi=2$\,eV at 3000\,{\AA} and $\mathrm{[Fe/H]}=-3.0$. At 4000\,{\AA}, the effects of scattering on line abundances are much smaller and only significant for the strongest lines; at 5000\,{\AA}, scattering can be neglected. Transitions of the CH, NH and OH molecules behave in an analogous way to neutral low-excitation atomic lines; the strongest 3D$-$3D abundance corrections appear at the shortest wavelengths. Scattering effects weaken towards higher metallicity as continuous absorption opacity becomes increasingly important; 3D$-$3D abundance corrections at solar metallicity ($\mathrm{[Fe/H]}=0.0$) only reach $-0.1$\,dex for the strongest high-excitation \ion{Fe}{II} lines at 3000\,{\AA}, while they are practically negligible for weak lines and at longer wavelengths.

The importance of velocity fields for line formation in the given wavelength range is tested by comparing the results with a 3D atmosphere with zero velocities and artificial microturbulent line broadening. In the absence of Doppler-shifts through the granulation flow, the line profiles become symmetric and saturate earlier for smaller microturbulence. The desaturating effect of scattering is weaker as granules dominate the emission, where photon destruction probabilities are larger, resulting in smaller abundance corrections. We repeat the curve of growth computation for 1D hydrostatic \texttt{MARCS} models with the same stellar parameters as the 3D models and find that scattering leads to a significant downward adjustment of the 3D$-$1D abundance corrections for the strongest Fe lines at the shortest wavelengths compared to 3D$-$1D corrections that were computed treating scattering as absorption; the deviation reaches $-0.35$\,dex for high-excitation \ion{Fe}{II} lines at 3000\,{\AA}, $\mathrm{[Fe/H]}=-3.0$ and microturbulence $\xi=2.0$\,km\,s$^{-1}$. Weak lines are much less affected, as 3D$-$3D and 1D$-$1D scattering abundance corrections are very similar. The scattering corrections for molecules are small compared to the 3D effects as they remain below $0.1$\,dex in all cases; our absolute 3D$-$1D corrections are offset from the results of \citet{Colletetal:2007} as we use fixed molecular equilibrium populations.

Chemical abundance analyses of metal-poor giant stars that include spectral lines in the UV and blue regions should take background continuum scattering into account. Convective velocity fields change the profile shapes, which shift towards slightly longer wavelengths and produce larger scattering abundance corrections for stronger lines compared to 1D hydrostatic models; it is therefore important to conduct abundance analyses based on 3D hydrodynamical model atmospheres.

Our investigation assumes coherent isotropic scattering, neglecting redistribution of radiation through thermal Doppler-shifts, which should only have a small effect on the synthetic line profiles. The LTE approximation is a more severe limitation: non-LTE effects, such as photoionization, are known to be important in metal-poor stars and need to be considered for a quantitative analysis of stellar line profiles \citep[for a discussion, see][]{Asplund:2005}.

\acknowledgements{
The authors would like to thank R. Trampedach and T. M. D. Pereira for their contributions to the \texttt{SCATE} line formation code, opacity tables and atomic data sets. M. Carlsson and P. S. Barklem are thanked for their advice and fruitful discussions. The research leading to these results has received funding from the European Research Council under the European Community's Seventh Framework Programme (FP7/2007-2013 Grant Agreement no. 247060).
}

\bibliographystyle{aa}
\bibliography{linfor}

\afterpage{\clearpage}

\appendix

\section{The \texttt{SCATE} code}\label{sec:code}

The \texttt{SCATE} spectral line formation code for stellar atmospheres computes 3D radiative transfer for blended spectral lines and a background continuum with coherent isotropic scattering. Local thermodynamic equilibrium (LTE) is assumed to compute level populations for lines and continuous opacity sources, using Saha-Boltzmann equilibrium calculations. A lookup-table is used for monochromatic continuum opacities, continuum photon destruction probabilities, Saha ionization equilibria, molecular equilibria (for hydrogen, carbon, nitrogen and oxygen) and partition functions to speed up the 3D line formation computations, in particular when a time-series of profiles is required.

\subsection{Numerical solution of the radiative transfer equation}

Computation of a line profile is divided into two steps: a short-characteristics-based radiative transfer solver with Gauss-Seidel-type convergence acceleration \citep{TrujilloBuenoetal:1995} computes the monochromatic radiation field at each frequency point $\nu$ across the profile to obtain the continuum source function with coherent scattering. The Gauss-Seidel method uses an upper/lower triangular approximate $\Lambda^{*}$ operator, which computes source function corrections during the formal solution and yields fast convergence; see \citet{Hayeketal:2010} for a detailed description of the implementation. The solver takes the anisotropy of the combined continuum and line source function in the observer's frame into account, which is due to the angle-frequency coupling of line opacity through Doppler-shifts. Linear or higher-order local interpolation is available for translating opacities, source functions and upwind intensities onto the characteristics grid. The discretized source function integral is computed using linear interpolation or second-order interpolation, which is required at high optical depths to correctly reproduce the diffusion approximation.

A second solver delivers angle-resolved surface intensities and surface fluxes for obtaining the line profiles. \texttt{SCATE} offers the differential \citet{Feautrier:1964} scheme as well as an integral scheme which is based on the formal solution of the transfer equation \citep{Olsonetal:1987}. The integral scheme computes the source function integral using the same interpolation methods as the scattering solver. The radiative transfer equation is solved on long characteristics that begin in the top layer of the hydrodynamical mesh and span across the simulation domain, until they reach the optically thick diffusion region. Before each calculation of surface intensities, the 3D model atmosphere is tilted into the ray direction using local cubic interpolation.

Both solution steps are parallelized over angle using OpenMP directives for shared-memory architectures, allowing for fine resolution of the angle-frequency coupling in the scattering computation and for integrating surface intensities over a large number of rays when radiative fluxes are needed. A modern 8-core cluster node delivers a line profile with scattering (spatial resolution $120\times120\times230$ points, frequency resolution 40 points, angular resolution for the scattering solver 24 ray directions, angular resolution for flux profile computation 16 ray directions) in $\approx15$\,min per atmosphere snapshot, while the radiative transfer problem without scattering is solved within $<1$\,min.

\texttt{SCATE} uses partial grid refinement on the vertical axis to improve the resolution of continuum optical surfaces. Additional layers are automatically inserted into the 3D model where vertical optical depth steps are large using cubic spline interpolation. The vertical axis is truncated when the optically thick diffusion region is reached and the radiation field is entirely thermal.

\subsection{Line opacities in the LTE approximation}

The line opacity at frequency $\nu'$ in the local rest frame of a gas parcel in the stellar atmosphere is computed using the expression
\begin{equation}
\chi_{\nu'}=\frac{\pi e^{2}}{m_{\mathrm{e}}c}f_{\mathrm{lu}}n_{\mathrm{l}}\left[1-e^{-\frac{h\nu'}{kT}}\right]\psi\left(\nu'-\nu_{0}'\right),
\label{eqn:LTEprofile}
\end{equation}
with the electron charge $e$, the electron mass $m_{\mathrm{e}}$, the speed of light $c$, the Planck constant $h$, the Boltzmann constant $k$ and the temperature $T$. Literature values of the oscillator strength $f_{\mathrm{lu}}$ are usually combined with the statistical weight $g_{\mathrm{l}}$ of the lower level of the transition, which enters the opacity through Boltzmann excitation factors for the level population density $n_{\mathrm{l}}$. The code computes Saha-Boltzmann equilibria with tabulated partition functions that were calculated using the NIST database \citep{NIST}. Molecule formation is taken into account for 12 important species by computing LTE equilibrium populations (see Table~\ref{tab:moldata}). The term in brackets in Eq.~(\ref{eqn:LTEprofile}) corrects the line opacity for stimulated emission, assuming atomic level populations in LTE.

The line profile $\psi$ is given by a Voigt function around the laboratory wavelength $\nu'_{0}$ of the transition. Doppler-shifts through macroscopic velocity fields $\mathbf{u}$ in the observer's frame are included in the non-relativistic approximation, which yields the line profile
\begin{equation}
\psi\left(\nu-\nu_{0}\right)=\psi\left(\nu-\nu_{0}'-\nu_{0}'\frac{\vec{\hat{n}\cdot u}}{c}\right)
\end{equation}
in the observer's frame, for a ray in direction $\vec{\hat{n}}$ and gas velocity $\vec{u}$. The vectorizable recipe of \citet{Huietal:1978} is used to evaluate the Voigt profile numerically.

Line profiles include Doppler broadening to account for shifts through random thermal gas motion. Microturbulent broadening may be included in the computations, but it is generally unnecessary for line formation with 3D hydrodynamical atmospheres \citep[see the discussion in][]{Asplundetal:2000b}.

Natural line broadening through radiative de-excitation is set by the profile width parameter $\Gamma^{\mathrm{rad}}$ using laboratory data if available, or it may be estimated using the expression
\begin{equation}
\Gamma^{\mathrm{rad}}\approx\frac{8\pi^{2}e^{2}}{m_{\mathrm{e}}c}\frac{1}{\lambda^{2}}\frac{g_{\mathrm{l}}}{g_{\mathrm{u}}}f_{\mathrm{lu}},
\end{equation}
which assumes a two level atom; $\lambda$ is the transition wavelength and $g_{\mathrm{u}}$ is the statistical weight of the upper level.

Pressure broadening through collisions of neutral atoms and ions with neutral hydrogen atoms is taken into account in the impact approximation. Profile widths are based on quantum mechanical calculations by \citet{Ansteeetal:1995,Barklemetal:1997,Barklemetal:1998,Barklemetal:2000a,Barklemetal:2001,Barklemetal:2005}. Collisional cross-sections are estimated using the power law
\begin{equation}
\sigma(v)=\sigma(v_{0})\left(\frac{v}{v_{0}}\right)^{-\alpha},
\end{equation}
where $v$ is the relative velocity of the particle and the perturbing \ion{H}{I} atom, $\sigma(v_{0})$ and $\alpha$ are tabulated, and $v_{0}$ is a reference velocity. The width parameter of the Lorentzian broadening profile is obtained by integrating over a Maxwell-Boltzmann distribution for $v$, which yields
\begin{equation}
\gamma^{\mathrm{H\,I}}=2\left(\frac{4}{\pi}\right)^{\frac{\alpha}{2}}\Gamma\left(\frac{4-\alpha}{2}\right)\bar{v}\sigma(v_{0})\left(\frac{\bar{v}}{v_{0}}\right)^{-\alpha}n_{\mathrm{H\,I}},
\end{equation}
with the gamma function $\Gamma$, the number density $n_{\mathrm{H\,I}}$ of neutral hydrogen, and the average velocity $\bar{v}=\sqrt{8kT/\pi\mu}$, where $\mu$ is the reduced mass of the particle and the perturbing \ion{H}{I} atom. The contribution from \ion{He}{I} atoms to pressure broadening is estimated through
\begin{equation}
\gamma^{\mathrm{H\,I}+\mathrm{He\,I}}\approx\gamma^{\mathrm{H\,I}}\left(1.0+c_{\mathrm{He\,I}}\frac{n_{\mathrm{He\,I}}}{n_{\mathrm{H\,I}}}\right),
\end{equation}
using an approximate scaling factor $c_{\mathrm{He\,I}}\approx0.41$. In the absence of quantum mechanical data, the classical \citet{Unsoeld:1955} approximation is used in the formulation found in \citet{Gray:2005}:
\begin{equation}
\log\gamma^{\mathrm{H\,I}}=20.0+0.4\log C_{6}+\log P-0.7\log T,
\end{equation}
for quantities given in cgs units; $P$ is the gas pressure. The interaction constant $C_{6}$ is approximated through the expression
\begin{equation}
C_{6}=0.3\cdot10^{-30}\left[\left(\chi_{\mathrm{u}}^{\mathrm{ion}}\right)^{-2}-\left(\chi_{\mathrm{l}}^{\mathrm{ion}}\right)^{-2}\right],
\end{equation}
with the ionization potentials $\chi_{\mathrm{u}}^{\mathrm{ion}}$ and $\chi_{\mathrm{l}}^{\mathrm{ion}}$ from the upper and lower level of the transition, both given in eV. For the more complicated case of hydrogen lines, broadening recipes of \citet{Barklemetal:2000b} are available.

\begin{table}[htdp]
\caption{Molecular data included in the equilibrium calculations; partition functions were computed using the NIST database \citep{NIST}.}
\centering
\begin{tabular}{ll}
\hline\hline
Molecule & Dissociation energy\\
\hline
H$_{2}$ & \citet{Sauvaletal:1984}\\
H$_{2}^{+}$ & \citet{Sauvaletal:1984}\\
H$_{2}$O & \citet{Ruscicetal:2002}\\
CH & \citet{Sauvaletal:1984}\\
C$_{2}$ & \citet{Urdahletal:1991}\\
CN & \citet{Huangetal:1992}\\
CO & \citet{Eidelsbergetal:1987}\\
OH & \citet{Ruscicetal:2002}\\
O$_{2}$ & \citet{Sauvaletal:1984}\\
NH & \citet{Marquetteetal:1988}\\
N$_{2}$ & \citet{Sauvaletal:1984}\\
NO & \citet{Sauvaletal:1984}\\
\hline
\end{tabular}
\label{tab:moldata}
\end{table}

\end{document}